\documentclass[journal]{IEEEtran}

\usepackage{graphicx}
\usepackage{longtable}
\usepackage{subcaption}
\usepackage{tikz}
\usepackage{ctable}
\usepackage{cite}
\usepackage[utf8]{inputenc}
\usepackage{blindtext}
\usepackage{color}
\usepackage{colortbl}
\usepackage{textgreek}
\usepackage{algorithm}
\usepackage[T1]{fontenc}
\usepackage{algpseudocode}

\usepackage{fancyhdr}
\usepackage{kantlipsum}

\definecolor{Gray}{gray}{0.9}
\definecolor{LightCyan}{rgb}{0.88,1,1}

\iffalse

\else

\fi

\begin{document}

\widowpenalty=10000
\clubpenalty=10000

\title{Survey on Enterprise Internet-of-Things Systems (E-IoT): A Security Perspective}

\author{
    \IEEEauthorblockN{Luis Puche Rondon, Leonardo Babun, Ahmet Aris, Kemal Akkaya, and A. Selcuk Uluagac\\}
    \IEEEauthorblockA{Cyber Physical Systems Security Lab\\
    Department of Electrical and Computer Engineering\\
    Florida International University, Miami, Florida\\
    Email:\{lpuch002, lbabu002, aaris, kakkaya, suluagac\}@fiu.edu}
}

\maketitle


\IEEEpeerreviewmaketitle

\newcommand{\pie}[1]{%
\begin{tikzpicture}
 \draw (0,0) circle (1ex);\fill (1ex,0) arc (0:#1:1ex) -- (0,0) -- cycle;
\end{tikzpicture}%
}

\ifCLASSOPTIONcaptionsoff
  \newpage
\fi

\begin{abstract}

As technology becomes more widely available, millions of users worldwide have installed some form of smart device in their homes or workplaces. These devices are often off-the-shelf commodity systems, such as Google Home or Samsung SmartThings, that are installed by end-users looking to automate a small deployment. In contrast to these ``plug-and-play'' systems, purpose-built Enterprise Internet-of-Things (E-IoT) systems such as Crestron, Control4, RTI, Savant offer a smart solution for more sophisticated applications (e.g., complete lighting control, A/V management, security). In contrast to commodity systems, E-IoT systems are usually closed source, costly, require certified installers, and are overall more robust for their use cases. Due to this, E-IoT systems are often found in expensive smart homes, government and academic conference rooms, yachts, and smart private offices. However, while there has been plenty of research on the topic of commodity systems, no current study exists that provides a complete picture of E-IoT systems, their components, and relevant threats. As such, lack of knowledge of E-IoT system threats, coupled with the cost of E-IoT systems has led many to assume that E-IoT systems are secure. To address this research gap, raise awareness on E-IoT security, and motivate further research, this work emphasizes E-IoT system components, E-IoT vulnerabilities, solutions, and their security implications. In order to systematically analyze the security of E-IoT systems, we divide E-IoT systems into four layers: E-IoT Devices Layer, Communications Layer, Monitoring and Applications Layer, and Business Layer. We survey attacks and defense mechanisms, considering the E-IoT components at each layer and the associated threats. In addition, we present key observations in state-of-the-art E-IoT security and provide a list of open research problems that need further research. 

\end{abstract}

\section{Introduction}\label{sec:introduction}

\begin{figure}
        \centering
        \begin{subfigure}[b]{0.23\textwidth}
            \centering
            \includegraphics[width=\textwidth]{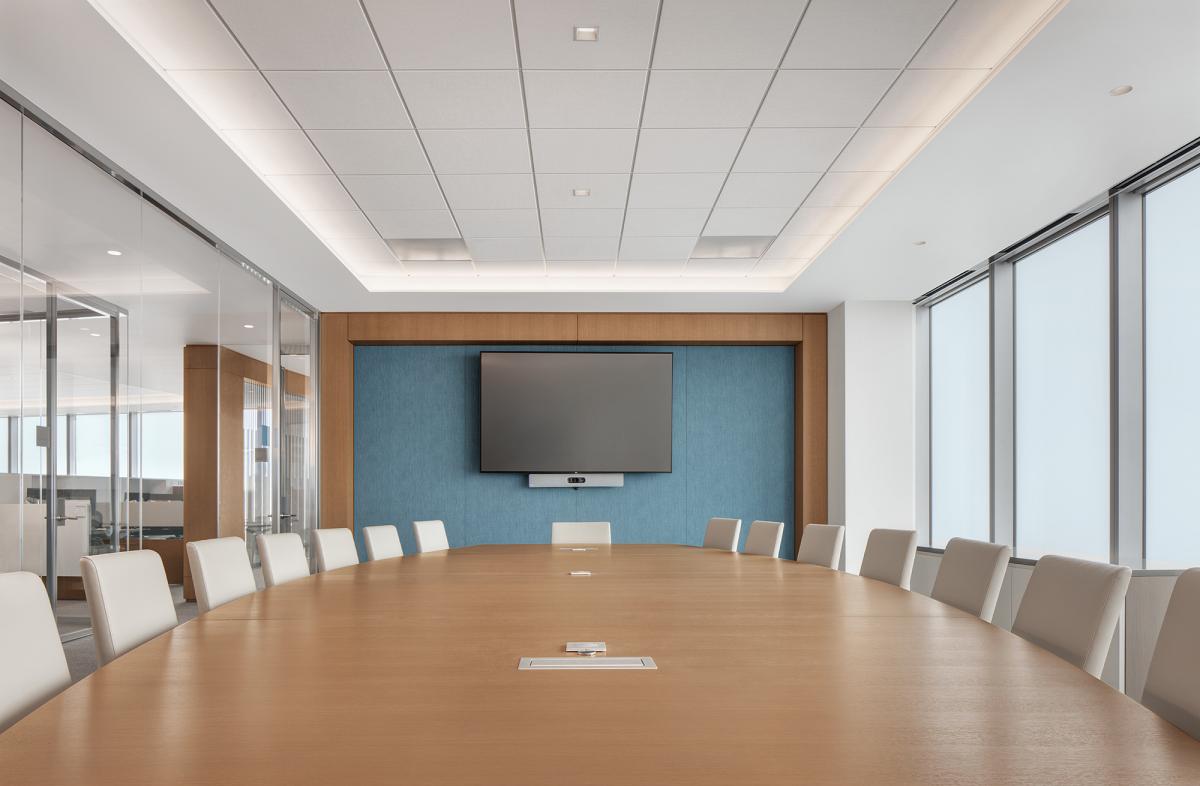}
            \caption{Conference room with multiple displays and E-IoT system for control.}
            \label{fig:locations_1}
        \end{subfigure}
        \hfill
        \begin{subfigure}[b]{0.23\textwidth}  
            \centering 
            \includegraphics[width=\textwidth]{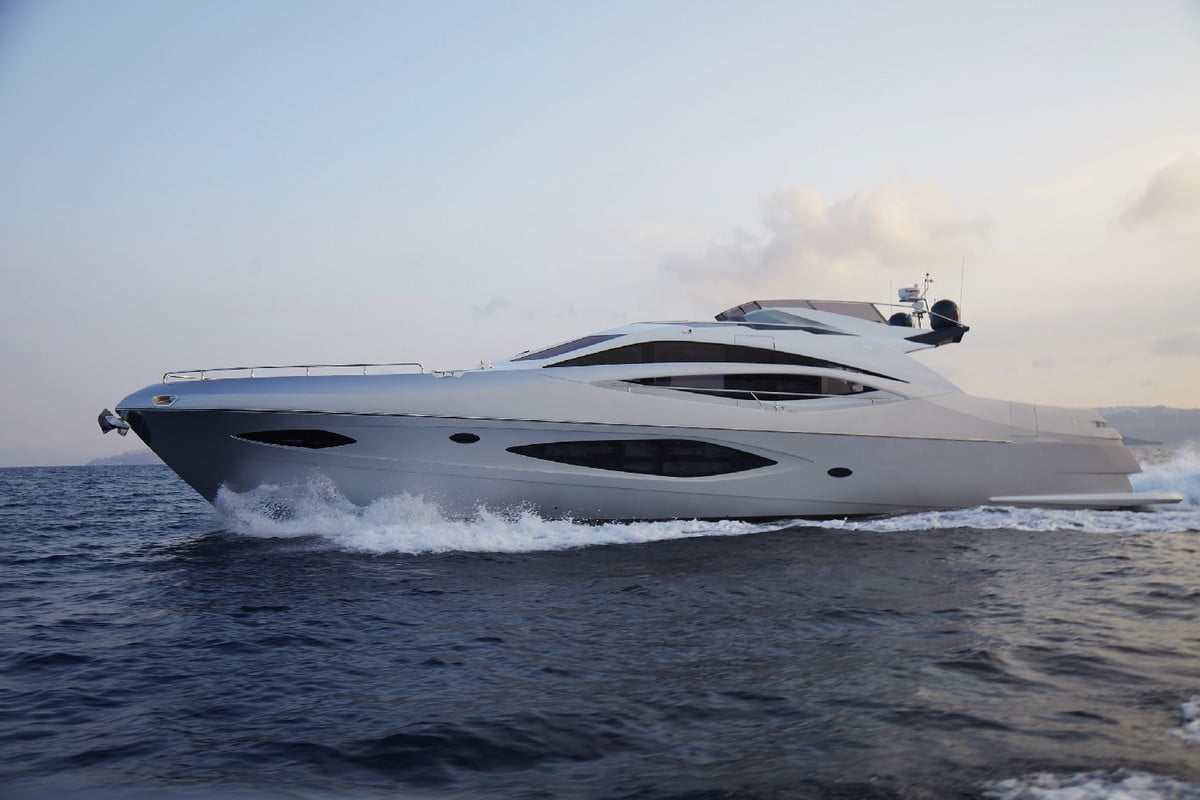}
            \caption{Yacht installations which use E-IoT smart systems to control lights and A/V.}
            \label{fig:locations_2}
        \end{subfigure}
        \vskip\baselineskip
        \begin{subfigure}[b]{0.23\textwidth}   
            \centering 
            \includegraphics[width=\textwidth]{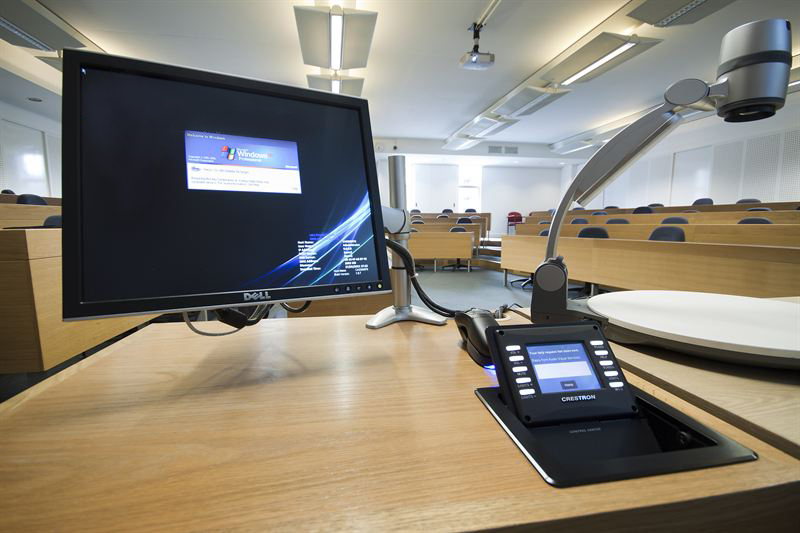}
            \caption{Smart classrooms to control all A/V equipment from a touchscreen.}
            \label{fig:locations_3}
        \end{subfigure}
        \hfill
        \begin{subfigure}[b]{0.23\textwidth}   
            \centering 
            \includegraphics[width=\textwidth]{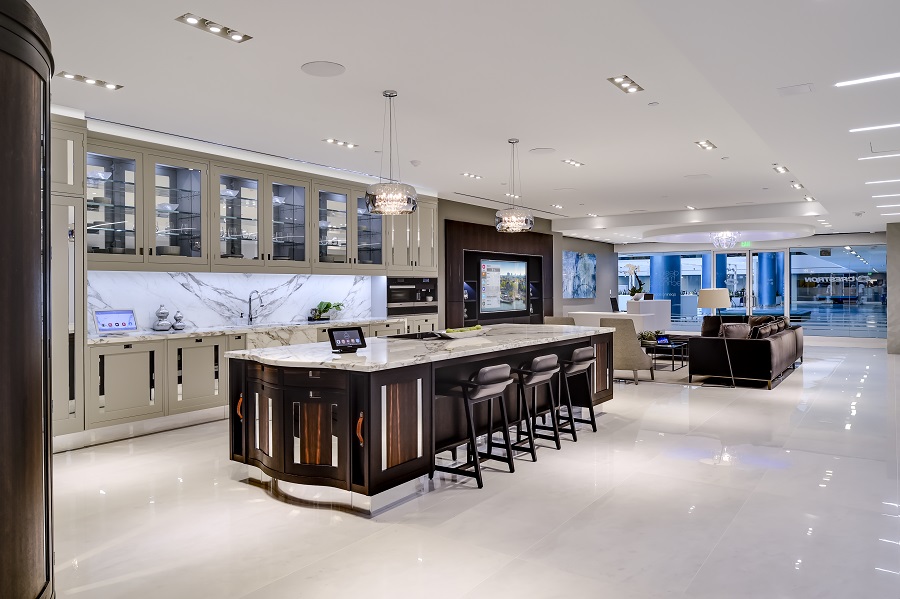}
            \caption{E-IoT homes, which usually include home theaters and smart systems.}
            \label{fig:locations_4}
        \end{subfigure}
        \vspace{-0.15in}
        \caption{Use cases of E-IoT systems, in more specialized applications.}
        \vspace{-0.3in}
        \label{fig:locations}
\end{figure}

The introduction of modern smart consumer electronics has led to the widespread adoption of smart devices, with over 45 million smart home components sold worldwide \cite{magazine, SmartHomesUSEurope}. Most users are familiar with commodity systems, off-the-shelf smart systems that are easily installed by the average end-user without specialized training (e.g., Samsung SmartThings, Google Home) \cite{SmartTS, iotdots}. However, in more complex installations, where robust, secure, and reliable smart solutions are needed, Enterprise Internet-of-Things (E-IoT) systems (e.g, Crestron, Control4, Savant, RTI) are accepted solutions. In contrast to commodity systems, E-IoT offers customized deployments, with more use-cases and applications. Offering users a broad set of compatible devices devices (e.g., sensors, Audio/Video equipment, interfaces), protocols (e.g., Zigbee, Z-wave, IP, proprietary protocols), custom programmed behavior, and system User Interface (UI) customization. As such, E-IoT systems are found in locations such as smart offices, smart buildings, luxury smart homes, yachts, and secure conference rooms (as illustrated in Figure \ref{fig:locations}).

While the security of many emerging commodity systems is well-understood due to prior research and mainstream knowledge, the security of E-IoT systems has been largely overlooked \cite{patent, patent2, smartgridjournal, usbposter, daint, denney2019usb, lopez2017survey, convolution, aegis, saint, smarthomesidechannel, acar2018peekaboo, waka, contextawaresensor, sikder2018survey, 6997498, saint-taint-analysis, iqtidar1, iqtidar2, iqtidar3, berkaymagazine, kratos, ziot, madiot, usbjournal, newaz2020adversarial, newaz2020survey, newaz2020heka, newaz2019healthguard}. As such, the lack of research and awareness coupled with the cost of devices and installation of E-IoT has led many users to mistakenly assume that E-IoT systems are completely secure. As E-IoT systems follow a unique design with specialty devices, proprietary software, and a large number of compatible protocols, there is a need to research unique threats and security of E-IoT systems. Further, E-IoT systems have been increasingly popular in smart installations, with Crestron growing to 1.5 billion dollars of annual revenue in 2018 and Control4 deploying over 15 million smart products in over 400,000 installations worldwide \cite{crestronorigins, control4about}. With many of these systems present in high-profile locations, understanding threats and defense strategies for E-IoT systems should be of great importance. However, no survey focuses on E-IoT system components, attacks, threats, and relevant defenses of E-IoT systems. We believe that this research gap in the literature is notable considering the prevalence of E-IoT deployments and ever-increasing attacks against smart systems. To address this research gap and analyze the security of E-IoT systems, we first divide E-IoT into four distinct layers: E-IoT Devices Layer, Communications Layer, Monitoring and Applications Layer, and Business Layer. As such, we consider E-IoT components at each layer, the associated threats, attacks, and defense mechanisms. Additionally, we present key observations in E-IoT security and provide a list of open issues that require further research. To the best of our knowledge, this is the first survey focusing solely on E-IoT security and proving a comprehensive review of threats, attacks, and defenses. With this work, we aim to raise awareness on E-IoT system security and motivate further research in this topic.

Although there are existing studies on IoT systems, this survey focuses solely on relevant threats and solutions to E-IoT systems. This study aims to provide users with adequate information on E-IoT system components, vulnerabilities, attacks, and defenses. With this work, we also aim to encourage further research and development from the research community on the topic of E-IoT systems. For instance, our survey highlights widely-used E-IoT proprietary technologies that have seen no security scrutiny and thus have relied on security through obscurity for decades. This survey may be valuable to researchers, E-IoT vendors, users, installers, and manufacturers that want to improve their security practices. Further, users who do not know about E-IoT concepts may find this study a beneficial resource. Ultimately, this survey sheds light on the security implications of E-IoT systems and raises awareness of security practices, protocols, and viable threats against E-IoT systems.

\textbf{Summary of Contributions: } The contributions of this work are as follows:

\begin{itemize}
    \item We highlight popular E-IoT system platforms and identify security challenges in these systems.
    \item We categorize and analyze E-IoT components, threats, attacks, and defenses by dividing E-IoT systems into four distinct layers.
    \item We present the need for further research in E-IoT systems and a number of proprietary technologies used in E-IoT. 
    \item We open discussion on the security of E-IoT systems, and related defense mechanisms.
\end{itemize}

\textbf{Organization:} This work is structured as follows, we begin with the background information of E-IoT, relevant protocols, and the E-IoT layers in Section \ref{sec:background}. Section \ref{sec:scope} summarizes the scope of this survey. In Section \ref{sec:deviceslayer} we cover the E-IoT devices layer threat taxonomy, vulnerabilities, and defenses. Similarly, in Section \ref{sec:communicationlayer} we address the communication layer, Section \ref{sec:applicationslayer} addresses the monitoring and applications layer, and Section \ref{sec:businesslayer} covers the business layer of E-IoT systems. In Section \ref{sec:lessonslearned} we highlight the lessons learned from this work and open issues. Related work is presented in Section \ref{sec:relatedwork}. Finally, we conclude the survey in Section \ref{sec:conclusion}.

\section{Background}\label{sec:background}

In this section, we highlight background information of E-IoT systems and the layered architecture of E-IoT systems.

\begin{figure}[t]
\centering{\includegraphics[width=0.40 \textwidth]{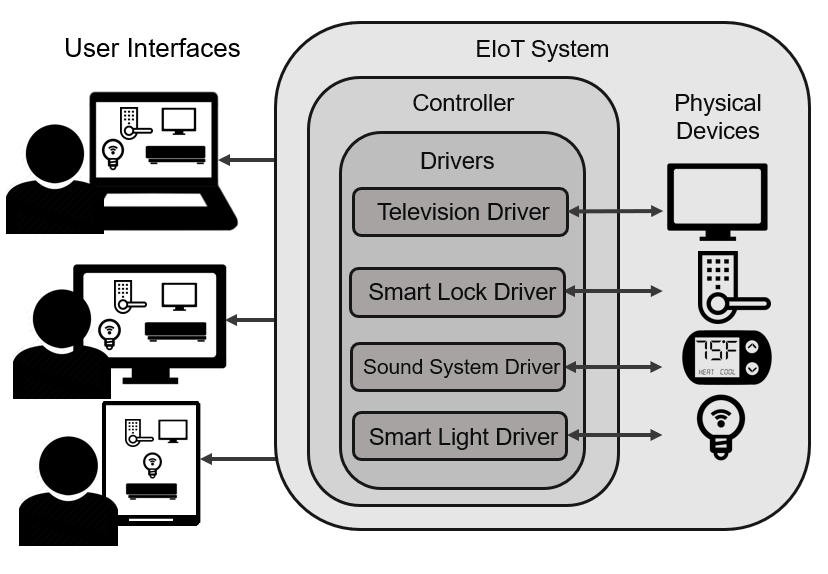}}
    \caption{Architecture of a typical E-IoT solution with user interfaces, controller, and physical devices.}
\label{fig:architecture}
\vspace{-0.2in}
\end{figure}

\subsection{Enterprise Internet of Things (E-IoT)}

We use the term \textit{E-IoT solution} to describe a fully-functioning E-IoT system deployment. Figure \ref{fig:architecture} depicts the general architecture of an E-IoT solution. E-IoT systems have unique design and deployment practices that discriminate them from regular consumer IoT systems. In its most basic form, the E-IoT solution contains four core components: the physical devices, the controller, user interfaces, and drivers. As all installations are custom-made, E-IoT system deployments vary from system to system. The first component of E-IoT systems is the \textit{physical devices}, which include any device integrated into the central system (e.g., sensors, televisions, lighting modules). To integrate physical devices, E-IoT systems use \textit{drivers}, which provide the system with all the necessary information to integrate a device to an E-IoT system. Drivers contain information such as model number, protocol type, code, commands, and physical connections. Each device requires a driver to be integrated. In E-IoT systems, the \textit{controller} serves as the central processing unit and stores all the drivers as well as user-specific custom programming required for the E-IoT system (e.g., scheduled events). Finally, \textit{user interfaces} serve as the main point of interaction between users and the E-IoT system. After any third-party devices are integrated, the end-user can use user interfaces such as tablets, phones, and remotes to control integrated devices. For instance, if an E-IoT user wants to turn a light on, he/she may use a phone app as the interface to communicate with the controller. The controller then uses a smart light driver of an integrated E-IoT light to toggle the light at a user's request. As such, with any E-IoT actions many components are involved (e.g., hardware, networking, drivers, proprietary, wireless). 

\begin{figure}[t]
\centering{\includegraphics[width=0.44\textwidth]{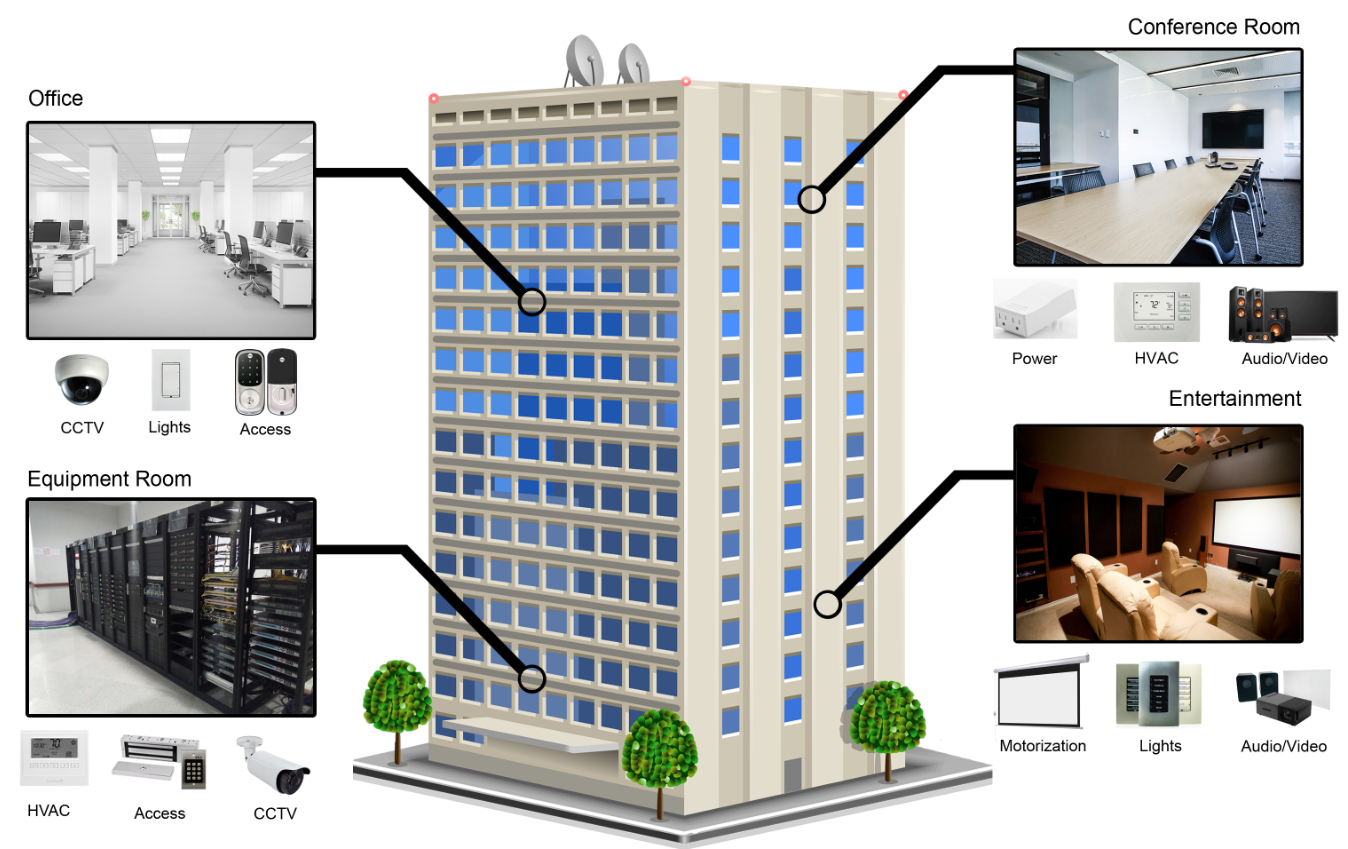}}
    \caption{Smart building with use-cases in different locations.}
\vspace{-0.3in}
\label{fig:smartbuilding}
\end{figure}

As designed, E-IoT systems may fulfill different purposes. One such purpose is specialization, such as centralized lighting control systems designed to control electrical loads in locations such as yachts or offices \cite{centralizedlighting, doelightingsystems}. Another purpose of E-IoT systems is integrating previously separate components into a smart system (e.g., Savant, Crestron, and Control4); components integrated can then work together and interact as a single system \cite{smartsystems}. For instance, integrating an alarm system with a lighting system allows a use case such as turning off all the lights when the alarm is activated. As such, E-IoT systems require trained installation and come at a higher cost than standard off-the-shelf systems. This added functionality over commodity systems has led E-IoT systems to become popular in expensive locations such as yachts, classrooms, smart offices, conference rooms, and luxury smart homes as shown in Figure \ref{fig:smartbuilding}. Further, the installation of an E-IoT system is done by an \textit{integrator}, a certified installer that performs the physical and software configuration for such a system. The configuration process requires specialized training and tools, which are provided by the system vendor to the integrator \cite{crestrondeploy, control4deploy}. However, hardware and software (e.g., integrated devices and drivers) used in E-IoT may also come from unverified third-party vendors and sources \cite{blackwiredrivers}.

\subsection{Consumer IoT vs. E-IoT}

As commodity IoT smart systems have some limitations (e.g., scale, compatibility), E-IoT offers a solution for complex and reliable deployments. In this subsection we highlight the differences and benefits of E-IoT and why E-IoT solutions are chosen over commodity IoT. As such, E-IoT has some unique security concerns and threats. We outline some of these differences in Table \ref{tbl:iotdiff}.

\noindent\textbf{Compatibility.} As smart systems grow in scale, a user must determine the best solution to easily control many different devices. While commodity systems are limited in scale and compatible products, there are fewer limitations on what can be integrated into E-IoT. As E-IoT vendors offer components such as drivers, which are used to integrate third-party devices with E-IoT systems, many third-party devices are compatible with E-IoT systems. However, from a security standpoint, broad support of protocols can pose a threat as an attacker may be able to attack through many available protocols. This is true as more diverse systems have more possible points of failure.

\noindent\textbf{Complexity.} Commodity smart systems are designed to handle small deployments of IoT devices. While this use case is sufficient for most consumers, commodity smart systems are not a viable solution for large, complex deployments. For instance, multi-room video and audio distribution is one of the more complex applications of E-IoT. With audio/video switchers that can control up to 164 inputs and outputs, E-IoT becomes a reliable way to manage large systems and deployments. 

E-IoT systems also allow for a high degree of flexibility and customization. A number of protocols and modes of communication are supported with drivers and expandable hardware components \cite{DriverCentral}. As a result, E-IoT can integrate more devices than consumer systems. All in all, the unprecedented level of complexity can mean that more vulnerabilities may occur at more stages and sectors of the E-IoT system in comparison to commodity IoT systems.

\begin{table*}[t]
    \centering
    \caption{Commodity IoT vs E-IoT Solutions.}
    \scalebox{1.00}{
    \begin{tabular}{c|c}
    \hline
       \textbf{Commodity IoT Solutions} & \textbf{E-IoT Solutions}\\ 
       \hline
        Simpler, easily deployable solutions & More complex, diverse smart solutions \\
        \rowcolor{Gray}
        Less compatible, approved devices & More compatible 3rd-party devices\\
        Lower cost of installation and maintenance &  High cost of installation, maintenance, and programming\\
        \rowcolor{Gray}
        User-deployed and maintained smart systems & Installer-deployed and maintained smart systems \\
        More often open-sourced, with technical documentation available publicly & Closed-source systems, with no technical documentation available\\
        \rowcolor{Gray}
        Often cannot be deployed as completely offline systems & Must be deployed as always-offline systems in some use-cases\\
    \hline
    \end{tabular}}
     \vspace{-0.1in}
    \label{tbl:iotdiff}
\end{table*}

\noindent\textbf{Delegation.} As the installation of E-IoT components is often complex; many users opt to have installation and maintenance of E-IoT systems delegated to a dedicated contractor. As such, in a similar manner to electricians, plumbers, and other specialists, E-IoT integrators are contracted only for the installation and maintenance of E-IoT systems. In effect, the end-user does not need to understand the technical details of the E-IoT system, the user only needs to know how to operate the system, removing layers of complexity for any visitors. The delegation of installation and maintenance of E-IoT means that in addition to technical expertise, integrators must consider the security aspects of E-IoT systems. Thus, clients depend on their hired integrators for the security of their systems. As such, if an integrator is careless, or does not keep security in mind, the E-IoT system will be insecure without the owner's knowledge.

\noindent\textbf{Offline Systems.} Some smart systems (e.g., Google home, Samsung SmartThings) have inherent limitations in their design as they rely on a constant Internet connection to function properly. Some E-IoT deployments may need offline systems to operate where an Internet connection is not desirable or too costly. For instance, a secure conference room may want to have all the equipment isolated from all the network, or yacht installations where Internet availability is costly or sporadic. From a security perspective, beyond losing vendor support, isolated systems may not receive security updates and may be prone to exploits if systems are not patched due to limited connectivity. However, it can also protect such isolated E-IoT environments from the attacks originated from the Internet.

\noindent\textbf{Cost.} As E-IoT requires specialized integrators, custom programming, proprietary hardware, and dedicated technical support, the systems come at a higher cost. Further, the physical installation of E-IoT often involves fully rack-mounted, cable-managed systems throughout a building or home. While consumer IoT solutions are designed be affordable by end-users, E-IoT installations may be valued at hundreds of thousands of dollars depending on the complexity \cite{HighEndAutomation}. The high cost of E-IoT systems may lead some users to wrongly assume their systems are secure.

\subsection{Architecture of E-IoT Systems}

\begin{figure}[t]
\centering{\includegraphics[width=0.45 \textwidth]{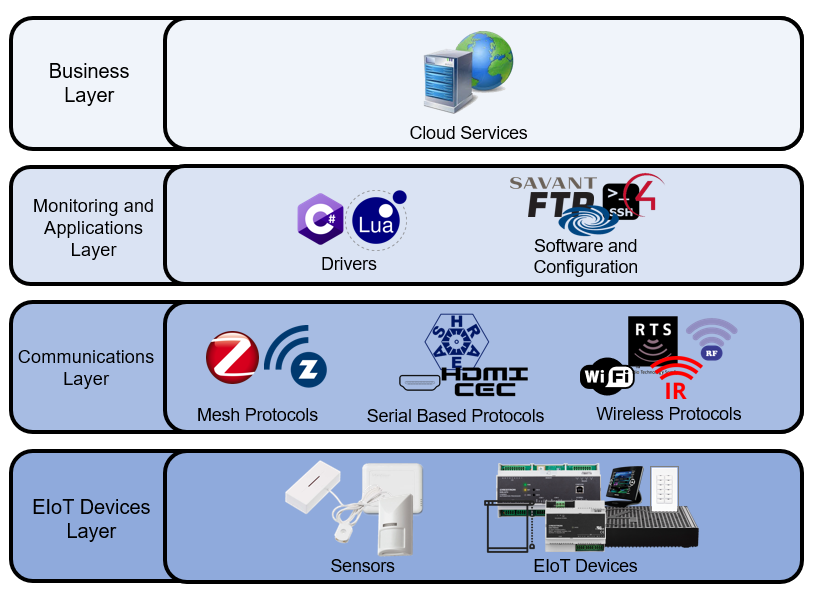}}
    \caption{E-IoT system four-layer model used in this survey.}
\label{fig:layers}
\vspace{-0.2in}
\end{figure}

To better analyze E-IoT solution components and relevant threats, we divide E-IoT systems into separate layers as depicted in Figure \ref{fig:layers}. Specifically, the layered E-IoT solution as described includes four distinct layers: (1) E-IoT Devices Layer, (2) Communications Layer, (3) Monitoring and Applications Layer, and (4) Business Layer. The lowest layer, \textit{E-IoT devices layer} includes the integrated E-IoT devices, physical interfaces used by devices, sensors, and any physical components of E-IoT systems. Next comes the \textit{communication layer}, which possesses all the communication protocols (e.g., open-source and proprietary) used by integrated devices in the E-IoT devices layer. To manage communication, configuration, software, and programmed events in E-IoT systems, the \textit{monitoring and applications layer} contains all software-based components (e.g., drivers, E-IoT applications, and configuration software) of E-IoT systems used by integrators and users. Finally, the \textit{business layer} includes cloud components of an E-IoT system, for instance, remote services or remote storage used by an E-IoT system. The combination of these layers creates a unique technology solution that is highly customizable to any user's need. For instance, with an E-IoT system, a user can configure events such as a good morning timer which simultaneously plays a specific song, opens the shades, and turns on the lights every morning or a panic button to call the police, blare the alarms, and flash all the lights integrated to a system. Additional details on the four layers are as follows: \smallskip
 
\noindent\textbf{E-IoT Devices Layer.} The E-IoT devices layer consists of all physical components of E-IoT systems. A physical component may be physical wiring, sensors, physical interfaces, or connection endpoints. E-IoT systems use many physical devices as part of their systems (e.g., motorized lifts, HVAC, sensors). These devices may be integrated for different applications. In some cases, they may be simply controlled by the system such as motorized projector lifts. Other cases may be external sensors such as water leak sensors to automatically shut off water valves prevent flooding. 
  
\noindent\textbf{Communications Layer.} The communications layer contains all protocols, interfaces, and communication services used by E-IoT systems. This includes protocols in any component of E-IoT systems. To integrate a wide range of smart devices into the central system, E-IoT systems must support a multitude of communication protocols (e.g., Zigbee, Cresnet, Serial) used by smart devices. For instance, to integrate alarm systems to larger E-IoT systems, integrators will often use serial-based adapters, or an available IP interface \cite{integrateserial, integrateIP}.

\noindent\textbf{Monitoring and Applications Layer.} The applications and monitoring layer contains software-based components of an E-IoT system

For instance, E-IoT system configuration, drivers, firmware, or programmable behavior all can be considered part of the application layer. E-IoT systems must have the capability to be customized for every installation. As all deployments may be different and fit for different purposes, custom applications are a large part of E-IoT systems.

\noindent\textbf{Business Layer.} The topmost layer for E-IoT systems is the business layer, which handles all external cloud services used by E-IoT solutions. While not used in all implementations, some E-IoT systems rely on cloud computing and online services for features and integration. For instance, some E-IoT system use-cases require 'always offline' configuration after being deployed (e.g., yachts, remote locations, secure locations). Cloud services provide E-IoT systems with expanded capabilities, remote connections, and other services. For instance, E-IoT systems with CCTV components may use cloud storage services to store video feed in case the local video recorder is damaged or stolen \cite{camio}.

\section{Scope of The Survey}\label{sec:scope}

In this section, we highlight the topics covered in this survey. While covering the topics that are closely related to E-IoT, we do not consider the topics that are not directly related to E-IoT or that are common to general computer systems in the scope of this study.

\noindent\textbf{Scope of E-IoT Devices Layer.} For the E-IoT devices layer section, we cover attacks (e.g., sensory attacks, node theft, battery exhaustion) and defense mechanisms that target components at the E-IoT devices layer. Included in these topics are E-IoT devices, supply chain attacks, and physical access attacks relevant to E-IoT systems. Topics outside of the scope of E-IoT devices layer are attacks on chipsets (e.g., processor side-channel attacks) and other physical devices that are either widely researched or unique to E-IoT. 

\noindent\textbf{Scope of Communications Layer.} This survey covers communication interfaces, publicly-documented protocols, proprietary protocols, and other relevant communication components as part of the communications layer. As such, this survey covers jamming attacks and other well-known attacks against public and proprietary protocols used in E-IoT. Finally, communication protocols such as TCP/IP, cellular communication, long-range radio protocols such as LoRaWAN, and their respective attacks are outside of this survey's scope as they are not common in E-IoT use-cases. 

\noindent\textbf{Scope of Monitoring and Applications Layer.} Topics in the monitoring and applications layer include E-IoT software, configuration, and software services. Topics outside of this layer's scope are operating systems as they are a common topic of research and not exclusive to E-IoT systems. For instance, Linux-based operating systems are common in E-IoT and other smart systems, making Linux a common topic of research. Also, outside of this layer's scope, web-based DDoS attacks, mobile application threats, ransomware, firmware attacks, and common software vulnerabilities.

\noindent\textbf{Scope of E-IoT Business Layer.} Relevant topics to the E-IoT business layer include remote access cloud services, maintenance services, and CCTV data storage. As the E-IoT business layer is not employed by all E-IoT systems, and cloud security is a diverse field, some topics are not covered. Topics outside of this survey's scope are encrypted storage access, computation of stored E-IoT content in cloud environments, online microservices, advanced persistent threats, virtualization technologies, general data storage, and other cloud concepts that are uncommon for E-IoT.

\begin{table*}[t]
    \centering
    \caption{Overview of E-IoT Devices Layer Attacks, Threats, and Mitigations.}
    \scalebox{1.00}{
    \begin{tabular}{c|c|c}
    \hline
       \textbf{Component} & \textbf{Attack/Threat} & \textbf{Mitigations}\\ 
    \hline
        Supply Chain & Supply Chain Attacks \cite{miller2013supply, supchaindi, supchainsw, fuentes2018securing, supchainCSO, yousefnezhad2020security, farooq2019iot, supplychainattacks, gorman2012counterfeit, bhasin2015survey, tehranipoor2010survey, king2008designing, robertson2018big}
            & RFID tagging during manufacturing and transport \cite{yangrfid, yangrfid2,yangrfid3}.\\  
            \rowcolor{Gray}
            & & Novel secure supply change architectures \cite{chamekhsc}.\\
            & & Other supply chain attacks \& defenses are summarized in Table \ref{tbl:supplychaindefenses} \\
    \hline
        \rowcolor{Gray}
        Physical Attacks 
            & Reset Sequence Attacks \cite{crestronreset, control4reset} & Limiting device access, device monitoring, and access control. \\
            & Node Capture Attacks \cite{nodecaptureattack, nodecaptureattackwang, nodecapturepw1, nodecapturepw2, nodecapturepw3, nodecapturepw4, nodecapturepw5, nodecapturepw6} & Limiting device access, device monitoring, access control, and loss prevention practices \cite{lossprevention}. \\ 
    \hline
        \rowcolor{Gray}
        Side and Sensory Channels 
            & EM Attacks \cite{standaert2010introduction, RS232Radiation, hwu1988electromagnetic,van1985electromagnetic, kuhn2004electromagnetic} & Best installation practices, directional antennas \cite{control4DSsecurity}. \\
            & Video Side-Channel \cite{savage2015visualizing, xu2014watching} & Software-based solutions may help mitigate several video-based side-channel attacks \cite{sun2014design, wang2015novel, enck2014taintdroid, wu2014droiddolphin, xu2015semadroid}.\\
            \rowcolor{Gray}
            & Sensory Side-Channel Attacks \cite{schwittmann2017identifying, schwittmann2016video, maiti2019light, zhou2018irexf, guri2016optical, guri2019air, uk2014light, loughry2002information, ronen2016extended} & Solutions to sensory attacks include dedicated machine-learning frameworks \cite{contextawaresensor, sikder2019context, sikder20176thsense}.\\
    \hline
        Battery Exhaustion 
        & Battery Exhaustion Attacks \cite{ShakhovBattery, batterybauernansa, MoyersBattery, MartinBattery}
            & IDS frameworks such as MVP-IDS and B-SIPS \cite{MoyersBattery, bsips}.\\
            \rowcolor{Gray}
            & & IDS-based solutions designed to protect against battery exhaustion \cite{nashintrusion, UpadhyayBattery}.\\
            & & Rate Limiting solutions \cite{hristozovBattery}\\
    \hline
    \end{tabular}}
     \vspace{-0.1in}
    \label{tbl:deviceslayersum}
\end{table*}

\section{E-IoT Devices Layer: Components and Security}\label{sec:deviceslayer}

In this section, we cover the E-IoT Devices layer, threats, defenses, and their implications. First, we introduce components of the E-IoT devices layer, and then cover threats and attacks. Finally, we give an overview of possible defense and mitigation mechanisms. Table \ref{tbl:deviceslayersum} provides an overview of E-IoT Device-layer components, threats, attacks, and mitigation strategies covered in this section.

\subsection{Elements of the E-IoT Devices Layer}

\noindent\textbf{E-IoT Devices.} Many devices, such as sensors and lighting controllers, are integrated into E-IoT systems to expand the use cases and functionality. Integrated devices may serve specific purposes (e.g., television, media player) or be a part of larger use-cases such as power control modules for lighting control systems. 

\noindent\textbf{Sensors.} E-IoT and E-IoT-integrated devices will very often have sensors used to trigger programmed actions in an E-IoT system. Sensors may play a role in E-IoT in several different ways. For instance, individual sensors (e.g., glass break, motion, contact) can be integrated directly into an E-IoT system thanks to the official support of E-IoT vendors for several protocols (e.g., ZigBee, Bluetooth, Z-wave) \cite{control4Sensors, savantSensors, crestronSensors}. In addition, an external system, such as an alarm system, can be configured to work with an E-IoT deployment. For instance, an E-IoT deployment with water leak detection sensors and automated valves can be configured to close a leak, inform the user via text, and display a message on E-IoT interfaces about the issue \cite{WaterLeak}.  

\subsection{Threat Model for E-IoT Devices Layer}

For this layer, Mallory compromises the E-IoT system solely through physical access to interfaces, devices, cabling, and unattended equipment. To compromise an E-IoT system through the E-IoT devices, Mallory is assumed to have physical access to devices during the manufacturing, installation, operation, or maintenance stages of the E-IoT system. Mallory is capable of this, as security for device-layer components in E-IoT environments relies on the specific devices and the integrator's installation practices (e.g., directional antennas, access restriction, tamper-proofing). We explain Mallory's possible actions at different stages of the E-IoT devices as follows:

\noindent\textit{Manufacturing and Transportation.} In the manufacturing or transportation stages of E-IoT equipment, Mallory may have several opportunities to compromise a device. During these stages, insiders (e.g., manufacturing workers, delivery drivers, packaging personnel) all have direct access to the E-IoT device before a device is installed, making supply chain attacks possible for Mallory, who may be in the role of an insider attacker. Further, Mallory could be an employee of outsourced manufacturing, and as such, it may be particularly difficult to prosecute Mallory during the manufacturing and transportation stages. In this role, Mallory may target E-IoT devices specifically if she has prior knowledge that E-IoT components may be installed in sensitive locations (e.g., secure conference room, access control, enterprise network). 

\noindent\textit{Deployment, Operation, and Maintenance.} E-IoT installations may see visitors such as presenting guests or maintenance workers that have direct access to E-IoT equipment. As such, Mallory as a visitor may perform a node capture attack and further compromise an E-IoT system. Additionally, if Mallory is a more knowledgeable attacker, she may perform sensory channel and side-channel attacks. In other roles, such as a role where Mallory is an IT professional, she could compromise devices in the same manner.

\subsection{E-IoT Devices Layer: Attacks and Vulnerabilities}

In the following subsection, we cover attacks and vulnerabilities relevant to the E-IoT devices layer. These attacks can be performed by Mallory as highlighted in the threat model.

\noindent\textbf{Supply Chain Attacks.} Even before E-IoT devices reach integrators, installers, and consumers, devices may be compromised during manufacturing and distribution stages. Several articles have highlighted supply chain threats and provided examples of how systems in different industries (e.g., medical, banking) have been targeted and compromised through supply chain attacks \cite{miller2013supply, supchaindi, supchainsw, fuentes2018securing, supchainCSO, yousefnezhad2020security}. As specific industries have been targeted, it is reasonable to assume that E-IoT systems may be a future target for supply chain attacks. With the price-point of E-IoT systems and high-profile clients, attackers may find E-IoT systems an attractive target for supply-chain attacks. Work by Farooq et al. analyzed the risks and research challenges in IoT supply chain security \cite{farooq2019iot}. This work highlighted three types of interactions in the supply chain: device-supplier interactions, supplier-supplier interactions, and device-device interactions. In device-supplier interactions, a supplier provides maintenance, security patches, and upgrades to devices. Supplier-supplier interactions are when suppliers use different companies to distribute devices. Finally, device-device interactions occur due to the inter-connectivity of devices in the supply chain, that is, communication between devices (e.g., configuration) in the supply chain. As such, an attacker could compromise a device at any of these interactions. 
The UK's National Cyber Security Centre highlighted several attacks that can occur from supply chain interactions \cite{supplychainattacks}. For instance, malware inserted into vendor websites or devices can ``trojanize'' devices before the devices leave the supply chain. As compromised software is very difficult to detect at the source, target companies may not suspect the software is altered or illegitimate. Supply chain threats also extend to embedded hardware such as chipsets, unauthenticated parts, and counterfeit components inserted in the supply chain. These counterfeit components may impact systems by being of lower quality \cite{gorman2012counterfeit}. In other cases, hardware threats extend to hardware trojans, which have been an ongoing topic of research \cite{bhasin2015survey, tehranipoor2010survey, king2008designing}. In this case, malicious chipsets and electronic components are inserted into devices, usually during manufacturing stages, compromising the integrity of the device. These types of attacks have been observed, in a notable case where Chinese manufacturers infiltrated 30 large U.S companies using malicious hardware components embedded in networking devices \cite{robertson2018big}. As such, E-IoT can easily become a target to a variety of supply chain attacks, as distribution, manufacturing, and installation stages of E-IoT provide ample opportunity to compromise E-IoT devices.

\noindent\textbf{Physical Attacks.} In any E-IoT deployment, E-IoT devices will be found throughout the location or establishment. Some of these devices may be installed in private, unsupervised areas (e.g., a keypad in a closet, an empty conference room). As such, it may be possible for visiting attackers to interact with physical devices integrated into E-IoT systems. As several vulnerabilities against physical devices rely on physical access to E-IoT devices and interfaces (e.g., node capture, tampering, button resets, theft). Physical access to devices and E-IoT components may allow an attacker to perform malicious actions on E-IoT devices, enabling programming mode, hard resets, or otherwise, change the configuration in E-IoT devices that can render them inoperable. For instance, ``button sequences'' may present a vulnerability to E-IoT devices. Reset sequences are used for purposes such as changing a device's configuration, resetting a device to factory settings, or even gain information about devices \cite{crestronreset, control4reset}. As such, an attacker can use these sequences to alter physical devices' configuration, gather information, or otherwise cause E-IoT components to become unavailable to the E-IoT system. 

Physical access to E-IoT devices allows malicious actors to perform \textit{node capture attacks}, where devices are physically captured (or stolen) to gather sensitive information about a system \cite{nodecaptureattack}. Although there is no study on node capture attacks in E-IoT, attacks applied on related domains may be applicable to E-IoT as well. In this respect, work by Wang et al. covered the implications of node capture attacks in wireless sensor networks (WSNs), which are relevant to wireless E-IoT devices (e.g., sensors, interfaces, remotes) as they often share the similar communication technologies \cite{nodecaptureattackwang}. The authors of the work identified ten unique vulnerabilities that can be exploited through node capture attacks affecting session keys, users, sensor nodes, gateways, and availability of the network. As such, the attacks could acquire communication keys, eavesdrop on messages, impersonate devices, track user activity, and impersonate users. Several other pieces of literature have discussed node capture attacks that exploit vulnerabilities to gather keys from connected devices \cite{nodecapturepw1, nodecapturepw2, nodecapturepw3, nodecapturepw4, nodecapturepw5, nodecapturepw6}. The work of Lin et al. focused more on the efficiency of node capture attacks and introduced the full graph attack (FGA), with two optimal algorithms for this attack \cite{lin2015maximizing}. The attack specializes in compromising relationships between nodes and paths. As such, the attacks reportedly increased the efficiency by 50\%  compared to previously proposed attacks.

\noindent\textbf{Side-Channel and Sensory Channel Attacks.} \textit{Side-channel attacks} are threats against the implementation of computer systems, rather than inherent weaknesses. These attacks allow attackers to compromise a system or component through an indirect channel (e.g., timing information, power consumption, electromagnetic leaks, auditory channels) \cite{standaert2010introduction}. A number of E-IoT components may be vulnerable to side-channel attacks through electromagnetic (EM) approaches. For instance, a study by Smulders et al. on serial-based communication suggested that electromagnetic radiation can be used to eavesdrop on physical cables and serial-based communication as a type of side-channel attack \cite{RS232Radiation}. These methods take advantage of a known fact that most electronic equipment emits electrical radiation, and bit amplitude in serial-based communication is relatively larger than other signals \cite{hwu1988electromagnetic}. Their tests performed with a standard AM/FM receiver antenna allowed intercepting and reading signals going through the wire. The work concluded that data signals transmitted over serial-based communication could be intercepted from several meters away. Further, this work noted that the equipment required to perform these scans is inexpensive and readily available, as such, similar attacks may be possible in similar unsecured networks with improved equipment and techniques. Legacy systems, or systems without authentication or encryption may be especially vulnerable to these or similar attacks. Electromagnetic attacks are not limited to wiring, as work published as early as 1986 by Eck et al. noted that electromagnetic radiation eavesdropping attacks are possible in video display units \cite{van1985electromagnetic}. Further work by Kuhn et al. noted that while technology has changed, electromagnetic eavesdropping can work on more modern LCD displays \cite{kuhn2004electromagnetic}. Researchers have found other ways to compromise systems that may be relevant to E-IoT. For instance, Savage et al. showed that with recorded video (e.g., from a CCTV system, intercom systems), an attacker could use passive sound recovery to eavesdrop on conversations \cite{savage2015visualizing}. Further work by Davis et al. demonstrated that an attacker could also use vibrations on object surfaces for eavesdropping under certain conditions (e.g., visible glass or water)\cite{davis2014visual}. 

As E-IoT may control smart lights, light-emitting devices, and light sensors, threats posed by visible-light side-channels may affect E-IoT deployments. Information leakage through optical side-channels has been an active topic of research. For instance, Xu et al. created a video recognition attack where they were able to identify a video being watched on a television using the light emitted by the television through a window \cite{xu2014watching}. Similar works as presented by Schwittmann et al. used ambient light sensors on smartphones and smartwatches to perform similar attacks \cite{schwittmann2017identifying, schwittmann2016video}. Alternatively, Light Ears, presented by Maiti et al., proposed a new attack vector designed to infer a user's private data and preferences from smart lighting media visualization features \cite{maiti2019light}. Based on this research, researchers used the light and sound intensity of smart lights to infer ongoing audio and video. Alternatively, covert optical channels have been researched, with Loughry et al. providing the first call of attention to possible information exfiltration attacks on air-gapped systems by using LED light indicators \cite{loughry2002information}. Similar data-exfiltration attacks have been demonstrated using LCD displays, security camera infrared lights, scanners, and smart lights \cite{zhou2018irexf, guri2016optical, guri2019air, uk2014light, ronen2016extended}.

As E-IoT systems rely on sensors for accurate measurements and to trigger pre-programmed events, physical sensor threats are a concern for E-IoT. Sensor threats and security have been an active topic of research with multiple surveys. However, most of these surveys focus on sensor communication and wireless sensor networks
\cite{sensorsurvey1, sensorsurvey2, sensorsurvey3, sensorsurvey4, sensorsurvey5, sensorsurvey6, sensorsurvey7, sensorsurvey8, sensorsurvey9, sensorsurvey10, sensorsurvey11}. As sensors are a vast research topic, different attacks and vulnerabilities on sensors have been discovered that can be applicable to E-IoT. Analog threats such as sound waves can maliciously influence an accelerometer's output and cause unintended effects in an E-IoT system configured to respond to specific readings \cite{sensorphysics}. Other proposed attacks, such as \textsc{DolphinAttack}, target microphones through inaudible voice commands, can be effective against E-IoT systems that integrate voice recognition and microphones \cite{zhang2017dolphinattack}. With many sensors lacking security mechanisms, E-IoT systems may be particularly vulnerable to sensor attacks. Work presented by Uluagac et al. summarized several sensory channels in cyber-physical systems (CPS) and devices that can be targeted by an attacker \cite{uluagac2014sensory}. These channels are the light, seismic, acoustic, and infrared channels. The \textit{light channel} functions through light sensors and ambient light temperatures. The light channel may be used in E-IoT to trigger programmed events at nighttime. Seismic channels are vibrational channels that can be detected by devices such as accelerometers that detect the physical movements of a device. Acoustic channels are based on sound waves and can be comparable to sonar technologies. Finally, infrared channels use infrared emitters for navigation assistance and can present a covert side-channel for attacks as it is not visible to the human eye. Further, this work highlighted that these sensory channels can all be used to trigger existing malware and that traditional security mitigation strategies do not defend against sensory channel attacks.

Other physical attacks on sensors rely on multiple sensors to function. One of the most researched examples is keystroke inference on devices with unprotected sensors \cite{spreitzer2014pin, cai2012practicality, al2013keystrokes, huang2019risk, owusu2012accessory, marquardt2011sp, narain2014single, lin2019motion, xu2012taplogger, miluzzo2012tapprints, nguyen2015using, hodges2018reconstructing, liang2018deep, roy2016Motors, vuagnoux2009compromising}. While keystroke inference research centers around mobile devices, it may be relevant to E-IoT. Many E-IoT interface devices (e.g., dedicated touchscreens, keypads, remotes) have similarities with mobile devices as they possess several sensors and receive user input. Many of these keystroke inference attacks rely on multiple sensors in different sensor channels to infer sensitive information (e.g., what a user is typing from sensor activity). For instance, PitchIn, a work presented by Han et al. proved that exploiting non-acoustic sensors used in smart environments can allow an attacker to perform speech reconstruction attacks \cite{han2017pitchin}. Multiple sensors (e.g., geophones, accelerometers, gyroscopes) were used to reconstruct audio and perform word recognition in the mentioned work. 

\noindent\textbf{Battery Exhaustion Attacks.} As a number of E-IoT devices are battery-powered (e.g., remotes, interfaces, sensors, etc.), an attacker could use battery exhaustion attacks to impact the operation of E-IoT systems negatively. Battery Exhaustion attacks are a type of Denial-of-Service (DoS) attack that aims to deplete the batteries of devices by forcing the device to perform an excess amount of tasks \cite{ShakhovBattery, batterybauernansa}. Moyers et al. presented the effects of wireless and Bluetooth battery depletion attacks on mobile devices \cite{MoyersBattery}. This work classified three distinct types of battery exhaustion implementations, \textit{service request power attacks}, \textit{benign power attacks}, and \textit{malignant power attacks}. For service request power attacks, attackers target devices by making repeated requests to these devices and exhaust power through the wireless network interface card. In benign power attacks, victims are forced to perform repeated tasks (e.g., data processing, diagnostics) and consume large amounts of power. Finally, malignant power attacks are usually implemented with malware designed to increase power consumption in a device (e.g., increasing the CPU clock). Other work by Martin et al. highlighted the effects of these attacks on wireless devices, noting that damage caused by battery exhaustion attacks may also cause long-term damage to battery life in addition to a denial-of-service condition when a device becomes unavailable \cite{MartinBattery}.

\begin{table*}[t]
    \centering
    \caption{Supply Chain Defenses Suggested by ENISA \cite{supplychaindefenses}.}
    \scalebox{1.00}{
    \begin{tabular}{c|c|c}
    \hline
       \textbf{Stages} & \textbf{Topic} & \textbf{Description}\\ 
       \hline
        Product Design & Secure Building Blocks &  Ensuring usage of accepted security standards (e.g., cryptography, software).\\
                    \rowcolor{Gray}
                    & Sabotage Prevention & Monitoring for security flaws created by insider threats.\\
                    & Recovery Plan & Consider security mechanisms, fail safes, and a recovery plan for the future.\\
                    \rowcolor{Gray}
                    & Combined Security Controls & Security mechanisms must consider HW and SW interactions. Security controls (e.g., secure boot) require the usage \\
                    \rowcolor{Gray}
                    && of tamper resistant hardware.\\
                    & Chain of Trust Definition & A clearly defined chain of trust is necessary to ensure trust on HW and SW elements. \\
                    \rowcolor{Gray}
                    & Resource Constraints & Purposely developing devices so that current and future security measures can be implemented.\\
        \hline
        Component Manufacturing 
                & Counterfeit Components & Mitigating security threats from counterfeit components through hardware authentication.\\
                \rowcolor{Gray}
                & Defective Components & Quality control and usage of tested parts to avoid security and device degradation. \\
        
        Component Assembly + & Firmware Access Control & Ensure access control mechanisms exist to software, firmware updates and other maintenance operations.\\
        Embedded Software && \\
                    \rowcolor{Gray}
                    & Backdoors & Monitoring on suspicious behaviours and backdoors implanted in hardware or low-level code.\\
        \hline
        Device Programming & Secure Provisioning & Ensuring the use of end-to-end provisioning mechanisms guaranteeing\\
                    && the security of credentials and cryptographic information.\\
                    \rowcolor{Gray}
                    & Coding Practices & Adoption of best practices such as code reviews and continuous integration of cybersecurity checks in the\\
                    \rowcolor{Gray}
                    && software development cycle.\\
        \hline
        Platform Development & Development Focus & Placing a development approach to focus both on functionality and security.\\ 
                    \rowcolor{Gray}
                    & Dependencies Management & Checks and review process to ensure that dependencies and libraries are available and conform to security practices.\\
                    & Network Security & Ensure that local network policies minimize the risk of intrusion.\\
        \hline
        \rowcolor{Gray}
        Distribution and & Value-added Resellers & Certification of resellers and third-party distributors to prevent tampering and unauthorized distribution of devices.\\
        \rowcolor{Gray}
        Logistics  && \\
                    & Theft and Counterfeit & Additional security measures to protect against theft and insertion of counterfeit or malicious components.\\   
                    & Protection &\\
                    \rowcolor{Gray}
                    & Device Identity & Enabling the ability to identify devices during the fabrication and distribution stages using different HW and.\\
                    \rowcolor{Gray}
                    && SW components.\\
                    & Registration Tracking & Ensuring proper device registration and onboarding into smart platforms such that devices can be tracked.\\
        \hline
        \rowcolor{Gray}
        Technical Support 
            & OTA Control Tools & Adoption of remote Over-The-Air control tools used for maintenance are properly managed and secured\\
        \rowcolor{Gray}
        and Maintenance && through a chain of trust.\\
                    & Software Patches & Usage of software versions that mitigates threats exposed in latest security disclosures.\\
        \hline
        \rowcolor{Gray}
        Device Recovery & Data Removal & Adopting secure data removal techniques to avoid sensitive information remaining on devices.\\
        \rowcolor{Gray}
        and Repurpose && \\
    \hline
    \end{tabular}}
     \vspace{-0.1in}
    \label{tbl:supplychaindefenses}
\end{table*}

\subsection{Mitigation of E-IoT Devices Layer Attacks}

In this subsection, we highlight possible mitigations to E-IoT devices-layer threats.

\noindent\textbf{Supply Chain Defenses.} A few solutions were proposed in the literature to defend against supply chain attacks. In order to secure the device endpoints, Yang et al. proposed an RFID-based solution that authenticates devices once they are deployed  \cite{yangrfid}. This work was taken further with the introduction of ReSC by the same authors, a solution proposed to defend against the theft of authentic smart devices, and the insertion of counterfeit malicious devices \cite{yangrfid2, yangrfid3}. Another approach by Chamekh et al, proposed the use of a Merkle tree management framework applied to supply chain architecture to provide a more trusted system and defend against supply chain attacks \cite{chamekhsc}. During transportation stages, tamper-proof and tamper-evident packages and equipment may also prevent unauthorized attackers from tampering with devices before they reach a client \cite{ehrensvard2007tamper, tamperevidentpack}. The European Union Agency for Cybersecurity (ENISA) provided comprehensive guidelines for IoT supply chain security \cite{supplychaindefenses}. These guidelines divide defense strategies into several relevant stages relevant to E-IoT: product design, component assembly and embedded software, device programming, platform development, distribution and logistics, technical support \& maintenance, and device recovery \& repurpose. For product design, guidelines dictate that secure software libraries and cryptographic practices, sabotage prevention, tamper-resistant software and hardware, and chain of trust are design practices that may prevent supply-chain attacks. These guidelines are detailed in Table \ref{tbl:supplychaindefenses}. Further the ENISA guidelines highlight that vendors can take some preventative measures such as, working with suppliers that provide security guarantees, maintaining transparency, having a skilled workforce, promoting security awareness, and developing novel trust models. 

One of the largest topic of research is counterfeit components inserted in the supply chain, as such best practices and solutions have been proposed. For instance, ENISA guidelines highlight that parts used during manufacturing should be authenticated to prevent counterfeit components from entering the supply chain. Further, to prevent defective components, ENISA also advises for quality control and testing of parts to prevent defective components \cite{supplychaindefenses}. Surveys conducted on the topic of counterfeit devices and hardware Trojans have suggested several solutions \cite{bhasin2015survey, tehranipoor2010survey}. First, optical inspection based detection relies on reverse engineering to detect Trojans. As such, techniques such as scanning optical microscopy, scanning electron microscopy, and pico-second imaging circuity analysis are used. Images captured with these techniques are then compared to benign chipsets provided by the designer. Testing-based detection techniques use functional testing to detect Trojans. As such, a functional set of vectors need to be designed for each chipset. Side-channel detection approaches rely on factors such as power consumption, EM emissions, and time delays to detect anomalies. Such approaches can also be used to detect trojans. For instance Agarwal et al. used Principle Component Analysis to create a side-channel fingerprint of a circuit and compare it to a known, benign model \cite{agrawal2007trojan}. Run-time detection approaches are also used, usually combining hardware and software to detect trojans. For instance, DEFENSE is a proposed monitoring framework that operates at device run-time to detect hardware anomalies and trojans \cite{abramovici2009integrated}. Finally, invasive techniques modify integrated circuit's structures to avoid the insertion of hardware Trojans. Authors have shown that hardware obfuscation methods can prevent Trojan insertion and assist other detection methods \cite{chakraborty2008hardware, chakraborty2009harpoon, chakraborty2009security}.

\noindent\textbf{Physical Security.} Physical security of cabling and devices is an important part of E-IoT deployments as E-IoT devices can be stolen, tampered with, or otherwise damaged. Vendors implement some physical mitigations and best practices for many of their devices. Additionally, E-IoT systems make an effort towards tamper-proofing their systems and offer suggestions on physical installation. For instance, Control4 released an exterior installation security best practices document \cite{control4DSsecurity}. This document highlights several important points on exposed devices such as door stations used for gate access and intercom. First, installers are encouraged to use standard tamper-resistant security screws shipped with devices to prevent opportunists from stealing or tampering with devices. Second, relays used to open security gates should not be connected at the door station itself and instead to a relay inside the building.  Relays' endpoints should be in a secure location as physical attackers may compromise devices by tampering with relays and gain unauthorized access to locations. Finally, they acknowledge the risks associated with the network cable running to public interfaces (e.g., door stations, intercoms) and highlight solutions such as network isolation, MAC address filtering, and wireless door station access as possible solutions. In some instances, E-IoT components may only be removed with custom tools to prevent theft and tampering. For instance, touchscreens may come with a special tool so that an unprepared attacker cannot easily remove the interface \cite{control4TPRemove}. Finally, integrators and users should take advantage of monitoring tools (e.g., wireless monitoring, IP monitoring) to identify devices that fall offline to know if they have been tampered with. Practices used for loss prevention may also be useful for E-IoT. Concepts such as beacons, smart tags, and geo-fencing may prevent node capture attacks and alert integrators before an attack occurs \cite{lossprevention}. Integrators may also take certain steps in the installation to make sure that E-IoT devices are secure. For instance, installers should follow best practices, place sensors in places where they are not easily reachable and do not leave any exposed wiring in installations. As noted earlier, physical access to exposed wiring and devices would make it trivial for an attacker to compromise an E-IoT system in public and unmonitored areas. Further, installers and users should consider physical access control to prevent access by unauthorized users.

\noindent\textbf{Side Channel and Sensory Channel Defenses.} There exist a number of defense solutions against side-channel attacks. For instance, for EM and many side-channel eavesdropping attacks, physical security and encryption provides a level of defense. For attacks that rely on sound, AuDroid is a policy-based framework for smart devices proposed by Petracca et al. \cite{petracca2015audroid}. AuDroid controls information flow in audio channels and notifies users when audio access is requested. Access control frameworks such as these may present a viable solution for side-channel attacks where sensory and audio channels can be abused. A number of defense mechanisms proposed for sensors and wireless sensor networks may be applicable to E-IoT against side-channel attacks. For instance, for sensors in mobile devices such as phones, security mechanisms have been an ongoing topic of research \cite{sun2014design, wang2015novel, enck2014taintdroid, wu2014droiddolphin, xu2015semadroid}. However, many of these proposed solutions rely on software-based approaches to defend against sensor-based attacks. Alternatively, solutions such as frameworks and intrusion detection systems have been proposed for wireless sensor networks and may apply to large E-IoT deployments configured to rely more heavily on sensors for programmed events \cite{strikos2007full, ioannis2007towards, ioannis2007towards, farooqi2013novel, pongaliur2008securing, yu2008framework}. One example, 6thSense, a sensor-based defense mechanism by Sikder et al. takes a machine learning approach to detect malicious behavior occurring in smart devices \cite{contextawaresensor, sikder2019context, sikder20176thsense}. The proposed solution relies on sensor co-dependence, sensor sampling, and real-time monitoring. Since E-IoT systems may share some similarities to proposed solutions (e.g., multiple sensors, centralized design), these defense mechanisms may apply to E-IoT against side-channel and sensory channel attacks. While many of these solutions may protect against side-channels, some side-channel attacks (e.g., LightEars) do not have direct solutions proposed beyond physical security and require future research.

\noindent\textbf{Battery Exhaustion Defenses.} A number of mitigation strategies have been proposed to combat battery exhaustion attacks on wireless devices. The solution for E-IoT may be entirely dependent on the type of the system. For instance, battery exhaustion defenses may be different in a Zigbee vs another wireless-based deployment. Buennemeyer et al. proposed Battery-Sensing Intrusion Protection System (B-SIPS) that focuses on small mobile hosts and correlates power consumption with wireless activity \cite{bsips}. Moyers et al. proposed an intrusion detection system (IDS) to protect against malicious activities \cite{MoyersBattery}. The proposed Multi-vector Portable Intrusion Detection System (MVP-IDS) works by monitoring electrical current changes and correlating this with malicious traffic. Other IDSs have been developed, such as the one proposed by Nash et al. that uses CPU load and disk access to estimate power consumption and detect if battery exhaustion attacks are occurring \cite{nashintrusion}. In situations where devices may be homogeneous, defenses against battery exhaustion attacks can be based on comparing these devices to create a realistic baseline and find anomalies that may be effective in wireless sensors and interfaces \cite{UpadhyayBattery}. Finally, work by Hristozov et al. using rate limiting approaches to defend against battery exhaustion attacks reported to be successful for devices supporting RESTful services \cite{hristozovBattery}. 

\newcommand{\technical}{Technical:\xspace}
\newcommand{\role}{Role\xspace}

\begin{table*}[t]
    \centering
    \caption{Overview of E-IoT Communication Layer Attacks, Threats, and Mitigations.}
    \scalebox{1.00}{
    \begin{tabular}{c|c|c}
    \hline
       Component & Attack/Threat & Mitigations\\ 
    \hline
        Serial-based Protocols 
            & Proprietary Protocol Attacks \cite{cresnetsniffer}
                & Vendor updates, access control, and honeypots \cite{mays2017defending}.\\
            \rowcolor{Gray}
            & Modbus \& BACNet Attacks \cite{Volkova:2019, BACNetDOC, Reachable-Bacnet, ASHRAE-BACNET}
                & Protocol-specific defenses, network security, protocol improvements, product updates, \\
        \hline
        WiFi 
            & WEP Attacks \cite{borisov2001intercepting, lashkari2009survey}
                & Update the WiFi security protocol and update to WPA2 or WPA3 if possible \cite{ftcprotect,cisaprotect}.\\  
            \rowcolor{Gray}
            & WPA/WPA2/WPA3 Attacks \cite{lashkari2009survey, khasawneh2014survey, vanhoef2017key, wpahack, vanhoef2020dragonblood, kohlios2018comprehensive, lounis2019bad, lounis2019wpa3}
                & Wireless equipment patching, custom SSID names, WiFi best practices, strong passwords,\\
            \rowcolor{Gray}
                & & firewalls, disabling WPS functionality, network security \cite{wpahack,wang2010practical,ftcprotect,cisaprotect}.\\
        \hline
        ZigBee and Z-Wave
            & ZLL Attacks, Ghost-in-Zigbee 
                & Disable nonce reuse, software/firmware updates, recommended configuration, protocol \\
            & Zigbee DoS, KillerBee, and 
                & changes, message authentication \cite{lounis2020attacks, whitehurst2014exploring, benzaid2016fast, ZwaveHowSecure}.\\ 
            & other Zigbee Attacks \cite{IOTWireless, wang2013zigbee, zillner2015zigbee, Zigbee-Ghost, ZigBee-Nuclear, Zigbee-PracticalAttacks, KillerBee}
                & \\
            \rowcolor{Gray}
            & Z-force, rogue controllers 
                & Always enable encryption, software/firmware updates, recommended configuration, protocol \\
            \rowcolor{Gray}
            & controller duplication, and 
                & changes, authentication on all messages \cite{lounis2020attacks, whitehurst2014exploring, benzaid2016fast, ZwaveHowSecure}.\\ 
            \rowcolor{Gray}
            & other Z-wave attacks \cite{fouladi2013security, IOTWireless, ZwaveRogue}
                & \\
        \hline
        Bluetooth 
            & Bluesniping, Bluechopping 
                & Software updates, setting Bluetooth to non-discoverable mode, directional antennas, disabling\\
            & Bluecutting, Bluedepriving, and 
                & Bluetooth when not in use \cite{bluetoothdefense}.\\
            & other Bluetooth Attacks \cite{lounis2019bluetooth, shaked2005cracking, darroudi2017bluetooth, minar2012bluetooth, dunning2010taming, hypponen2007nino, sun2018man, haataja2010two, haataja2008practical, haataja2008man, barnickel2012implementing, hering2004bluetooth, spill2007bluesniff, lounis2020attacks, lounis2018connection, alsaidi2018security}
                & \\
        \hline
        \rowcolor{Gray}
        IR 
            & IR Attacks \cite{IRHacking, IRTrolling}
                & Access control and CCTV monitoring. \\
        \hline
        General Wireless
            & Jamming \cite{jamming1, jamming2, jamming3, jamming4, reactivejamming, jammingzb}
                & Defenses vary by wireless protocol used and implementation \cite{jamming1, jamming2, jamming3, jamming4, jamming5, jamminganalysis, jammingdetection, jammingstatisticaldef}. \\
        \hline
        \rowcolor{Gray}
        HDMI-Based Protocols 
            & CEC-Based Attacks \cite{hdmi-walk, nccgroup, smithcec, davisblackhat}
                & Access control, CEC-less cables/adapters, and IDS such as HDMI-Watch \cite{hdmi-watch}. \\
    \hline
    \end{tabular}}
     \vspace{-0.1in}
     \label{tbl:communicationlayersum}
\end{table*}

\section{Communications Layer}\label{sec:communicationlayer}

In this section, we firstly cover components of the E-IoT Communications layer such as interfaces and protocols. We follow up with the threat model for this layer. Moreover, we introduce E-IoT communication layer threats and attacks. Finally, we highlight mitigations and security mechanisms applicable to the E-IoT communications layer. Table \ref{tbl:communicationlayersum} provides an overview of components, attacks, threats, and mitigation strategies at the E-IoT communications layer. 

\subsection{Elements of the E-IoT Communications Layer}

\noindent\textbf{Ethernet.} Internet Protocol (IP) communication has become one of the most widely deployed standards in internal and external networks. Often, modern homes and offices already have the physical Ethernet wiring and infrastructure for Internet Protocol. As such, an E-IoT system installer can use both standards (IPv4 or IPv6) for Ethernet-based communication \cite{ipversions}. Additionally, with IP, integrators have the flexibility to divide traffic flow of connected devices with subnetting and virtual LANs (VLANs). For instance, an integrator can divide a larger network into segmented sections with subnetting, determining the maximum number of devices in each segment through network configuration \cite{subnetting}. Similarly, VLANs are used to improve information flow, security and better manage an IP network \cite{vlan}. For instance, Pakedge, a vendor of E-IoT-centered network solutions, encourages VLANs for E-IoT installations and network segmentation \cite{pakedgevlans}. As IP is popular and widely supported by many vendors, E-IoT systems will often use IP communication in some of their components. Ethernet provides the advantage of a superior level of reliability and speed compared to the wireless counterpart. Further, Ethernet can power devices through Power-over-Ethernet (PoE) technology \cite{poweroverethernet}. As such, integrators only need to cable a PoE-capable connection to a device, such as a touchscreen, to provide data and power through a single connection. Physical cabling has proven to be a reliable communication method between smart components and remains popular for high-bandwidth, high-reliability applications. For instance, Ethernet may be used to control devices in the equipment rack such as IP-capable A/V receivers, Ethernet-powered IP cameras, or hardwired touchscreens \cite{receiverIP}. Moreover, Ethernet offers different networking topologies (e.g., star, ring, single-switch), which grant integrators the flexibility needed for custom E-IoT installations \cite{iptopologies}.  

\noindent\textbf{WiFi.} Wireless Fidelity (WiFi) is a frequently used communication protocol for smart devices where Ethernet cabling endpoints are not viable. Various modes within IEEE 802.11 have allowed for increased speeds and frequencies. The main advantage of wireless communication is that E-IoT devices (e.g., thermostats, controllers, A/V) may use a wireless connection without requiring an extra physical connection to integrate into an existing system. Similarly to Ethernet Category cables: 802.11 generations b, a, g, n provide different levels of data rates, as well as operate in 2.4 GHz or 5.0 GHz\cite{wifi}. In many E-IoT systems, WiFi serves different purposes due to its widespread nature. Many smart device vendors enable wireless network connections natively on their devices, making such devices easy to integrate into E-IoT systems. Examples of WiFi usage in E-IoT systems may include interfaces (e.g., phones, touch screens, tablets) and physical devices (e.g., displays, receivers, projectors). In terms of WiFi security, a number of configurations are available for accepted WiFi security standards, such as the Wireless Equivalent Standard (WEP) which is obsolete now or WiFi Protected Access (WPA), with the latest release being WPA3 security \cite{baek2004survey, al2020analyzing}. Furthermore, in larger and more complex network deployments, enterprise solutions exist and are usually installed by trained integrators \cite{enterprisewifi}. As such, a number of different configurations are possible with WiFi communication dependent on the equipment, level of security, and installation requirements of an E-IoT deployment.

\noindent\textbf{ZigBee and Z-wave.} Two of the most popular mesh-network protocols for smart devices are ZigBee and the proprietary Z-wave \cite{Zwave, Zigbee}. Various vendors have embedded radio communication hardware on their thermostats to connect their devices to more extensive mesh networks. While ZigBee and Z-wave are different protocols, they are used for similar purposes in E-IoT systems. For instance, these protocols are often used in low-bandwidth applications to integrate devices such as thermostats, light dimmers, relays, and sensors to a larger system. Mesh networking allows users to retrofit existing installation by replacing existing components such as light switches for wireless-enabled components. For Zigbee, usually, there are three types of devices within the ZigBee mesh network: a coordinator, routers, and end devices \cite{ramya2011study}. The ZigBee coordinator is the root of the ZigBee network and manages components necessary for ZigBee to operate (e.g., security keys, access control, security policies, stack profile). The ZigBee Router relays information and routes ZigBee packets among devices. Some ZigBee routers may also have the functions of end-devices. Finally, the end-devices send and receive communication from parent nodes and are usually designed for a specific purpose (e.g., door locks, light bulbs, sensors). Z-wave follows a similar device architecture with three basic device types, controllers, routers, and slaves. These devices fulfill similar purposes as their ZigBee counterparts~\cite{zwaveguide}.

\noindent\textbf{Bluetooth} is a wireless standard for data exchange between portable and fixed devices. A short-wavelength protocol, Bluetooth operates from the 2.4 to 2.485 GHz range \cite{bluetoothspec}. Additionally, Bluetooth may operate as Bluetooth Low-Energy (BLE) or Bluetooth Mesh, which allow for more varied applications to the protocol \cite{bluetoothoperations}. With the number of Bluetooth devices in the market, E-IoT systems are compatible with the protocol for different purposes. For instance, Savant may use Bluetooth Low Energy for their smart lighting solutions, while other systems use Bluetooth for connecting mobile devices and stream music to the central system \cite{savantbluetooth}. Bluetooth networks, commonly known as piconets, follow a master and slave architecture where up to seven active slave devices can be connected to a master device \cite{scarfone2008guide}. 

\noindent\textbf{IR} Infrared (IR) is a wireless optical communication medium used to control devices over short, line-of-sight ranges \cite{IRIntro}. While limited, as it cannot penetrate through walls and the short transmission rate, IR remains popular in many consumer devices (e.g., A/V, televisions remotes, motorized components). As such, because of this widespread support, IR sees common use in many E-IoT systems that need to integrate these devices into centralized E-IoT systems. E-IoT systems integrate these devices using IR flashers placed on physical devices; these flashers relay messages directly to the receiving device \cite{IRFlashers}. As some devices can only be controlled through IR, E-IoT makes widespread use of IR communication.

\noindent\textbf{Proprietary Wireless} Not all protocols used by E-IoT systems are well-known or open-source. Proprietary wireless communication protocols are often used in E-IoT systems and have not seen much research. For instance, the Radio Technology Somfy (RTS), is used by Somfy, one of the major vendors of E-IoT motorized blinds \cite{somfyRF}. Similarly, popular system vendors such as Lutron, Levitron, Legrand, and Crestron also use proprietary wireless protocols that have remained mostly unexplored \cite{lutronRF,crestronRF, legrandRF, levitronRF}. Table \ref{tbl:proprietaryrf} highlighted some proprietary wireless protocols used by E-IoT systems and their usage in E-IoT.

\begin{figure}[t]
\centering{\includegraphics[width=0.35\textwidth]{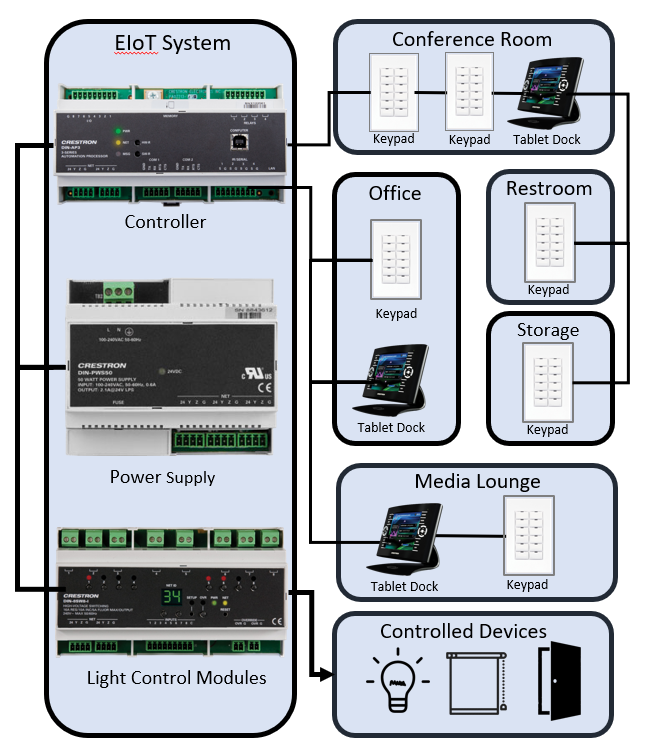}}
    \caption{E-IoT System with, daisy-chain topology configuration with two lines and controlled devices.}
\label{fig:daisychainsystem}
\vspace{-0.1in}
\end{figure}

\noindent\textbf{Serial-based.} Serial-based communication is a precursor to several modern device communication standards. While many may consider the use of serial-based communication as deprecated, various E-IoT systems and connected devices officially support serial-based communication for system-to-device integration. Further, some E-IoT systems have built their systems on top of existing serial-based communication for proprietary devices. For instance, since the accepted inception in 1969 \cite{RS232}, Recommended Standard 232 (RS-232) has been a well-known medium for device-to-device communication. This standard is often used in E-IoT environments for communication between devices. Some of these devices include thermostats, projectors, A/V receivers, A/V switchers, motorized lifts, displays, pool controllers, motorized drapery, and alarm systems that interface with other devices directly through serial-based links. A more specific example is the Carrier Infinity Series systems module for HVAC units. This module allows an E-IoT system to communicate with Carrier HVAC systems through serial interface or allow for remote access using a physical Ethernet connection \cite{SAM-Module}. In many cases, serial-based communication is wired in a ``daisy-chain'' bus configuration where the cabling goes from device-to-device instead of each device is individually wired to the E-IoT controller as shown in Figure \ref{fig:daisychainsystem}. Such a wiring configuration is a common practice in E-IoT, as daisy-chain is easier to wire and saves the integrators and users in labor and wiring costs.

The use of serial-based protocols for a variety of use-cases is widespread among E-IoT vendors. For instance, Crestron's Cresnet has become a ubiquitous name in residential, marine, and commercial installations \cite{Cresnet}. Cresnet uses RS-485 half-duplex communication used for communication between devices (e.g., interfaces, components, keypads) and the controller\cite{Crestron}. Similarly, vendors such as Control4, LiteTouch, and Savant use proprietary serial-based protocols to communicate with interfaces \cite{litetouchmanual, control4keypad, savantsmartdeploy}. These connections usually are daisy-chained together and work with multiple lines. In addition to these examples, many product vendors manufacture devices with native serial communication to where many devices and systems are integrated into E-IoT through serial communication. For instance, advanced audio receivers, televisions, and alarm systems can all be integrated into E-IoT using serial communication \cite{somfyrts, receiverIP, serialtv, integrateserial}. The technical specifications of many of these highlighted serial-based communication protocols are not publicly available, and thus any security mechanisms remain largely unknown. 

Another type of serial-based communication are building automation protocols such as BACnet. BACnet was designed specifically to meet the requirements for automation and control within corporate offices, buildings, and other commercial establishments. BACnet can be integrated into some E-IoT systems, with many devices available. The protocol is also used for communication in sensors, security systems, energy management, lighting control, physical access systems, and elevator controls \cite{ASHRAE-BACNET}. BACnet operates on top of RS-485 and RS-232 to provide application and networking layers for device operation. BACnet implements four layers: Application Layer, Network Layer, Data Link Layer, and Physical Link Layer. In this protocol, RS-232 is used for point to point communication while RS-485 handles Master/Slave Token Passing \cite{TI-BACNET}. Since BACnet is an open protocol, it has been adopted by various device vendors and manufacturers as a form of external control.

\noindent\textbf{HDMI.} The High Definition Multimedia Interface (HDMI) is one of the core components of audio/video systems. It acts as the main physical connection between multiple devices (e.g., televisions, projectors, video players, receivers). As such, HDMI is one of the most common interfaces used worldwide, with billions of compatible devices in the wild \cite{HDMIBil, HDMIBil2, hdmistandard}. Per HDMI design, communication transmitted is not limited to audio and video, as HDMI transmits control and information signals through the cabling through the 19-pin connector \cite{HDMIPinout}. Further, HDMI can be a part of distribution networks with switchers, splitters, and other interconnects that allow multiple HDMI-enabled devices to share A/V signals and communicate. As A/V distribution is an important part of E-IoT, HDMI serves a major role in E-IoT systems \cite{control4wha, crestronwha, savantwha}. Further, some E-IoT systems use communication protocols embedded in HDMI to control and integrate devices into an E-IoT system \cite{crestronCEC}. For instance, the HDMI connection includes the Consumer Electronics Control (CEC) to expand the functionality of HDMI systems \cite{HDMISpec}. The CEC protocol is a component of HDMI communication and was developed to enable interoperability between HDMI devices. CEC is a low-bandwidth protocol with a maximum of 16 devices and functions in a bus architecture. Some E-IoT systems use CEC to control A/V devices such as receivers, televisions, and projectors. Thus, many vendors implement CEC features on their devices under different trade names, including Anynet+ (Samsung), Aquos Link (Sharp), BRAVIA Link/Sync (Sony), CEC (Hitachi), CE-Link and Regza Link (Toshiba), SimpLink (LG), VIERA Link (Panasonic), EasyLink (Philips), Realink (Mitsubishi) \cite{tradenames}. 


\begin{table}[t]
    \centering
    \caption{Examples of E-IoT system proprietary RF protocols.}
    \scalebox{0.90}{
    \begin{tabular}{c|c|c}
    \hline
       Vendor & Protocol & Product Lines\\ 
       \hline
       Lutron & Clear Connect Technology RF \cite{lutronRF} &  Lighting, Shades, Interfaces\\
       \rowcolor{Gray}
       Somfy & Radio Technology Somfy \cite{somfyRF} & Lighting, Shades, Interfaces \\
       Levitron & LevNet RF \cite{levitronRF} & Lighting, Shades, Interfaces \\
       \rowcolor{Gray}
       Legrand & TopDog RF \cite{legrandRF} & Lighting, Shades, Interfaces\\
       Crestron & infiNET EX/ER \cite{crestronRF} & HVAC, Lighting, Shades, Interfaces\\
    \hline
    \end{tabular}}
     \vspace{-0.1in}
    \label{tbl:proprietaryrf}
\end{table}

\subsection{Threat Model for E-IoT Communications Layer}
In this layer, we consider Mallory compromising an E-IoT system through the communications layer, targeting the confidentiality, availability, and integrity of the system. Thus, Mallory compromises the E-IoT system through communication components, often without the need of physical access. Attacks on this layer may benefit weak protocols, protocol vulnerabilities, flaws in implementation, and other similar factors. As such, Mallory, in this case, is knowledgeable in communication vulnerabilities and has the equipment necessary to compromise E-IoT. For instance, Mallory may carry sniffers and the software necessary to eavesdrop on communication channels and inject messages into E-IoT communication. We explain Mallory's possible roles in attacking E-IoT communication layer as follows:

\noindent\textit{Visitors and Unprivileged Users.} Some users (e.g., visitors, insiders) may not have sufficient privileges to interact with all of the components of a deployed E-IoT system. Mallory, as a malicious unprivileged user, may use protocol vulnerabilities to gain unauthorized access to devices near her. As such, attacks on serial-based protocols, short-range wireless, and HDMI are feasible. An unprivileged user may just need some preliminary knowledge of the protocols used.

\noindent\textit{IoT Hackers.} Malicious actors such as hackers may target E-IoT systems specifically in public locations (e.g., presentation rooms, bars, campuses). In this scenario, Mallory as a malicious hacker, may choose to perform reconnaissance of an E-IoT deployment without direct physical access to the system. Additionally, more sophisticated attackers may attempt to compromise a system, gain unauthorized access, cause Denial-of-Service attacks, or otherwise disrupt E-IoT operations through the communications layer. In this case, Mallory only has unauthorized access to all E-IoT system components. 

\subsection{Communication Layer: Attacks and Vulnerabilities}

In this subsection, we give an overview of the attacks and vulnerabilities of the E-IoT Communications Layer. Specifically, we cover attacks to serial-based protocols, wireless protocols, HDMI-based protocols, and building automation protocols.

\noindent\textbf{Serial-based Protocol Attacks.} One of the challenges of properly evaluating serial-based protocols in E-IoT is the proprietary nature of many of these protocols. Many proprietary protocols are long-lived and do not advertise any form of security mechanism to the communication. Information on many of these protocols (e.g., Cresnet) is sparse. These protocols rely largely on security through obscurity as many of these protocols were designed for functionality but not security in mind. Even with this lack of research, online communities and integrators have explored E-IoT protocols and managed to create sniffers to capture serial-based communication for debugging \cite{cresnetsniffer}. As such, these sniffers work without any form of authorization beyond physical access and expose possible threats to E-IoT serial-based protocols. One can draw parallels to industrial control protocols, a comprehensive review of the security channels in industrial protocols can be found in a study by Volkova et al.~\cite{Volkova:2019}. This study highlighted that aging serial-based communication technologies such as Modbus can be attacked (e.g., credential theft attacks, replay attacks, Man-in-the-middle attacks) by a knowledgeable attacker. 

An analysis in 2003 by the Department of Commerce found some threats to building automation protocols such as BACnet. While most systems were not connected to the Internet, there was still backdoor access via modem connections to controllers \cite{BACNetDOC}. The study also noted various attacks on passwords, confidentiality, integrity, Denial of Service, spoofing, and eavesdropping within a BACnet installation. Gasser et al. discussed research on Internet-exposed BACnet systems \cite{Reachable-Bacnet}. BACnet is often an integral part of connected Industrial Control Systems (ICS); these are critical infrastructural systems for any size business and offices \cite{ICS}. BACnet operates on UDP ports 47808-47823 on default\cite{ASHRAE-BACNET}. Researchers used a pre-made BACnet payload in conjunction with Zmap\cite{Zmap} to scan for devices in the IPv4 address space for valid responses. Using this methodology, researchers managed to confirm a total of 15,429 exposed BACnet devices on the Internet. A notable characteristic of BACnet/IP UDP protocol is that it is both stateless and does not require handshake nor authentication. The previously mentioned characteristics of BACnet make it susceptible to Amplification Attacks, a Denial-of-Service attack where a response payload is larger than the request payload \cite{Reachable-Bacnet}.

\noindent\textbf{WiFi Attacks.} WiFi communication has been an active topic of research due to its broad appeal and uses in many connected devices. A myriad of WiFi attacks have been covered in different publications, surveys, and technical documents. Additionally, attacks may be dependent on installed hardware, firmware, security used (e.g., WEP, WPA, WPA2, WPA3), and specific implementation. A survey by Lashkari et al. highlighted weaknesses to security mechanisms in WiFi communication \cite{lashkari2009survey}. Specifically, this work notes that WEP is susceptible to attacks (e.g., packet forgery, replay attacks, de-authentication) and vulnerabilities such as improper key-management and problems with the RC-4 algorithm. Other work from Borisov et al. goes further into the insecurities of the WEP protocol and how poor security practices (e.g., keystream reuse, key management) allows an attacker to compromise WiFi with WEP security \cite{borisov2001intercepting}. Specifically, WEP is vulnerable to eavesdropping attacks, message modification, message injection, message decryption, authentication spoofing, and reaction attacks against WEP. While considered more secure, WPA vulnerabilities also exist. Lashkari et al. note that WPA/WPA2 has definite security improvements over WEP, such as the use of the Advanced Encryption Standard (AES) and the Temporal Key Integrity Protocol (TKIP) \cite{lashkari2009survey}.

However, even with improvements, WPA and WPA2 can be susceptible to attacks (e.g., brute force attacks, dictionary attacks). A related attack for WPA/WPA2 is a handshake capture attack. An attacker can capture the communication handshake and attempt to perform brute force attacks or dictionary attacks against the captured handshake \cite{khasawneh2014survey}. An attack proposed by Vanhoef et al. introduces key re-installation attacks against WPA/WPA2 where attackers can force a WiFi network to reuse old keys and compromise confidentiality in the network \cite{vanhoef2017key}. As such, key re-installation attacks would allow Mallory to perform actions such as packet replay, decryption, and forging in some implementations, severely impacting the confidentiality and integrity of WiFi communications. Other attacks such as the Reaver and Pixie-Dust attacks also target WPA-based security, specifically exploiting the WiFi Protected Setup (WPS) protocol in routers \cite{wpahack}. Finally, as a newer security mechanism, some weaknesses have been found in WPA3. As such, denial-of-service attacks, connection deprivation attacks, and handshake attacks can compromise WiFi communication with WPA3 security \cite{vanhoef2020dragonblood, kohlios2018comprehensive, lounis2019bad, lounis2019wpa3}. As many E-IoT devices use WiFi communication, any WiFi attacks could compromise the confidentiality, integrity, and availability of E-IoT and E-IoT-integrated components.

\noindent\textbf{ZigBee and Z-Wave Attacks.} Wireless technologies are common in E-IoT systems in many different use-cases and have been an active topic of research in the security community. As described in \cite{IOTWireless}, various communication protocols (e.g., ZigBee, Z-Wave) can be attacked, negatively impacting E-IoT systems by directly affecting user interfaces. There have been known security breaches in ZigBee devices.  The ZigBee Light Link (ZLL) standard was designed with easy client integration, and installation in mind \cite{wang2013zigbee}. One known breach in 2015 involved the leakage of the master key for light-based ZigBee devices. This leak rendered ZLL devices insecure \cite{zillner2015zigbee}. It must be noted that there are variations between ZigBee systems, software, hardware, and chipsets; not all attacks may be effective on all ZigBee Systems even if the ZigBee stack is an accepted standard. 

Energy depletion attacks such as Ghost-in-ZigBee \cite{Zigbee-Ghost} may prove to be effective against battery-powered E-IoT components. In addition to depleting ZigBee devices' power, it can facilitate threats such as DoS and replay attacks on a ZigBee network. The attack method involves sending false messages to nodes within a ZigBee network to trigger processor-intensive computations (e.g., cryptographic operations). The damage of these unnecessary computations is both power-based and performance based on the affected device. Ghost-in-ZigBee attacks also demonstrate three unique types of DoS attacks. First is a computational load attack, which can be done by sending numerous messages at the same time to trigger the depletion of a node's energy. However, such an attack could be easily detected with abnormality detection. The second type of DoS is referred to as MAC misbehavior, which takes advantage of ZigBee's channel sensing. When a targeted node receives continuous traffic, all nodes within that region will not communicate through that node. The third is a replay attack in which a malicious attacker may use frame counters greater than valid values in their message. Since ZigBee keeps an Access Control List (ACL) table, this table will be updated to match the malicious counter values. Any legitimate node trying to make contact after the alteration will be rejected due to their frame counter values being less than the altered values, leading to a malicious spoofing attack. The article \cite{IOTWireless} mentions a third attack on ZigBee spanning from hardware implementations. Going further in-depth, in \cite{ZigBee-Nuclear}, researchers attacked an implementation of an Atmel chip used with Phillips Hue bulbs and ZigBee Light Link (ZLL) mode. In this attack, the researchers created a custom circuit board to target the ZigBee chipset used with smart bulbs and created a worm to spread the infection among light nodes. 

In \cite{Zigbee-PracticalAttacks} three types of attacks on ZigBee were demonstrated using the KillerBee toolkit \cite{KillerBee}. The first attack takes advantage of ZigBee's discovery process and mimicked a legitimate device to gather information about other devices within the ZigBee network. This information spans various channels and will yield responses from ZigBee nodes within a channel. The second attack is the interception of packages. This attack functions on the basis that some ZigBee networks use weak or no encryption. As such, an attacker can eavesdrop on communication using the toolset and a USB adapter to capture traffic on a given channel. As the third proposed attack, if the previous two attacks are successful, an attacker can intercept and record ZigBee traffic. As such, an attacker can replay previously recorded packets and have ZigBee devices accept sent messages. Z-Wave vulnerabilities may depend on implementation practices, firmware, and hardware. Using reverse engineering methods, Fouladi et al. in \cite{fouladi2013security} provided some examples of available exploits that could compromise entire devices. The attack used Z-force, a packet interception, and injection tool, to reset the established network key and take advantage of the protocol's steps. The researchers describe the issue as a lack of 'state validation' in some Z-wave devices. An attacker can use packet injection to force Z-Wave devices to overwrite their current shared network key with an attacker-specified key. They demonstrated a successful attack on a connected door lock. While follow-up publications note that some of the attacks described have been patched, devices that have not been updated and usage of older firmware may be vulnerable to these attacks \cite{IOTWireless}.

The research by Fuller et al. explored vulnerabilities of rogue controllers within Z-Wave established networks ranges \cite{ZwaveRogue}. This work introduced an attack that used a malicious Z-Wave controller to attack unsecured devices. To begin, the researchers established a Home Automation Network (HAN) using Z-wave devices such as connected door locks, smart lights, and connected water valves. The attacker must first gain access to the local WLAN network to perform this attack, assuming the network is improperly secured. Once access to the network has been granted, an attacker can scan the network and retrieve the address of the Z-Wave gateway and any other gateways. The researchers then took advantage of known gateway vulnerabilities and, in this case, attacked a VeraEdge Z-Wave controller. Further, they retrieved and saved a backup file for the entire system. With this information, the researchers could then duplicate a legitimate Z-Wave controller with a malicious one in the same network. This rogue controller could then communicate to Z-wave devices within that network, compromising all of the available devices. The researchers also noted that with this backup file, there is the possibility that sensitive information and activity can be retrieved. Further, log files could also prove valuable to an attacker gathering information in usage or future attacks. 

\noindent\textbf{Bluetooth Attacks.} Another popular short-range wireless solution is Bluetooth. As mentioned earlier, Bluetooth is used by some E-IoT systems during standard operation and device configuration. Due to mobile devices, IoT, and other common use-cases of Bluetooth, attacks on Bluetooth have been widely documented, with a number of surveys written on the topic of Bluetooth security \cite{lounis2019bluetooth, shaked2005cracking, darroudi2017bluetooth, minar2012bluetooth, dunning2010taming}. Relevant to E-IoT, attacks highlighted in these surveys include man-in-the-middle attacks that can occur by compromising Bluetooth's Secure Simple Pairing (SSP) to impersonate trusted parties \cite{hypponen2007nino, sun2018man, haataja2010two, haataja2008practical, haataja2008man, barnickel2012implementing}. Further, another attack relevant to E-IoT is Bluesniping, which uses specialized antennas to sniff Bluetooth communication beyond the expected Bluetooth range \cite{hering2004bluetooth}. Bluesniffing attacks may also be a concern, as attackers may be able to infer E-IoT activity from sniffing packets coming from Bluetooth-based interfaces and devices \cite{spill2007bluesniff}. Disruption attacks such as Bluechopping, Bluecutting, and Bluedepriving may also affect the availability of E-IoT devices as these attacks all work to disrupt Bluetooth communication through different approaches \cite{lounis2020attacks}. For instance, for bluechopping, an attacker spoofs the identity of a connected Bluetooth device to cause a DoS condition. Bluecutting, an attack that disrupts Bluetooth communication by spoofing a Bluetooth device and requesting a target device begins re-pairing. As attacker then discards the stored link key and pairing can't be performed \cite{lounis2018connection}. Finally, bluedepriving interrupts Bluetooth communication by causing a conflict between a spoofed device and a legitimate device so that this legitimate device cannot pair through Bluetooth connection \cite{alsaidi2018security}. It must be noted that similar to other protocols, many Bluetooth attacks are dependent on implementations, software versions, and use-cases of Bluetooth devices.

\noindent\textbf{IR Attacks.} IR communication is used in E-IoT in the form of IR flashers to control integrated devices (e.g., displays, projectors, blinds). As such, most of these systems use simple, line-of-sight receivers without any form of authentication from the remote. Many of the controlling codes are available from online sources in websites such as remote central \cite{remotecentral}. As such, it is trivial for an attacker to capture or emit IR commands through line-of-sight \cite{IRHacking}. A malicious attacker could simply use an IR blaster to control IR-enabled devices and disrupt the operation of E-IoT systems \cite{IRTrolling}. In other cases, attackers may be able to reconfigure IR-enabled devices as if they had the original device remote. In terms of E-IoT, if a device is reconfigured or reset, an E-IoT system may not be able to communicate with these devices.

\noindent\textbf{General Wireless Attacks.} In this category, we cover any attacks that can apply to wireless in the The Industrial, Scientific, and Medical (ISM) frequency bands and is not unique to any communication protocol. Jamming attack can negatively impact E-IoT system communication in multiple modes of communication and fall under a specialized Denial-of-Service attack. Specifically, jamming presents a major threat to wireless networks and any E-IoT device that uses wireless networks (e.g., interfaces, sensors, relays), causing the devices to fall offline. Several works and surveys have covered jamming attacks against wireless communication that can be relevant to E-IoT systems \cite{jamming1, jamming2, jamming3, jamming4}. Specifically, these surveys highlight several proven jamming techniques against wireless networks (e.g., spot jamming, sweep jamming, barrage jamming, deceptive jamming). Further, jamming attacks are often cheap, easy to perform, and difficult to mitigate. The capabilities of more elaborate jamming attacks such as reactive jamming are covered by Wilhelm et al., highlighting the dangers of reactive jamming in wireless networks, where jamming techniques can target specific packets in wireless communication \cite{reactivejamming}. While reactive jamming may have limitations due to cost, demonstrations of jamming attacks show that an attacker can target specific wireless communication (e.g., ZigBee) with some technical knowledge and widely-available low-cost devices \cite{jammingzb}. 

\noindent\textbf{HDMI-Based Protocol Attacks.} HDMI is one of the core connections of video distribution and contains various protocols that can pose a threat to E-IoT systems. In HDMI-Walk, Puche et al. demonstrated that the CEC protocol can be used to gain arbitrary control of CEC-supported device functions \cite{hdmi-walk}. Specifically, the authors demonstrated that CEC can be used with HDMI distributions to attack multiple HDMI devices. The HDMI-Walk attacks further showed that an attacker might control devices, transfer information, cause DoS conditions, eavesdrop, and otherwise harm HDMI networks through a single point of connection or compromised device. For all of the attacks, the researchers inserted a device into an HDMI-capable distribution. The first attack used the inserted device to gather information about all of the connected HDMI devices, returning details such as the language, model number, power state, and running version. Two more attacks proved that eavesdropping and facilitation of existing attacks are possible with CEC. The authors showed that CEC could be used for unauthorized data transfer  by transferring audio information and WPA handshakes from one end of the distribution to another rogue device. Finally, there were two DoS attacks demonstrated in HDMI-Walk. On the first attack, the attacker device was configured to identify televisions powering on through CEC broadcast and shutting the displays down before they initiated. The second DoS attack abused television input change and overwhelmed displays through CEC, causing them to become inoperable. Further, the authors of HDMI-Walk noted that CEC propagation is not obvious and difficult to mitigate, creating networks without the user's awareness. Other relevant work on HDMI sub-protocols was published by the NCC group identified on CEC-based fuzzing vulnerabilities through CEC, and other viable threats through HDMI \cite{nccgroup}. Specifically, the NCC Group identified that HDMI's HEC channel could be used for corporate boundary breach, endpoint protection circumvention, and unauthorized network extension. Similar work presented by Smith et al. contributed to further CEC-based fuzzing with the development of the tool CECSTeR, used to execute CEC-based fuzzing attacks on CEC-supported devices \cite{smithcec, davisblackhat}.

\subsection{Mitigation of Communication Layer Attacks. }

\noindent\textbf{Serial-based Communication Defenses.} While not specific to E-IoT, research in serial-based communication defense mechanisms may apply to E-IoT. Studies by Dudak et al.~\cite{Dudak:2019}, and Wilson et al.~\cite{Wilson:2018} provide insight into securing serial-based protocols and considerations that must be taken to design protocols securely. Further, as standardization may help secure serial-based communication in ICS, the IEEE 1711.2 working group's efforts have focused on creating the Secure SCADA communications protocol \cite{SSCP:2019}. A similar approach has not been taken for proprietary E-IoT communication protocols yet, but could guarantee interoperability and secure protocol design in the future. In a survey by Volkova et al. highlighting attacks and defenses \cite{Volkova:2019}, the authors noted that network security, best practices, and software updates may help mitigate threats to Modbus and similar serial-based protocols. However, the authors noted that even with existing mitigation strategies, there are vulnerabilities that have to be mitigated by the protocol specifications. Finally, many proprietary protocols may require physical access to compromise, so controlling physical access may be a viable mitigation strategy.

For building automation protocols, vulnerabilities are often dependent on the implementation and installation. ASHRAE, the compendium behind BACnet, has released a security architecture to its initial construction for the deployment of a security layer for BACnet networks \cite{BacnetSecurity}. In the addendum, ASHRAE acknowledges the need to update the 56-bit DES cryptographic standard used for communication since 2004 to AES-128 bits. As several threats have been found in DES encryption, protocol updates are needed. Further, the BACNet specification explicitly notes that BACnet security encryption is optional and dependent on an integrator to be implemented. To keep E-IoT systems and related components secure, integrators should configure systems to use available encryption. Further, entities that create and maintain communication standards must update their protocols to newer cryptographic standards.

\noindent\textbf{WiFi Defenses.} In a similar manner to many technologies, one of the best solutions to defend against WiFi and other wireless vulnerabilities is ensuring that the most secure protocol implementations are in place in E-IoT devices. For instance, attacks and vulnerabilities such as Reaver have been patched in many modern routers \cite{wpahack}. Literature also references other solutions such as experimental defense mechanisms (e.g., custom key generation practices, modified WiFi standards); however, as most vendors and integrators cannot realistically implement these mechanisms, they are outside of the scope of this survey \cite{wang2010practical}. In a similar manner to the individual network configuration of devices, integrators should follow accepted best practices when configuring WiFi security, such as the ones suggested by the United States Federal Trade Commission \cite{ftcprotect}. For instance, access to a network should be limited, and routers should be secured with strong passwords, custom SSID names, with management features. Strong passwords practices can help mitigate handshake cracking and brute force attacks on WiFi. Further, using WEP is considered insecure and outdated, and as such, it should be avoided unless completely necessary. Other best practices were also highlighted by the Cybersecurity \& Infrastructure Security Agency (CISA) \cite{cisaprotect}. Some defenses proposed include installing firewalls, maintaining anti-virus software, frequent networking equipment updates, and following wireless configuration recommendations from manufacturers. Several attacks can be prevented through best practices and proper configuration in WPA/WPA2 devices. For instance, disabling features such as WPS in routers may be a good practice to prevent threats such as the Reaver and Pixie-dust attacks \cite{sanatinia2013wireless}. Surveys conducted on WiFi security also suggest that if it is possible, users should update their systems to the latest WPA3 security standard, however acknowledge that this is not ideal in all cases \cite{lounis2020attacks, lashkari2009survey}. Further, these surveys note that proper configuration of WPA3 can prevent key cracking attacks.

\noindent\textbf{ZigBee/Z-Wave Defenses.} One of the best solutions to Zigbee and Z-wave protocol vulnerabilities is verifying that vendors use the latest and the most secure protocol implementations. Further, E-IoT integrators should follow the best practices offered by E-IoT vendors and manufacturers. A survey by Lounis et al. highlighted how updated protocols have resolved many attacks for short-range wireless protocols \cite{lounis2020attacks}. This survey also highlights that network administrators (integrators - in E-IoT systems) should monitor and verify that devices are properly configured and updated. However, as users and integrators of E-IoT systems rely on E-IoT device manufacturers and vendors, solutions for vulnerabilities will come from vendor updates and best practices. For instance, manufacturers of E-IoT controllers must make sure that short-range wireless nonces are not reused to prevent key generation attacks \cite{whitehurst2014exploring}. Additionally, a work by  Benzaid et al. highlighted that polling messages and responses should also be authenticated to prevent spoofing attacks on short-range wireless networks \cite{benzaid2016fast}. An article published in 2006 on Z-wave security highlighted the main differences between ZigBee and Z-wave security \cite{ZwaveHowSecure}. The article noted that Z-wave protocol encryption is optional and for that reason, encryption should always be enabled as a security measure. The study also noted that older Z-Wave systems are open to various attacks, especially if encryption has not been enabled. As such, maintaining systems properly updated and securely configured should be a priority for E-IoT communication.

\noindent\textbf{Bluetooth Communication Defenses.} In a similar manner to other wireless defenses, one of the best solutions for Bluetooth attacks are updates and making sure that best practices are followed in Bluetooth configuration. A set of Bluetooth-specific best practices have been proposed in \cite{bluetoothdefense}. For instance, disabling Bluetooth functions when they are not in use, disabling device ID broadcast, strong passwords, and verifying incoming transmissions have been suggested to mitigate Bluetooth-based threats. Further, a survey by Lounis et al.~\cite{lounis2020attacks} notes that Bluetooth software updates are necessary to defend against well-known Bluetooth attacks (e.g., bluechopping, bluecutting, bluedepriving). 

\noindent\textbf{IR Communication Defenses.} As IR communication is line-of-sight, physical security may be one of the best defense approaches. With very little literature on IR communication and defenses, it may be an idea for integrators to cover IR receivers when not in use to prevent attackers from tampering with devices. Further, access control may prevent unauthorized users from disrupting the operation of E-IoT-controlled devices using IR emitters. As IR requires line-of-sight, it may be easy to discern when an attacker is meddling with a device. Further, as CCTV can display the IR spectrum, it may be possible to use cameras to identify an attacker using IR to communicate with devices \cite{IRsee}.

\noindent\textbf{General Wireless Communication Defenses.} Securing wireless communication from jamming attacks has been a topic of research with a number of different approaches suggested. Numerous surveys are available on wireless jamming defenses and counter-measures \cite{jamming1, jamming2, jamming3, jamming4, jamming5, jamminganalysis, jammingdetection, jammingstatisticaldef}. As such, a solution for E-IoT deployments will depend on the wireless technology used and the particular wireless use-cases. A survey on this topic by Aristides et al. divides anti-jamming approaches into three different types: proactive, reactive, and mobile agent-based solutions \cite{jamming3}. Proactive counter-measures in the background cannot be initiated, stopped, or resumed on demand and require prolonged implementation time and high implementation cost. An example of proactive measures are software and hardware-based solutions that detect jamming attacks before they occur (e.g., DEEJAM) \cite{wood2007deejam}. Reactive anti-jamming approaches reduce computation energy costs compared to proactive counter-measures. Reactive jamming defenses rely on active jamming attacks and aim to mitigate attacks (e.g., JAM) \cite{wood2003jam}. However, in the case of some jamming attacks, reactive-anti-jamming may have some detection delays. Finally, mobile agent-based solutions employ anti-jamming agents that move between hosts to detect jamming attempts. For different protocols, there exist different jamming approaches, subject to surveys of their own. Vendors of E-IoT should consider the best defenses for their supported devices and implement them in their systems.

\noindent\textbf{HDMI-based Communication Defenses.} While HDMI sub-protocols are usually secured through restrictions to physical access; other options have been explored. For instance, in work proposed by Puche et al., the authors created a passive intrusion detection system framework designed to protect against HDMI-based threats \cite{hdmi-watch}. The framework uses features in CEC communication to build a machine learning classifier and does not require modification to the original protocol, as a modification to the protocol is problematic, with billions of HDMI devices distributed worldwide. Physical defenses against these attacks involve the use of CEC-less adapters, which can prevent CEC signal from propagating over large distributions \cite{ceclessadapter}. As such, an integrator may use a CEC-less adapter to prevent public, easily-reachable HDMI endpoints from receiving CEC communication.

\begin{table*}[t]
    \centering
    \caption{Summary of E-IoT Application Layer Attacks, Threats, and Mitigations.}
    \scalebox{1.00}{
    \begin{tabular}{c|c|c}
    \hline
       Component & Attack/Threat & Mitigations\\ 
    \hline
        E-IoT Drivers
            & Driver-based Attacks \cite{rondon2020poisonivy} & Treat drivers as untrusted software, avoid unverified drivers, user awareness. \cite{rondon2020poisonivy}\\
        \hline
        \rowcolor{Gray}
        E-IoT Software \& Services
            & Software Vulnerabilities \cite{LawshaeCrestron, CVECrestron, CVESavant, presentationvulns} 
                & Frequent updates, isolated E-IoT devices, legacy equipment considerations, vulnerability awareness  \cite{patchedvulns}.\\
            & Improper Encryption \cite{chovulns, rollingyourcrypto, cybergibbonsdualcom}
                & Accepted encryption methodology, software verification \cite{rollingyourcrypto}.\\
        \hline
        \rowcolor{Gray}
        E-IoT Configuration
            & Remote Hijacking \cite{miraibotnet, Dahua} 
                & Strong passwords, avoid port forwarding, VPNs, configuration best practices \cite{crestronSecurity, synacktips, williamsportfor} \\
    \hline
    \end{tabular}}
     \vspace{-0.1in}
     \label{tbl:applicationlayerover}
\end{table*}

\section{Monitoring and Applications Layer}\label{sec:applicationslayer}

In this section, we highlight the monitoring and application layer of E-IoT systems. First, we cover elements of the E-IoT monitoring and applications layer. We then introduce the threat model at this layer. Third, we cover monitoring and application-layer threats and attacks. Finally, we provide relevant defenses and mitigation mechanisms. An overview of attacks, threats, and mitigation strategies covered in this section are outlined in Table \ref{tbl:applicationlayerover}.

\subsection{Elements of the Monitoring and Applications Layer}

E-IoT Monitoring and Applications Layer consists of E-IoT drivers, E-IoT software and services, and E-IoT configuration.

\noindent\textbf{E-IoT Drivers.} As introduced in Section \ref{sec:background}, drivers are an important part for E-IoT system functionality. As a software-based component of E-IoT systems, drivers provide all the information necessary for an E-IoT system to integrate a device or web service into the system. As such, E-IoT drivers are not standardized from system-to-system and may be known under a different name (e.g., Crestron modules, Control4 Drivers) \cite{rondon2020poisonivy}. Drivers are inserted and configured in an E-IoT system during programming or maintenance by integrators. Thus, drivers can be obtained in three different ways. (1) Drivers may be acquired directly from the E-IoT configuration software. (2) Drivers may be acquired from a catalog of drivers provided by the main E-IoT vendor. (3) Drivers can be downloaded from third-party sites (e.g., forums, device vendors, third-party developers). However, while many vendors will validate drivers acquired from their software or repositories, drivers from third-party sources are often not checked for malicious content. Additionally, some drivers are not free which may tempt integrators to use a free, unverified driver with malicious code available online \cite{blackwiredrivers}.

\noindent\textbf{E-IoT Software Services.} E-IoT systems use several software services for configuration and maintenance. Beyond proprietary tools used by E-IoT vendors, such as Control4's composer and Crestron's Simpl, E-IoT uses common application services  \cite{control4composerrelease, crestronsoftware}. Available software services may vary from system to system. While E-IoT systems may have well-known, documented software services such as File Transfer Protocol (FTP), Secure Shell (SSH), and Telnet communication, E-IoT solutions may also run unknown proprietary services. With the closed-source nature of many E-IoT systems, documentation and details of these proprietary services remain mostly unavailable. As such, operating manuals available online and troubleshooting guides are among the few sources of information on these services. In contrast, well-known and commonly-used services are easier to identify. For instance, file transfer is necessary for E-IoT tasks such as firmware upgrades, image uploads, and vendor software configuration. As such, one of the accepted file-transfer services is FTP, and for more secure communication, Secure FTP (SFTP) \cite{whatisftp, ftpsftp}. Another requirement for E-IoT is diagnostics and configuration; thus, integrators need to communicate directly to the E-IoT system. Secure shell services may be used for diagnostics and configuration as integrators use secure shell clients such as PuTTy to connect to, diagnose, and configure E-IoT system and system components through services such as Telnet or SSH \cite{control4putty, putty}. Another use-case of software services is webservers and web interfaces using HTTP or HTTPS. E-IoT systems may host webservers and web interfaces to allow integrators to configure, diagnose, or monitor devices. For instance, CCTV systems host a web interface to configure cameras, view recordings, view a live feed, and manage CCTV systems \cite{CCTVweb}. Finally, software suites such as Busybox are common in IoT and E-IoT alike, as BusyBox provides many common UNIX utilities in a compact executable with size optimization and a modular design \cite{busybox, busyboxconfig}. Due to the convenient design, E-IoT vendors such as Control4 run BusyBox on their main controllers and devices \cite{control4busybox}. 

\noindent\textbf{E-IoT Configuration.} Beyond software, configuration of E-IoT systems can impact the overall security of the system. Some E-IoT users may need to access E-IoT system features remotely. Additionally, remote access aids integrators, as it allows them to provide remote technical support, especially in moving installations such as yachts. As such, E-IoT vendors and integrators permit remote access through a variety of different methods. While the configuration is different for each system, most E-IoT systems are accessed remotely through subscription services, virtual private networks (VPNs), or port forwarding. First, some vendors offer subscription services, creating a secure and easy way for clients and integrators to connect remotely to an E-IoT system (e.g., Control4's 4Sight) \cite{control4sight}. VPNs are another popular solutions recommended by many vendors, granting users remote access to the E-IoT network and equipment. For this reason, vendors will recommend routers with VPN functionality to integrators. Finally, as E-IoT devices (e.g., controllers, CCTV NVRs) often use ports for control and configuration, integrators often port forward these devices to allow remote access \cite{cctvsetup, CCTVweb, crestronSecurity}.

\subsection{Threat Model for E-IoT Monitoring and Applications Layer}

For monitoring and application layer threats, we consider Mallory, an attacker knowledgeable on configuration and software vulnerabilities of E-IoT systems. As such, an attacker on this layer compromises E-IoT functionality and may gain access to unauthorized resources without any physical contact with E-IoT systems. For this layer, an attacker needs technical knowledge of E-IoT systems and software-based attacks. Mallory can be in the roles of malicious users, integrators, or remote attackers. 

\noindent\textit{Users.} Mallory, in the role of a frequent or visiting malicious user, could attack E-IoT systems through the monitoring and applications layer. As malicious actors, these users may attempt privilege escalation, modify E-IoT systems, or otherwise try to cause unintended operation. As regular users are meant to operate E-IoT systems and not alter any configuration, vulnerabilities may allow Mallory to compromise E-IoT system components as an unprivileged user. Further, in an improperly configured network, if Mallory has network access and proprietary configuration software, she may modify E-IoT software, remote access configuration, or compromise an E-IoT system through software (e.g., malicious drivers).

\noindent\textit{Integrators.} Integrators often will have full access to E-IoT systems. As a malicious integrator, Mallory may become the attacker in certain situations (e.g., bribed or disgruntled employees \cite{ransomwareWired}). In this scenario, Mallory already has the proprietary tools and access to one or many E-IoT systems through maintenance software. Mallory could inadvertently compromise multiple systems using malicious drivers or remote access tools. Further, Mallory may target wealthy or famous clients and eavesdrop on information for personal gain or otherwise disrupt E-IoT system operation. 

\noindent\textit{Remote Attackers.} Mallory may be a remote attacker seeking systems to compromise. She may find E-IoT systems exposed to the Internet. If Mallory is a more capable attacker, she may use configuration tools and manuals used by E-IoT vendors to gain complete access to E-IoT systems, install malicious E-IoT drivers, and otherwise compromise exposed systems.

\subsection{Monitoring and Applications Layer Attacks and Vulnerabilities}

In this subsection, we cover monitoring and application layer attacks and vulnerabilities. 


\noindent\textbf{E-IoT Driver Attacks.} As explained in Section \ref{sec:background}, E-IoT drivers contain all the programming necessary for E-IoT systems to integrate third-party components such as devices, APIs, and web services \cite{control4composerrelease, control4driversearch}. 
Research on the topic of E-IoT drivers by Puche et al. demonstrated that drivers can be used to compromise E-IoT systems \cite{rondon2020poisonivy}. Specifically, the authors performed a DoS attack, maliciously expended system resources, and assumed control of the E-IoT controller's networking functions through malicious E-IoT drivers. The authors note that an integrator may inadvertently compromise an E-IoT system by downloading unverified drivers from third-party vendors, forums, or any external site. Since there is no verification mechanism for drivers in E-IoT controllers, an attacker can gain the control of the E-IoT system.

\noindent\textbf{E-IoT Software Service Attacks.} E-IoT systems will run a combination of proprietary services and well-known services in their devices. As such, some vulnerabilities have been exposed by researchers on E-IoT systems. For instance, in Defcon 26 (2017), Lawshae et al. presented several Crestron controller vulnerabilities \cite{LawshaeCrestron}. Specifically, Crestron controllers could be compromised through the CTP console, a Telnet-like interface for Crestron E-IoT systems used to configure and diagnose Crestron devices. This interface also allowed Lawshae to have direct chip communication, browser remote control, UI interaction, and microphone recording capabilities. Further, as of the time of this writing, CVE Details show over twenty vulnerabilities for Crestron devices and six for Savant systems \cite{CVECrestron, CVESavant}. For Savant and Crestron systems, these vulnerabilities include de-authentication code overflow, authentication bypass, remote code execution, directory traversal, cross-site request forgery, and DoS. Vulnerabilities have also been discovered in presentation devices and systems. For instance, Crestron presentation devices, Barco wePresent, and Extron ShareLink presentation systems have had numerous vulnerabilities discovered (e.g., stack overflows, unauthenticated command injection) as they all share underlying code  \cite{presentationvulns}. Vulnerability research by Synack, a company that specializes in security research, tested the now discontinued SR-250 Control4 controllers and found several unpatched vulnerabilities described as unauthenticated management vulnerabilities \cite{chovulns}. Moreover, improper implementation of encryption could threaten the confidentiality and integrity of E-IoT data. Practices such as 'rolling your own encryption' (e.g., implementing self-made cryptographic functions and algorithms) have left products from companies (e.g., Dualcom, Telegram) vulnerable to attackers \cite{rollingyourcrypto}. For instance, Dualcom alarm signaling products were demonstrated to be vulnerable and susceptible to cracking attacks due to improper use of encryption mechanisms \cite{cybergibbonsdualcom}. As such, improperly implemented encryption can open up E-IoT components to a great number of attacks (e.g., malicious sniffing, brute-force, man-in-the-middle, replay). 

\noindent\textbf{E-IoT Configuration Attacks.} One of the most notable examples of a failure in IoT security was made abundantly clear with the Mirai botnet, which overwhelmed high profile targets through DDoS attacks. The malware hijacked exposed IoT devices and used them to create a botnet. How the Mirai malware grew to a peak of six-hundred thousand infections so quickly is one of the reasons why users should be wary of the security of their connected E-IoT systems and other Internet-facing devices\cite{miraibotnet}. Research on this botnet revealed issues with the current state of exposed IoT and E-IoT devices. Mirai created a bank of targeted devices with 46 unique passwords. Most of these passwords targeted exposed systems such as security cameras, CCTV video recorders, routers, and printers. Initially, Mirai used this bank of default passwords to brute force through Telnet and SSH authentication. Future iterations of Mirai altered themselves to attack through known exploits in targeted systems. Attacks such as these could also take advantage of known backdoors, such as those seen in Dahua DVRs and IP cameras, where a firmware had to be released for all installed devices due to found vulnerabilities \cite{Dahua}. An attacker could compromise an E-IoT system through port forwarded devices. As of the writing of this paper, a search in Shodan.io, a search engine for exposed devices connected to the Internet reveals over 30,000 E-IoT devices exposed online from major vendors (Control4, Crestron, Savant, and Lutron)~\cite{Shodanio}.

\subsection{Mitigation of Monitoring and Applications Layer Attacks} 

\noindent\textbf{Driver Defenses.} E-IoT vendors will often provide a number of drivers or validate drivers developed by third parties. As such, integrators should try to use validated drivers to prevent driver-based threats. The work that presented driver-based attacks highlighted that vendors should approach drivers in a similar manner to untrusted software \cite{rondon2020poisonivy}. Further, without standardization, drivers are implemented differently in each system; thus, security mechanisms that are viable for one E-IoT system may not be viable for others. A proposed solution is a permission system for drivers, based on the function and what a driver should be allowed to do (e.g., a serial-based controlled device should not have a driver with network connectivity) \cite{rondon2020poisonivy}. Finally, many users and integrators may not be aware that malicious code could exist in drivers and thus, awareness of this possible threat is one of the best and only current defenses.

\noindent\textbf{E-IoT Software and Services Defenses.} In a similar manner to any smart system; vendors, users, and integrators should follow patching and firmware best practices. As E-IoT vendors will note and often patch vulnerabilities with later releases, integrators should install the latest software and firmware versions. Moreover, users should schedule frequent product updates \cite{patchedvulns}. Following frequent updates and patching in E-IoT systems can help mitigate known service vulnerabilities from software services. Further, overall awareness on running services and versions can help integrators gauge the risk of exposing E-IoT components to a network. It may be possible to anticipate unpatched vulnerabilities and prevent an attack before it occurs. As such, integrators may want to isolate E-IoT systems from other networked systems (e.g., guest-accessible networks) and enable proper network-based mechanisms to prevent unauthorized access. Additionally, legacy and discontinued equipment that cannot be upgraded or updated presents a major threat to many smart systems beyond E-IoT, especially Internet-facing systems. Integrators need to be aware of legacy equipment and make sure that their clients are aware of the risks of legacy equipment. Finally, E-IoT developers should avoid mistakes during development such as improper encryption mechanisms by using the latest libraries, avoiding custom encryption, and following verification processes \cite{rollingyourcrypto}.

\noindent\textbf{E-IoT Configuration Defenses.} E-IoT vendors will often release security best practices, and integrators should follow these best practices for configuring E-IoT systems~\cite{crestronSecurity}. Moreover, installers and users should avoid weak and insecure passwords as Internet-facing devices with weak password practices have allowed attackers to compromise devices in previous large-scale automated attacks \cite{miraibotnet}. A whitepaper published by Synack provided an outline relevant to E-IoT, and professionally-installed systems \cite{synacktips}. Proposed best practices from this guide highlighted that vendors and manufacturers should not rely on users for security. Basic password strength requirements should be enforced, as compromising a remote access account could give an attacker access to an E-IoT system. Users should also receive notifications when device statuses change or when sessions are initiated. Finally, the whitepaper notes that vendors should avoid SSL pinning, self-signed certificates, and custom encryption. One other source of vulnerabilities is port-forwarding, which exposes devices to the Internet. E-IoT vendors have always advised dealers and users not to port forward devices as some devices were not designed to be exposed directly to the Internet \cite{williamsportfor}. Instead, integrators and users should opt for VPN configuration or vendor remote access services.

\begin{table*}[t]
    \centering
    \caption{Overview of E-IoT Business Layer Attacks, Threats, and Mitigations.}
    \scalebox{1.00}{
    \begin{tabular}{c|c|c}
    \hline
       Component & Attack/Threat & Mitigations\\ 
    \hline
        Security and Vendor 
            & Data Breaches \cite{de2019cyber, liu2015survey, ryan2013cloud, shahzad2014state, subashini2011survey, grobauer2010understanding, modi2013survey, singh2016survey, fernandes2014security, polash2014survey, singh2017cloud, xiao2012security, ardagna2015security, hashizume2013analysis}
                & Storage encryption, strong access control, accepted authentication practices, server-side \\
        cloud services & & encryption.  \\ 
        \rowcolor{Gray}
            & Insecure APIs and Web Services \cite{de2019cyber, liu2015survey, ryan2013cloud, shahzad2014state, subashini2011survey, grobauer2010understanding, modi2013survey, singh2016survey, fernandes2014security, polash2014survey, singh2017cloud, xiao2012security, ardagna2015security, hashizume2013analysis}
                & Vulnerability scanning, malware protection, browser updates, data sanitation, web development\\
            \rowcolor{Gray}
                && best practices.\\
            & Unauthorized Intrusion \cite{de2019cyber, liu2015survey, ryan2013cloud, shahzad2014state, subashini2011survey, grobauer2010understanding, modi2013survey, singh2016survey, fernandes2014security, polash2014survey}
                & Network-based, host-based, distributed, and hypervisor IDS. \\
            \rowcolor{Gray}
            & Account Hijacking \cite{de2019cyber, liu2015survey, ryan2013cloud, shahzad2014state, subashini2011survey, grobauer2010understanding, modi2013survey, singh2016survey, fernandes2014security, polash2014survey, singh2017cloud, xiao2012security, ardagna2015security, hashizume2013analysis}
                & Strong access control and authentication, two-factor authentication, web development best \\
            \rowcolor{Gray}
                && practices \cite{fernandes2014security, kumar2019cloud}.\\
            & DoS Attacks \cite{de2019cyber, liu2015survey, ryan2013cloud, shahzad2014state, subashini2011survey, grobauer2010understanding, modi2013survey, singh2016survey, fernandes2014security, polash2014survey, singh2017cloud, xiao2012security, ardagna2015security, hashizume2013analysis}
                & Rate-limiting, structured schemes, crypto puzzles, proxy filtering, load balancing reputation-based  \\
                && access control, and flexible network configuration \cite{fernandes2014security, kumar2019cloud}.\\
    \hline
    \end{tabular}}
     \vspace{-0.1in}
     \label{tbl:businesslayerover}
\end{table*}

\section{Business Layer}\label{sec:businesslayer}

In this section, we highlight the E-IoT business layer and common security concerns. Specifically, we first address common business-layer components of E-IoT systems. Second, we highlight possible threats and attacks. Finally, we cover possible defense and mitigation mechanisms. An overview of components, threats, attacks, and mitigations discussed in this layer can be found in Table \ref{tbl:businesslayerover}.

\subsection{Elements of the E-IoT Business Layer}

In this subsection, we highlight common elements of the E-IoT Business Layer.

\noindent\textbf{Security Cloud Services.} E-IoT CCTV systems usually record camera footage in local hard disk storage in a Digital Video Recorder (DVR) or a Network Video Recorder (NVR) for analog or IP cameras respectively \cite{dvrnvr}. However, if a DVR or NVR is damaged or stolen in traditional systems, all video recordings are lost. Moreover, CCTV systems have limited storage space and will often overwrite old recordings with new footage once storage runs out. As a solution to this issue, vendors offer online cloud storage solutions designed specifically for security cameras and CCTV. In addition to cloud storage, security cloud services allow integrators and users to manage multiple E-IoT deployments from a single hub. For instance, services beyond cloud storage include health checks, remote access, and remote camera control for the end-user. Another feature of security cloud services is machine-learning-based tagging and human activity recognition on recorded video, with providers such as Camio providing these features \cite{camio}. With this feature, recorded video data can be labeled by events (e.g., van passing, pizza delivery, red shirt), allowing users to search for a specific events in stored recordings easily \cite{humanact1, humanact2}.

\noindent\textbf{Vendor Services.} E-IoT vendors will often offer cloud services for monitoring and maintenance of E-IoT systems. These services serve a variety of purposes as maintenance is an important part of E-IoT deployments. First, maintenance services monitor integrated devices in the E-IoT system network. As such, an integrator can know when a device falls offline and can address the issue before a client notices. Second, maintenance services can send integrators and clients phone and email notifications on needed updates, unplanned downtime, Internet failure, and ongoing network issues. Services such as Pakedge's Bakpak are designed to work with E-IoT and provide vendors with many features. As such, they allow integrators to provide both remote support, monitoring, and maintenance to multiple user E-IoT systems \cite{pakedgebakpak}. Cloud services are also used in E-IoT for secure remote access to E-IoT systems. While not all of E-IoT systems offer this service, major E-IoT vendors or device makers always offer some form of remote access solution. For instance, services such as Control4's 4sight offer users remote access through mobile apps and cloud support. Further, 4sight also allows integrators to service a specific E-IoT system remotely through a secure connection.

\subsection{Threat Model for E-IoT Business Layer}

For this layer, we consider Mallory compromising an E-IoT system through the E-IoT Business layer. As such, Mallory is knowledgeable about cloud and remote services. Specifically, Mallory knows how to use business layer services (e.g., security cloud, vendor maintenance) to compromise one or multiple E-IoT systems remotely. As an attacker, in addition to knowledge on integrated services, Mallory must possess an Internet connection,  knowledge on business services, and capabilities to perform phishing attacks, dictionary attacks, or web-based attacks. Mallory can be in the roles of hackers or integrators to target business layer of E-IoT. 

\noindent\textbf{Hackers.} In this scenario, Mallory may be a remote attacker with or without prior knowledge of E-IoT systems. Mallory may target E-IoT cloud and vendor services and disrupt these systems. If Mallory is a more knowledgeable attacker, she may be aware that E-IoT systems can be compromised through management services. Mallory may acquire sensitive information from E-IoT systems such as CCTV recordings, schedules, and E-IoT usage patters. Further, Mallory can also use phishing techniques (e.g., texting, email, apps) to obtain a user's or integrator's credentials and compromise one or multiple systems.

\noindent\textbf{Malicious Integrators.} Mallory may be a malicious integrator or an insider in an integrator company with access to user accounts and credentials for remote support. As such, Mallory can become a malicious actor (e.g., disgruntled employees, insiders) and compromise a user's E-IoT system to disrupt or for personal gain. Additionally, as Mallory is an integrator or an employee, she may know E-IoT user's financial or societal status. This may tempt Mallory to sell information (e.g., passwords, accounts, CCTV footage) of users to external attackers for financial gain.

\subsection{Business Layer Attacks and Vulnerabilities}

\noindent\textbf{Cloud Attacks.} As cloud services are a part of E-IoT, cloud service threats and attacks are relevant to E-IoT systems. While architectures may vary from system to system and service to service, threats to integrated cloud services could negatively impact E-IoT system security. As an active topic of research, several surveys have highlighted threats, attacks, and best practices in cloud computing \cite{de2019cyber, liu2015survey, ryan2013cloud, shahzad2014state, subashini2011survey, grobauer2010understanding, modi2013survey, singh2016survey, fernandes2014security, polash2014survey, singh2017cloud, xiao2012security, ardagna2015security, hashizume2013analysis}. Relevant to E-IoT, surveys by Liu et al., Ryan, and Shazad have highlighted several key challenges to cloud security \cite{liu2015survey, ryan2013cloud, shahzad2014state}. For instance, these studies suggest that cloud components are susceptible to DDoS attacks and that encryption solutions will not protect sensitive data if the cloud provider cannot be trusted due to insiders. A comprehensive survey by Fernandes et al. raised additional issues which may occur with cloud computing \cite{fernandes2014security}. Issues related to unreliable computing, data storage, availability, cryptography, sanitation, and malware can arise from systems that rely on cloud services. Further, this survey highlights how keyloggers, phishing, malicious redirects, URL-guessing attacks, browser vulnerabilities, cross-site scripting, XML/SAML wrapping attacks, and MitM attacks may impact cloud services. Finally, a survey by Kumar et al. highlights some of the common cloud security threats, such as data breaches, weak access control, insecure APIs, application vulnerabilities, account hijacking, malicious insiders, advanced persistent threats (APT), data loss, nefarious use of cloud services, DoS, and DDoS \cite{kumar2019cloud}. In terms of E-IoT, these threats could mean that E-IoT users may lose access to their accounts, face information theft, or experience system downtime from integrated or vendor-provided cloud services. 

\vspace{-0.05in}
\subsection{Mitigation of Business Layer Attacks}

\noindent\textbf{Cloud Defenses.} Several defenses have been proposed for threats that can impact cloud-based services. In this respect, several surveys have been conducted on the topic of cloud security, often highlighting attacks, defenses, and mitigation mechanisms relevant to E-IoT cloud services \cite{fernandes2014security, ryan2013cloud, singh2017cloud, ardagna2015security, modi2013survey, kumar2019cloud, singh2016survey}. Majority of these works note that there are many ways to attack different types of cloud systems. As such, different mitigation strategies exist for each threat. For instance, to defend against data breaches, properly implemented encryption should be used by cloud services. Vendors should require strong passwords and authentication practices in their cloud services to address weak access control that could compromise users, integrators, and E-IoT systems. Further, as accounts may be compromised, some articles have suggested that two-factor authentication may add an additional layer of security to cloud services \cite{twofactorcloud}. As browser vulnerabilities such as XSS and redirection attacks can impact web-based GUI interfaces, users should update browsers, have malware protection, and follow best practices to prevent web-based vulnerabilities. Surveys by Fernandes et al. and Kumar et al. specify mitigation strategies against cloud threats~\cite{fernandes2014security, kumar2019cloud}. For instance, APIs should be protected with good sanitation practices, secure development standards, signatures, and encryption. To prevent intrusion in cloud systems, the authors highlight that  network-based, host-based, distributed, and hypervisor intrusion detection systems can be helpful. DDoS mitigation can improve the overall reliability of cloud services in case of volumetric attacks. Specifically, DDoS mitigation strategies may take the form of rate-limiting, proxy filtering, load balancing, crypto puzzles, and flexible network configuration depending on the cloud system and use case. As such, E-IoT vendors and manufacturers should follow these practices to guarantee the security of cloud-based components used in E-IoT. 

Cloud hosting can also benefit from privacy-preserving techniques to protect a user's information. For instance, cloud service providers can provide an additional layer of mitigation by applying Homomorphic Encryption (HE) concepts to stored information \cite{acar2018survey}. With HE, concepts such as Partially Homomorphic Encryption (PHE), Somewhat Homomorphic Encryption (SWHE), and finally Fully Homomorphic Encryption (FHE) can be applied to improve data privacy. Specifically, PHE, SWHE, and FHE allow for a number of operations (depending on the type) on encrypted data without the need to decrypt the data for these operations. This allows users to store encrypted information, without the risk of exposing sensitive information to untrusted cloud providers. While these approaches are experimental and require further research, they should be considered for cloud services for E-IoT systems and storage.

\section{Lessons Learned and Open Issues}\label{sec:lessonslearned}

In this survey, we analyzed the threats and defenses concerning individual layers. However, an attacker can follow a cross-layer approach, which means he/she can attack multiple layers at once. So the security of E-IoT systems should be considered holistically. In this section, we cover lessons learned and open issues in state-of-the-art E-IoT security research.

\subsection{Lessons Learned}

\noindent\textbf{Customized Deployments.} E-IoT deployments are diverse and complex. There may be unique threats from deployment to deployment, especially with the numerous use cases in E-IoT. For instance, a lightning control E-IoT environment will be different from a smart media management system in terms of vulnerabilities. Specifically, a lighting system may rely more heavily on serial-based communication interfaces in every room than media management that relies on touchpanels and mobile interfaces. Further, even in a lighting system, a purely serial-based system will have different threats and vulnerabilities than a lighting control system with Zigbee/Z-wave interfaces. Many attacks (e.g., node-capture attacks, sensory channel attacks) may have unique system-to-system consequences. Namely, if safety or motorization devices rely on E-IoT sensors, an attacker may create a much bigger safety issue if these sensors are compromised. The degree of customization in E-IoT means that one deployment's solutions and security guidelines are not applicable for all deployments. 

\noindent\textbf{Legacy Systems.} Legacy systems are systems considered to be outdated, discontinued, and that no longer receive software or security updates. With companies such as Crestron established since the 70's, it is expected that there are legacy E-IoT systems installed worldwide \cite{crestronorigins}. As these systems may not receive updates for several reasons. For instance, an E-IoT system may simply be discontinued or the manufacturer may no longer exist (e.g., Litetouch lighting control systems) \cite{litetouchmanual}. Alternatively, systems have frequent updates and a user may choose not to update because of cost of the update (e.g., new devices, software, updated drivers). For example, if an entire building is wired to function with a legacy system, re-wiring and re-programming may be a costly endeavor as opposed to simply keeping an older E-IoT system. In other cases, discontinued devices may not work on newer systems (e.g., driver availability) and a user might choose to keep the current E-IoT system without updating to keep control of these integrated devices. E-IoT needs unique solutions that can provide protection to legacy E-IoT systems which cannot be updated.

\noindent\textbf{Reliance on Integrators.} In E-IoT systems, consumers rely upon integrators. This reliance on integrators may create attack scenarios where an E-IoT is compromised because of this trust. For instance, bad account management could allow attackers to compromise one or multiple systems purely due to integrators and remote support tooling. Additionally, as integrators handle devices before installation, they can be considered part of the supply chain. This adds another stage to the deployment process where devices may be compromised by an attacker or a malicious employee for an installation company. Integrators maintain and diagnose E-IoT devices in case of any issues, working directly with the client. As such, there is very little oversight on how well systems are configured. As poorly-configured devices pose a threat to E-IoT systems, a method for auditing E-IoT system security may be necessary to guarantee systems are properly configured. Further, more research into E-IoT security can assist vendors in evaluating existing best practices for integrators and developing new best practices. Finally, as E-IoT relies very heavily on integrators, new solutions are needed to protect end-users against attacks that may come from malicious insiders or poor configuration. 

\noindent\textbf{(Near) Future E-IoT Attacks.} Attackers are in constant search of new systems to attack, with nation-state attackers having the capabilities to perform attacks never thought to be viable before. Attacks have already been observed that target E-IoT devices among other devices with Mirai being one of the most well-known. In Mirai, research has shown that CCTV systems, specifically DVRs and NVRs were targeted in the password banks \cite{miraibotnet}. These devices were possibly configured with default passwords in many cases and reflects the need for auditing and better research on E-IoT. If research is not done, E-IoT systems may end up being used in large-scale attacks, such as DDoS. Attacks would not be limited to DDoS, ransomware attacks may be different in E-IoT, both rendering a system inoperable and requiring an integrator to repair the affected system. In more advanced attacks, it may be possible for an attacker to compromise touchscreens, keypads, controllers, and other devices for cryptomining through malicious firmware or a malicious controller. Another recent and notable example of an attack has been coined the ``SolarWinds'' attack, where thousands of devices were compromised through vendor tool updates \cite{oxfordfallout, solarwinds}. It may be possible for E-IoT to be affected by similar attacks in the future if precautions are note taken. This shows that trusted software must also be held to high scrutiny. Finally, there are the privacy aspects of E-IoT. The cost of E-IoT systems means that clients may be wealthy professionals, well-funded companies, or otherwise lucrative targets for attackers, especially in E-IoT smart homes. An attacker may predict the cost of an E-IoT system and attack a target that can benefit them more (e.g., businessmen, politicians, wealthy, companies), easily searching by vendor online and targeting an exposed, poorly configured system.

\subsection{Open Issues}

\noindent\textbf{Proprietary Communication.} E-IoT supports a diversity of protocols, from publicly-known to proprietary. However, proprietary protocols in E-IoT are often closed-source, with no public specification. Additionally, E-IoT hardware and software are often unavailable to the public. Researching E-IoT communication can be difficult without vendor cooperation as much of the protocols and practices need to be reverse-engineered. As a result, many E-IoT components and protocols have not been properly investigated, and many attacks have not been discovered and addressed. For instance, serial-based proprietary protocols such as Cresnet are used extensively in Crestron systems; however, little to no security research exists on this protocol. This case is also valid for many wireless protocols as highlighted in Section \ref{sec:communicationlayer} such as RadioRa, TopDog RF, and infiNET, where the security methodology used is unknown. It is a realistic assumption that vulnerabilities must exist with the age and lack of oversight of many of these proprietary protocols. The absence of known vulnerabilities is not due to strong design but through security through obscurity. Unfortunately, once adversaries find vulnerabilities in these protocols, it may lead to easily-compromised systems as security through obscurity provides very little legitimate protection. Much more research and collaboration with vendors are needed for assess the security of these protocols and develop security tools (e.g., monitoring, auditing).

\noindent\textbf{Proprietary Software.} In a similar manner to protocols, research on E-IoT software is a challenge as much of E-IoT development is closed-source with minimal external resources available. Further, even if research is done on one E-IoT system, different systems will follow different integration and configuration practices. For instance, components like drivers are different in every system, and the implementation may allow for completely different attacks in each system. We found that many E-IoT systems have operated under security through obscurity for their software in addition to communication protocols, a practice that is currently insufficient. As such, it may be necessary to find flaws in E-IoT software components and correct these flaws before malicious actors compromise E-IoT installations. It may be a good idea for vendors to cooperate with research and academic communities and provide closed-source software to prevent attacks before they occur. In comparison to more open industries, in E-IoT, an attacker that acquires source code for E-IoT devices may have a running start in compromising these devices before security researchers have even acknowledged the problem. In response to looming threats, vendors could also offer hackathons, bug bounties, and reward responsible disclosures of vulnerabilities that can affect their systems.

\noindent\textbf{Honeypots as a Defense Strategy.} It may be possible for honeypots to offer insight and warnings on possible attackers against E-IoT systems and complement existing security mechanisms. Litchfield et al. noted that honeypots can vary between different applications, highlighting that high-interaction honeypots are not suitable in some applications \cite{litchfield2016rethinking}. Other solutions may be possible, for instance, Conpot, a honeypot system developed by the Honeynet Project supporting industrial protocols such as BACnet, EtherNet/IP, and Modbus. These developments are applicable in E-IoT installations as honeypot frameworks may be expanded to work with less-known proprietary E-IoT protocols \cite{honeynetProject, compot}. For instance, Mays et al. proposed a solution to defend home and building automation systems using decoy networks \cite{mays2017defending}. In this work, researchers created a honeypot network on the smart automation Insteon protocol to hide communication using a dummy network and hide genuine network traffic from attackers. Such approaches could apply to E-IoT and other custom systems that rely heavily on physical components, and hopefully to understand the behavior of attackers, thus secure E-IoT systems.

\section{Related Work}\label{sec:relatedwork}

The security of IoT smart devices have been an ongoing topic of research in the recent years. As such, a number of IoT security surveys have been conducted \cite{ammar2018internet, lin2017survey, alaba2017internet, hassan2019current, hassija2019survey, oracevic2017security, deogirikar2017security, balte2015security, zhao2013survey, kraijak2015survey, yang2017survey, pawar2016survey}. Most of these surveys cover attacks, defenses, security challenges and general counter-measures in IoT, others are more specific.  For instance, a survey by Hassan et al., highlights current research trends in in IoT security \cite{hassan2019current}. Other work has focused on IoT security aspects, such as the survey by Deogirikar et al., which focused specifically on known IoT attacks \cite{deogirikar2017security}. Individually, as early as 2013, works have highlighted threats in smart devices, and how attackers always search for new, unexplored threat vectors \cite{saint, daint, lopez2017survey, convolution, smartgridjournal, patent, patent2, aegis}. However, very little ongoing research has focused on specific vulnerabilities targeting E-IoT systems or proprietary technologies. Often, vendors will only guarantee security and perform internal security analysis in their own devices \cite{crestronSecurity}. 

Some topics covered in this survey have had dedicated surveys examine attacks, defenses and threats for each topic. For instance, several surveys have covered jamming attacks, and defenses against wireless communication \cite{jamming1, jamming2, jamming3, jamming4}. Surveys on communication are also relevant, with a number of surveys covering attacks, risks, and defense mechanisms for Bluetooth communication \cite{lounis2020attacks, darroudi2017bluetooth, minar2012bluetooth, dunning2010taming}. Sensory channels are also an active subject of research and relevant to E-IoT. As such, related surveys on the security of sensory channels touch upon subjects beyond E-IoT applications such as WSNs and large-scale sensor deployments \cite{sensorsurvey1, sensorsurvey2, sensorsurvey3, sensorsurvey4, sensorsurvey5, sensorsurvey6, sensorsurvey7, sensorsurvey8, sensorsurvey9, sensorsurvey10, sensorsurvey11}. Further, as cloud computing is an active topic of research, a number of relevant surveys have also covered cloud computing threats \cite{de2019cyber, liu2015survey, ryan2013cloud, shahzad2014state, subashini2011survey, grobauer2010understanding, modi2013survey, singh2016survey, fernandes2014security, polash2014survey, singh2017cloud, xiao2012security, ardagna2015security, hashizume2013analysis}. Other works focus on cloud defense mechanisms, applying to different use cases beyond the scope of this survey and E-IoT applications \cite{fernandes2014security, ryan2013cloud, singh2017cloud, ardagna2015security, modi2013survey}.

\textit{Our work differs from previously discussed works as this survey focuses on the insecurities, possible threats, and defenses applicable to E-IoT. To the best of our knowledge, this is the first work that focuses solely on E-IoT systems and their security, categorizing E-IoT systems into four unique layers. Specifically, we categorize E-IoT components into four distinct layers, (1) the E-IoT Devices layer, (2) the communications layer, (3) the Monitoring and Applications layer, and (4) the Business layer. We take this approach as E-IoT system architecture differs from many IoT systems, as highlighted in Section \ref{sec:background}. Further, we present a threat model for each distinct layer of E-IoT, as each layer may present different threats and require different capabilities from attackers.}

\section{Conclusion}\label{sec:conclusion}

The rising popularity of smart systems has led to millions of users worldwide interacting with smart devices on a day-to-day basis.  Many of these devices are commodity, off-the-shelf systems (e.g., Google Home, Samsung SmartThings), easily maintained and installed by the average end-user in small deployments. However, while easy to install, commodity systems are limited and do not provide a viable solution for more sophisticated applications. For more extensive installations, E-IoT systems provide a custom-installed solutions to fit a client's needs. As such, systems such as Crestron, Control4, RTI, and savant offer a solution for more sophisticated applications (e.g., complete lighting control, A/V management, managed CCTV security), where commodity systems are insufficient. For this reason, E-IoT systems are common in locations such as high-end smart homes, government and private offices, yachts, and conference rooms. In contrast to commodity systems, E-IoT systems are usually proprietary, costly, closed source, and more robust for their configured use cases. However, even with their popularity, very little research has focused on the overall security of E-IoT systems. Namely, no study provides a complete overview of E-IoT systems, their components, threats, and relevant vulnerabilities in the literature. To address this research gap, motivate further research, and raise awareness on E-IoT insecurities, this work focused solely on E-IoT systems. Specifically, we discussed E-IoT components, vulnerabilities, and their security implications. To provide a better analysis of E-IoT, we divided E-IoT systems into four layers: E-IoT Devices Layer, Communications Layer, Monitoring and Applications Layer, and Business Layer. We considered E-IoT components at every layer, the associated threat models, attacks, and defense mechanisms. We also presented critical observations on E-IoT security and provided a list of open research problems that require further research. We believe this study will raise awareness on E-IoT and E-IoT security, and motivate further research.

\section*{Acknowledgments}
This work is partially supported by the US National Science Foundation (Awards: NSF-CAREER-CNS-1453647, NSF-1663051, and  NSF-1718116). The views are those of the authors only.

\bibliographystyle{unsrt}
\bibliography{references}

\begin{thebibliography}{100}

\bibitem{magazine}
H.~{Aksu}, L.~{Babun}, M.~{Conti}, G.~{Tolomei}, and A.~S. {Uluagac}.
\newblock {Advertising in the IoT Era: Vision and Challenges}.
\newblock {\em IEEE Communications Magazine}, 2018.

\bibitem{SmartHomesUSEurope}
The number of smart homes in europe and north america reached 45 million in
  2017.
\newblock
  {https://iotbusinessnews.com/2018/09/24/20413-the-number-of-smart-homes-in-europe-and-north-america-reached-45-million-in-2017/},
  Sept, 2018.
\newblock Online: Accessed 10-December-2019.

\bibitem{SmartTS}
Mohamed Sultan.
\newblock Smart to smarter: Smart home systems history, future and challenges.
\newblock Online: Accessed 10-December-2019.

\bibitem{iotdots}
Leonardo Babun, Amit~Kumar Sikder, Abbas Acar, and A.~Selcuk Uluagac.
\newblock Iotdots: {A} digital forensics framework for smart environments.
\newblock {\em CoRR}, 2018.

\bibitem{patent}
Leonardo Babun, Hidayet Aksu, and A.~Selcuk Uluagac.
\newblock Detection of counterfeit and compromised devices using system and
  function call tracing techniques.
\newblock (10027697), 7 2018.

\bibitem{patent2}
{Babun, Leonardo, Aksu, Hidayet, Uluagac, Selcuk A.}
\newblock {Method of Resource-limited Device and Device Class Identification
  using System and Function Call Tracing Techniques, Performance, and
  Statistical Analysis}.
\newblock (10242193), March 2019.

\bibitem{smartgridjournal}
Leonardo Babun, Hidayet Aksu, and A.~Selcuk Uluagac.
\newblock {A System-Level Behavioral Detection Framework for Compromised CPS
  Devices: Smart-Grid Case}.
\newblock {\em ACM Trans. Cyber-Phys. Syst.}, 4(2), nov 2019.

\bibitem{usbposter}
Kyle Denney, Enes Erdin, Leonardo Babun, and A.~Selcuk Uluagac.
\newblock {Dynamically Detecting USB Attacks in Hardware: Poster}.
\newblock In {\em Proceedings of the 12th Conference on Security and Privacy in
  Wireless and Mobile Networks}, page 328–329, 2019.

\bibitem{daint}
Leonardo Babun, Z.~Berkay Celik, Patrick McDaniel, and A.~Selcuk Uluagac.
\newblock Real-time analysis of privacy-(un)aware iot applications.
\newblock {\em Proceedings on Privacy Enhancing Technologies}, 2021(1):145 --
  166, 01 Jan. 2021.

\bibitem{denney2019usb}
Kyle Denney, Enes Erdin, Leonardo Babun, Michael Vai, and Selcuk Uluagac.
\newblock Usb-watch: A dynamic hardware-assisted usb threat detection
  framework.
\newblock In {\em International Conference on Security and Privacy in
  Communication Systems}, pages 126--146. Springer, 2019.

\bibitem{lopez2017survey}
Juan Lopez, Leonardo Babun, Hidayet Aksu, and A~Selcuk Uluagac.
\newblock A survey on function and system call hooking approaches.
\newblock {\em Journal of Hardware and Systems Security}, 1(2):114--136, 2017.
\newblock Accessed: 11-17-2018.

\bibitem{convolution}
C.~{Kaygusuz}, L.~{Babun}, H.~{Aksu}, and A.~S. {Uluagac}.
\newblock Detection of compromised smart grid devices with machine learning and
  convolution techniques.
\newblock In {\em 2018 IEEE International Conference on Communications (ICC)},
  pages 1--6, May 2018.

\bibitem{aegis}
Amit~Kumar Sikder, Leonardo Babun, Hidayet Aksu, and A.~Selcuk Uluagac.
\newblock Aegis: A context-aware security framework for smart home systems.
\newblock In {\em Proceedings of the 35th Annual Computer Security Applications
  Conference}, 2019.

\bibitem{saint}
Z.~Berkay Celik, Leonardo Babun, Amit~Kumar Sikder, Hidayet Aksu, Gang Tan,
  Patrick McDaniel, and A.~Selcuk Uluagac.
\newblock Sensitive information tracking in commodity iot.
\newblock In {\em 27th {USENIX} Security Symposium}, pages 1687--1704, 2018.

\bibitem{smarthomesidechannel}
M.~A.~N. {Abrishamchi}, A.~H. {Abdullah}, A.~{David Cheok}, and K.~S.
  {Bielawski}.
\newblock Side channel attacks on smart home systems: A short overview.
\newblock In {\em IECON 2017 = 43rd Annual Conference of the IEEE Industrial
  Electronics Society}, pages 8144--8149, Oct 2017.

\bibitem{acar2018peekaboo}
Abbas Acar, Hossein Fereidooni, Tigist Abera, Amit~Kumar Sikder, Markus
  Miettinen, Hidayet Aksu, Mauro Conti, Ahmad-Reza Sadeghi, and Selcuk Uluagac.
\newblock Peek-a-boo: I see your smart home activities, even encrypted!
\newblock In {\em Proceedings of the 13th ACM Conference on Security and
  Privacy in Wireless and Mobile Networks}, WiSec '20, page 207–218. ACM,
  2020.

\bibitem{waka}
A.~{Acar}, H.~{Aksu}, A.~S. {Uluagac}, and K.~{Akkaya}.
\newblock Waca: Wearable-assisted continuous authentication.
\newblock In {\em 2018 IEEE Security and Privacy Workshops (SPW)}, pages
  264--269, May 2018.

\bibitem{contextawaresensor}
A.~K. {Sikder}, H.~{Aksu}, and A.~S. {Uluagac}.
\newblock A context-aware framework for detecting sensor-based threats on smart
  devices.
\newblock {\em IEEE Transactions on Mobile Computing}, 19(2):245--261, Feb
  2020.

\bibitem{sikder2018survey}
Amit~Kumar Sikder, Giuseppe Petracca, Hidayet Aksu, Trent Jaeger, and A.~Selcuk
  Uluagac.
\newblock A survey on sensor-based threats to internet-of-things (iot) devices
  and applications.
\newblock {\em CoRR}, abs/1802.02041, 2018.

\bibitem{6997498}
A.S. Uluagac, V.~Subramanian, and R.~Beyah.
\newblock Sensory channel threats to cyber physical systems: A wake-up call.
\newblock In {\em IEEE Conference on Communications and Network Security (CNS),
  2014}, pages 301--309.

\bibitem{saint-taint-analysis}
Z.~Berkay Celik, Leonardo Babun, Amit~Kumar Sikder, Hidayet Aksu, Gang Tan,
  Patrick McDaniel, and A.~Selcuk Uluagac.
\newblock Sensitive information tracking in commodity iot.
\newblock In {\em 27th Security Symposium ({USENIX} Security 18)}, Baltimore,
  MD, 2018.

\bibitem{iqtidar1}
AKM~Iqtidar Newaz, Amit~Kumar Sikder, Mohammad~Ashiqur Rahman, and A~Selcuk
  Uluagac.
\newblock Healthguard: A machine learning-based security framework for smart
  healthcare systems.
\newblock In {\em 2019 Sixth International Conference on Social Networks
  Analysis, Management and Security (SNAMS)}, 2019.

\bibitem{iqtidar2}
AKM Newaz, Amit~Kumar Sikder, Mohammad~Ashiqur Rahman, and A~Selcuk Uluagac.
\newblock A survey on security and privacy issues in modern healthcare systems:
  Attacks and defenses.
\newblock {\em arXiv preprint arXiv:2005.07359}, 2020.

\bibitem{iqtidar3}
AKM~Iqtidar Newaz, Amit~Kumar Sikder, Leonardo Babun, and A~Selcuk Uluagac.
\newblock Heka: A novel intrusion detection system for attacks to personal
  medical devices.
\newblock In {\em IEEE Conference on Communications and Network Security
  (CNS)}, 2020.

\bibitem{berkaymagazine}
Z.~B. {Celik}, P.~{McDaniel}, G.~{Tan}, L.~{Babun}, and A.~S. {Uluagac}.
\newblock {Verifying Internet of Things Safety and Security in Physical
  Spaces}.
\newblock {\em IEEE Security Privacy}.

\bibitem{kratos}
Amit~Kumar Sikder, Leonardo Babun, Z.~Berkay Celik, Abbas Acar, Hidayet Aksu,
  Patrick McDaniel, Engin Kirda, and A.~Selcuk Uluagac.
\newblock Kratos: Multi-user multi-device-aware access control system for the
  smart home.
\newblock In {\em 13th ACM Conference on Security and Privacy in Wireless and
  Mobile Networks}, 2020.

\bibitem{ziot}
L.~{Babun}, H.~{Aksu}, L.~{Ryan}, K.~{Akkaya}, E.~S. {Bentley}, and A.~S.
  {Uluagac}.
\newblock Z-iot: Passive device-class fingerprinting of zigbee and z-wave iot
  devices.
\newblock In {\em 2020 IEEE International Conference on Communications (ICC)},
  pages 1--7, 2020.

\bibitem{madiot}
J.~{Myers}, L.~{Babun}, E.~{Yao}, S.~{Helble}, and P.~{Allen}.
\newblock Mad-iot: Memory anomaly detection for the internet of things.
\newblock In {\em 2019 IEEE Globecom Workshops (GC Wkshps)}, pages 1--6, 2019.

\bibitem{usbjournal}
K.~{Denney}, L.~{Babun}, and A.~S. {Uluagac}.
\newblock {USB-Watch: a Generalized Hardware-Assisted Insider Threat Detection
  Framework}.
\newblock {\em Journal of Hardware and Systems Security}, 2020.

\bibitem{newaz2020adversarial}
AKM Newaz, Nur~Imtiazul Haque, Amit~Kumar Sikder, Mohammad~Ashiqur Rahman, and
  A~Selcuk Uluagac.
\newblock Adversarial attacks to machine learning-based smart healthcare
  systems.
\newblock {\em arXiv preprint arXiv:2010.03671}, 2020.

\bibitem{newaz2020survey}
AKM Newaz, Amit~Kumar Sikder, Mohammad~Ashiqur Rahman, and A~Selcuk Uluagac.
\newblock A survey on security and privacy issues in modern healthcare systems:
  Attacks and defenses.
\newblock {\em arXiv preprint arXiv:2005.07359}, 2020.

\bibitem{newaz2020heka}
AKM~Iqtidar Newaz, Amit~Kumar Sikder, Leonardo Babun, and A~Selcuk Uluagac.
\newblock Heka: A novel intrusion detection system for attacks to personal
  medical devices.
\newblock In {\em 2020 IEEE Conference on Communications and Network Security
  (CNS)}, pages 1--9. IEEE, 2020.

\bibitem{newaz2019healthguard}
AKM~Iqtidar Newaz, Amit~Kumar Sikder, Mohammad~Ashiqur Rahman, and A~Selcuk
  Uluagac.
\newblock Healthguard: A machine learning-based security framework for smart
  healthcare systems.
\newblock In {\em 2019 Sixth International Conference on Social Networks
  Analysis, Management and Security (SNAMS)}, pages 389--396. IEEE, 2019.

\bibitem{crestronorigins}
{Mark N. Vena}.
\newblock How crestron paved the way for the smart home, and more.
\newblock
  {https://www.forbes.com/sites/moorinsights/2018/08/23/how-crestron-paved-the-way-for-the-smart-home-and-more/\#397001f141f8},
  2018.
\newblock Online: Accessed 18-May-2020.

\bibitem{control4about}
{Control4}.
\newblock Control4: About our company.
\newblock {https://www.control4. com/company/}, 2020.
\newblock Online: Accessed 18-May-2020.

\bibitem{centralizedlighting}
Control4.
\newblock Control4 panelized lighting: Reference guide for electricians.
\newblock
  https://www.control4.com/docs/product/panelized-lighting/professional-reference-guide/latest/panelized-lighting-professional-reference-guide-rev-b.pdf,
  Feb, 2014.
\newblock Online: Accessed 10-December-2019.

\bibitem{doelightingsystems}
U.S~Department of~Energy.
\newblock {Cyber Security for Lighting Systems}.
\newblock
  https://www.energy.gov/sites/prod/files/2018/06/f52/cyber\_security\_\\lighting.pdf,
  2018.
\newblock Online: Accessed 15-August-2020.

\bibitem{smartsystems}
AudioAdvice.
\newblock Which smart home system is best? control4 vs. crestron vs. savant.
\newblock
  https://www.audioadvice.com/videos-reviews/control4-vs-crestron-vs-savant/.
\newblock Online: Accessed 10-December-2019.

\bibitem{crestrondeploy}
Crestron.
\newblock Crestron technical institute.
\newblock https://www.crestron.com/trai- ning.
\newblock Online: Accessed 20-December-2019.

\bibitem{control4deploy}
Control4.
\newblock Getting started with composer pro.
\newblock https://www.control4.
  com/files/dealers/200-00168-ComposerProGettingStarted.pdf, Jun, 2010.
\newblock Online: Accessed 20-December-2019.

\bibitem{blackwiredrivers}
{Blackwire Designs}.
\newblock Control4 automation apps and drivers.
\newblock
  {https://www.blackwiredesigns.com/cat/automation-apps-and-drivers/control4\_drivers/},
  2008.
\newblock Online: Accessed 18-May-2020.

\bibitem{DriverCentral}
{drivercentral}.
\newblock Control4 drivers.
\newblock {https://drivercentral.io/platforms/control4-drivers/}, 2020.
\newblock Online: Accessed 20-May-2020.

\bibitem{HighEndAutomation}
{Audrey Noble}.
\newblock A look inside the amazing smart-home systems that rich people use.
\newblock
  {https://www.businessinsider.com/smart-home-tech-that-rich-people-use-2018-7},
  2018.
\newblock Online: Accessed 20-June-2020.

\bibitem{integrateserial}
{ADI}.
\newblock {HoneyWell Serial Interface Module}.
\newblock https://www.adiglobaldis tribution.us/Product/4100SM, {2020}.
\newblock Online: Accessed 25-September-2020.

\bibitem{integrateIP}
{Honeywell}.
\newblock {Ethernet Interface User Manual}.
\newblock https://www.honeywell
  process.com/library/support/Public/Documents/51-52-25-96.pdf, {2001}.
\newblock Online: Accessed 25-September-2020.

\bibitem{camio}
{Camio}.
\newblock Ai for cost-effective remote video monitoring.
\newblock {https://camio. com/}, 2020.
\newblock Online: Accessed 18-July-2020.

\bibitem{miller2013supply}
John~F Miller.
\newblock Supply chain attack framework and attack patterns.
\newblock Technical report, MITRE CORP MCLEAN VA, 2013.

\bibitem{supchaindi}
{Nate Lord}.
\newblock Supply chain cybersecurity: Experts on how to mitigate third party
  risk.
\newblock {https://digitalguardian.com/blog/supply-chain-cybersecurity}, 2020.
\newblock Online: Accessed 25-September-2020.

\bibitem{supchainsw}
{Edward Kovacs}.
\newblock Iot devices at major manufacturers infected with malware via supply
  chain attack.
\newblock
  {https://www.securityweek.com/iot-devices-major-manufacturers-infected-malware-supply-chain-attack},
  2020.
\newblock Online: Accessed 25-September-2020.

\bibitem{fuentes2018securing}
Mayra~Rosario Fuentes and Numaan Huq.
\newblock Securing connected hospitals.
\newblock
  {https://www.key4biz.it/wp-content/uploads/2018/04/rpt-securing-connected-hospitals.pdf},
  2018.

\bibitem{supchainCSO}
{Maria Korolov}.
\newblock What is a supply chain attack? why you should be wary of third-party
  providers.
\newblock {https://www.csoonline
  .com/article/3191947/what-is-a-supply-chain-attack-why-you-should-be-wary-of-third-party-providers.html},
  2019.
\newblock Online: Accessed 22-September-2020.

\bibitem{yousefnezhad2020security}
Narges Yousefnezhad, Avleen Malhi, and Kary Fr{\"a}mling.
\newblock Security in product lifecycle of iot devices: A survey.
\newblock {\em Journal of Network and Computer Applications}, page 102779,
  2020.

\bibitem{farooq2019iot}
Muhammad~Junaid Farooq and Quanyan Zhu.
\newblock Iot supply chain security: Overview, challenges, and the road ahead.
\newblock {\em arXiv preprint arXiv:1908.07828}, 2019.

\bibitem{supplychainattacks}
{National Cyber Security Centre}.
\newblock {Supply chain security guidance}.
\newblock
  https://www.ncsc.gov.uk/collection/supply-chain-security/supply-chain-attack-examples,
  {2020}.
\newblock Online: Accessed 10-November-2020.

\bibitem{gorman2012counterfeit}
Celia Gorman.
\newblock Counterfeit chips on the rise.
\newblock {\em IEEE Spectrum}, 49(6):16--17, 2012.

\bibitem{bhasin2015survey}
Shivam Bhasin and Francesco Regazzoni.
\newblock A survey on hardware trojan detection techniques.
\newblock In {\em 2015 IEEE International Symposium on Circuits and Systems
  (ISCAS)}, pages 2021--2024. IEEE, 2015.

\bibitem{tehranipoor2010survey}
Mohammad Tehranipoor and Farinaz Koushanfar.
\newblock A survey of hardware trojan taxonomy and detection.
\newblock {\em IEEE design \& test of computers}, 27(1):10--25, 2010.

\bibitem{king2008designing}
Samuel~T King, Joseph Tucek, Anthony Cozzie, Chris Grier, Weihang Jiang, and
  Yuanyuan Zhou.
\newblock Designing and implementing malicious hardware.
\newblock {\em Leet}, 8:1--8, 2008.

\bibitem{robertson2018big}
Jordan Robertson and Michael Riley.
\newblock The big hack: How china used a tiny chip to infiltrate us companies.
\newblock {\em Bloomberg Businessweek}, 4, 2018.

\bibitem{yangrfid}
K.~{Yang}, D.~{Forte}, and M.~M. {Tehranipoor}.
\newblock Protecting endpoint devices in iot supply chain.
\newblock In {\em 2015 IEEE/ACM International Conference on Computer-Aided
  Design (ICCAD)}, pages 351--356, 2015.

\bibitem{yangrfid2}
Kun Yang, Domenic Forte, and Mark Tehranipoor.
\newblock Resc: An rfid-enabled solution for defending iot supply chain.
\newblock 23(3), February 2018.

\bibitem{yangrfid3}
Kun Yang, Domenic Forte, and Mark~M. Tehranipoor.
\newblock Cdta: A comprehensive solution for counterfeit detection,
  traceability, and authentication in the iot supply chain.
\newblock 22(3), April 2017.

\bibitem{chamekhsc}
M.~{Chamekh}, M.~{Hamdi}, S.~{El Asmi}, and T.~{Kim}.
\newblock Secured distributed iot based supply chain architecture.
\newblock In {\em 2018 IEEE 27th International Conference on Enabling
  Technologies: Infrastructure for Collaborative Enterprises (WETICE)}, pages
  199--202, 2018.

\bibitem{crestronreset}
Crestron.
\newblock Tsw model touchscreen manual.
\newblock https://bit.ly/3mcRNrI, 2020.
\newblock Online: Accessed 20-June-2020.

\bibitem{control4reset}
Control4.
\newblock Control4 zigbee: The definitive magic button press guide.
\newblock
  https://technet.genesis-technologies.ch/control4-zigbee-the-definitive-guide/,
  2013.
\newblock Online: Accessed 20-June-2019.

\bibitem{nodecaptureattack}
M.~V. {Bharathi}, R.~C. {Tanguturi}, C.~{Jayakumar}, and K.~{Selvamani}.
\newblock Node capture attack in wireless sensor network: A survey.
\newblock In {\em 2012 IEEE International Conference on Computational
  Intelligence and Computing Research}, pages 1--3, 2012.

\bibitem{nodecaptureattackwang}
C.~{Wang}, D.~{Wang}, Y.~{Tu}, G.~{Xu}, and H.~{Wang}.
\newblock Understanding node capture attacks in user authentication schemes for
  wireless sensor networks.
\newblock {\em IEEE Transactions on Dependable and Secure Computing}, pages
  1--1, 2020.

\bibitem{nodecapturepw1}
Tamara Bonaci, Linda Bushnell, and Radha Poovendran.
\newblock Probabilistic analysis of covering and compromise in a node capture
  attack.
\newblock {\em Network Security Lab (NSL), Seattle, WA, Techical Report}, 1,
  2010.

\bibitem{nodecapturepw2}
T.~{Bonaci}, L.~{Bushnell}, and R.~{Poovendran}.
\newblock Node capture attacks in wireless sensor networks: A system theoretic
  approach.
\newblock In {\em 49th IEEE Conference on Decision and Control (CDC)}, pages
  6765--6772, 2010.

\bibitem{nodecapturepw3}
Pradip De, Yonghe Liu, and Sajal~K Das.
\newblock Deployment-aware modeling of node compromise spread in wireless
  sensor networks using epidemic theory.
\newblock {\em ACM Transactions on Sensor Networks (TOSN)}, 5(3):1--33, 2009.

\bibitem{nodecapturepw4}
D.~S. {Kim}, Y.~K. {Suh}, and J.~S. {Park}.
\newblock Toward assessing vulnerability and risk of sensor networks under node
  compromise.
\newblock In {\em 2007 International Conference on Computational Intelligence
  and Security (CIS 2007)}, pages 740--744, 2007.

\bibitem{nodecapturepw5}
A.~K. {Mishra} and A.~K. {Turuk}.
\newblock Adversary information gathering model for node capture attack in
  wireless sensor networks.
\newblock In {\em 2011 International Conference on Devices and Communications
  (ICDeCom)}, pages 1--5, 2011.

\bibitem{nodecapturepw6}
A.~{Ramos}, B.~{Aquino}, M.~{Lazar}, R.~H. {Filho}, and J.~J. P.~C.
  {Rodrigues}.
\newblock A quantitative model for dynamic security analysis of wireless sensor
  networks.
\newblock In {\em GLOBECOM 2017 - 2017 IEEE Global Communications Conference},
  pages 1--6, 2017.

\bibitem{lossprevention}
{Ingram Micro}.
\newblock 4 innovations in theft and loss prevention.
\newblock
  {https://imaginenext.ingrammicro.com/iot/4-innovations-in-theft-and-loss-prevention},
  2019.
\newblock Online: Accessed 22-September-2020.

\bibitem{standaert2010introduction}
Fran{\c{c}}ois-Xavier Standaert.
\newblock Introduction to side-channel attacks.
\newblock In {\em Secure integrated circuits and systems}, pages 27--42.
  Springer, 2010.

\bibitem{RS232Radiation}
Peter Smulders.
\newblock The threat of information theft by reception of electromagnetic
  radiation from rs-232 cables.
\newblock {\em Computers \& Security}, 9(1):53 -- 58, 1990.

\bibitem{hwu1988electromagnetic}
Shian-Uei Hwu and Donald~R Wilton.
\newblock Electromagnetic scattering and radiation by arbitrary configurations
  of conducting bodies and wires.
\newblock Technical report, San Diego State Univ Foundation CA, 1988.

\bibitem{van1985electromagnetic}
Wim Van~Eck.
\newblock Electromagnetic radiation from video display units: An eavesdropping
  risk?
\newblock {\em Computers \& Security}, 4(4):269--286, 1985.

\bibitem{kuhn2004electromagnetic}
Markus~G Kuhn.
\newblock Electromagnetic eavesdropping risks of flat-panel displays.
\newblock In {\em International Workshop on Privacy Enhancing Technologies},
  pages 88--107. Springer, 2004.

\bibitem{control4DSsecurity}
{Control4}.
\newblock Door station - exterior: Security best practices.
\newblock
  {https://www.control4.com/docs/product/security/best-practices/english/latest/security-best-practices-rev-a.pdf},
  2012.
\newblock Online: Accessed 22-September-2020.

\bibitem{savage2015visualizing}
Neil Savage.
\newblock Visualizing sound.
\newblock {\em Commun. ACM}, 58(2):15–17, January 2015.

\bibitem{xu2014watching}
Yi~Xu, Jan-Michael Frahm, and Fabian Monrose.
\newblock Watching the watchers: Automatically inferring tv content from
  outdoor light effusions.
\newblock In {\em Proceedings of the 2014 ACM SIGSAC Conference on Computer and
  Communications Security}, pages 418--428, 2014.

\bibitem{sun2014design}
Mingshen Sun, Min Zheng, John~CS Lui, and Xuxian Jiang.
\newblock Design and implementation of an android host-based intrusion
  prevention system.
\newblock In {\em Proceedings of the 30th annual computer security applications
  conference}, pages 226--235, 2014.

\bibitem{wang2015novel}
Xiaolei Wang, Yuexiang Yang, Yingzhi Zeng, Chuan Tang, Jiangyong Shi, and Kele
  Xu.
\newblock A novel hybrid mobile malware detection system integrating anomaly
  detection with misuse detection.
\newblock In {\em Proceedings of the 6th International Workshop on Mobile Cloud
  Computing and Services}, pages 15--22, 2015.

\bibitem{enck2014taintdroid}
William Enck, Peter Gilbert, Seungyeop Han, Vasant Tendulkar, Byung-Gon Chun,
  Landon~P Cox, Jaeyeon Jung, Patrick McDaniel, and Anmol~N Sheth.
\newblock Taintdroid: an information-flow tracking system for realtime privacy
  monitoring on smartphones.
\newblock {\em ACM Transactions on Computer Systems (TOCS)}, 32(2):1--29, 2014.

\bibitem{wu2014droiddolphin}
Wen-Chieh Wu and Shih-Hao Hung.
\newblock Droiddolphin: a dynamic android malware detection framework using big
  data and machine learning.
\newblock In {\em Proceedings of the 2014 Conference on Research in Adaptive
  and Convergent Systems}, pages 247--252, 2014.

\bibitem{xu2015semadroid}
Zhi Xu and Sencun Zhu.
\newblock Semadroid: A privacy-aware sensor management framework for
  smartphones.
\newblock In {\em Proceedings of the 5th ACM Conference on Data and Application
  Security and Privacy}, pages 61--72, 2015.

\bibitem{schwittmann2017identifying}
Lorenz Schwittmann, Christopher Boelmann, Viktor Matkovic, Matth{\"a}us Wander,
  and Torben Weis.
\newblock Identifying tv channels \& on-demand videos using ambient light
  sensors.
\newblock {\em Pervasive and Mobile Computing}, 38:363--380, 2017.

\bibitem{schwittmann2016video}
Lorenz Schwittmann, Viktor Matkovic, Torben Weis, et~al.
\newblock Video recognition using ambient light sensors.
\newblock In {\em 2016 IEEE International Conference on Pervasive Computing and
  Communications (PerCom)}, pages 1--9. IEEE, 2016.

\bibitem{maiti2019light}
Anindya Maiti and Murtuza Jadliwala.
\newblock Light ears: Information leakage via smart lights.
\newblock {\em Proceedings of the ACM on Interactive, Mobile, Wearable and
  Ubiquitous Technologies}, 3(3):1--27, 2019.

\bibitem{zhou2018irexf}
Zheng Zhou, Weiming Zhang, and Nenghai Yu.
\newblock Irexf: Data exfiltration from air-gapped networks by infrared remote
  control signals.
\newblock {\em arXiv preprint arXiv:1801.03218}, 2018.

\bibitem{guri2016optical}
Mordechai Guri, Ofer Hasson, Gabi Kedma, and Yuval Elovici.
\newblock An optical covert-channel to leak data through an air-gap.
\newblock In {\em 2016 14th Annual Conference on Privacy, Security and Trust
  (PST)}, pages 642--649. IEEE, 2016.

\bibitem{guri2019air}
Mordechai Guri and Dima Bykhovsky.
\newblock air-jumper: Covert air-gap exfiltration/infiltration via security
  cameras \& infrared (ir).
\newblock {\em Computers \& Security}, 82:15--29, 2019.

\bibitem{uk2014light}
Yuval~Elovici Mordechai~Guri.
\newblock Bridgeware: The air-gap malware.
\newblock 2018.

\bibitem{loughry2002information}
Joe Loughry and David~A Umphress.
\newblock Information leakage from optical emanations.
\newblock {\em ACM Transactions on Information and System Security (TISSEC)},
  5(3):262--289, 2002.

\bibitem{ronen2016extended}
Eyal Ronen and Adi Shamir.
\newblock Extended functionality attacks on iot devices: The case of smart
  lights.
\newblock In {\em 2016 IEEE European Symposium on Security and Privacy
  (EuroS\&P)}, pages 3--12. IEEE, 2016.

\bibitem{sikder2019context}
Amit~Kumar Sikder, Hidayet Aksu, and A~Selcuk Uluagac.
\newblock Context-aware intrusion detection method for smart devices with
  sensors, September~17 2019.
\newblock US Patent 10,417,413.

\bibitem{sikder20176thsense}
Amit~Kumar Sikder, Hidayet Aksu, and A~Selcuk Uluagac.
\newblock 6thsense: A context-aware sensor-based attack detector for smart
  devices.
\newblock In {\em 26th $\{$USENIX$\}$ Security Symposium ($\{$USENIX$\}$
  Security 17)}, pages 397--414, 2017.

\bibitem{ShakhovBattery}
V.~{Shakhov}, I.~{Koo}, and A.~{Rodionov}.
\newblock Energy exhaustion attacks in wireless networks.
\newblock In {\em 2017 International Multi-Conference on Engineering, Computer
  and Information Sciences (SIBIRCON)}, pages 1--3, 2017.

\bibitem{batterybauernansa}
Michael Bauer, Mark Coatsworth, and Justin Moeller.
\newblock Nansa: A no-attribution nosleep battery exhaustion attack for
  portable computing devices.
\newblock http://pages.cs.wisc.edu/~bauer/CS736Final.pdf, 2015.
\newblock Online: Accessed 11-Feb-2021.

\bibitem{MoyersBattery}
B.~R. {Moyers}, J.~P. {Dunning}, R.~C. {Marchany}, and J.~G. {Tront}.
\newblock Effects of wi-fi and bluetooth battery exhaustion attacks on mobile
  devices.
\newblock In {\em 2010 43rd Hawaii International Conference on System
  Sciences}, pages 1--9, 2010.

\bibitem{MartinBattery}
T.~{Martin}, M.~{Hsiao}, {Dong Ha}, and J.~{Krishnaswami}.
\newblock Denial-of-service attacks on battery-powered mobile computers.
\newblock In {\em Second IEEE Annual Conference on Pervasive Computing and
  Communications, 2004. Proceedings of the}, pages 309--318, 2004.

\bibitem{bsips}
Timothy~K Buennemeyer, Michael Gora, Randy~C Marchany, and Joseph~G Tront.
\newblock Battery exhaustion attack detection with small handheld mobile
  computers.
\newblock In {\em 2007 IEEE International Conference on Portable Information
  Devices}, pages 1--5. IEEE, 2007.

\bibitem{nashintrusion}
Daniel~Charles Nash.
\newblock {\em An Intrusion Detection System for Battery Exhaustion Attacks on
  Mobile Computers}.
\newblock PhD thesis, Virginia Tech, 2005.

\bibitem{UpadhyayBattery}
R.~{Upadhyay}, S.~{Khan}, H.~{Tripathi}, and U.~R. {Bhatt}.
\newblock Detection and prevention of ddos attack in wsn for aodv and dsr using
  battery drain.
\newblock In {\em 2015 International Conference on Computing and Network
  Communications (CoCoNet)}, pages 446--451, 2015.

\bibitem{hristozovBattery}
Stefan Hristozov, Manuel Huber, and Georg Sigl.
\newblock Protecting restful iot devices from battery exhaustion dos attacks.
\newblock {\em arXiv preprint arXiv:1911.08134}, 2019.

\bibitem{control4Sensors}
Control4.
\newblock {Control4 Sensors}.
\newblock https://www.control4.com/solutions/\\products/sensors/, 2020.
\newblock Online: Accessed 20-June-2020.

\bibitem{savantSensors}
Savant.
\newblock {Savant Climate Control}.
\newblock https://www.savant.com/climate, 2020.
\newblock Online: Accessed 20-June-2020.

\bibitem{crestronSensors}
Crestron.
\newblock {Crestron Lightning and Environment Sensors}.
\newblock https://www.crestron.com/Products/Lighting-Environment/Sensors, 2020.
\newblock Online: Accessed 20-June-2020.

\bibitem{WaterLeak}
Control4.
\newblock {Keep Your House Dry, DAM-it!}
\newblock https://www.control4.com /blog/363/keep-your-house-dry-damit/, 2015.
\newblock Online: Accessed 20-June-2020.

\bibitem{lin2015maximizing}
Chi Lin, Guowei Wu, Chang~Wu Yu, and Lin Yao.
\newblock Maximizing destructiveness of node capture attack in wireless sensor
  networks.
\newblock {\em The Journal of Supercomputing}, 71(8):3181--3212, 2015.

\bibitem{davis2014visual}
Abe Davis, Michael Rubinstein, Neal Wadhwa, Gautham~J. Mysore, Fr\'{e}do
  Durand, and William~T. Freeman.
\newblock The visual microphone: Passive recovery of sound from video.
\newblock {\em ACM Trans. Graph.}, 33(4), July 2014.

\bibitem{sensorsurvey1}
Kai Xing, Shyaam Sundhar~Rajamadam Srinivasan, Major Jose, Jiang Li, Xiuzhen
  Cheng, et~al.
\newblock Attacks and countermeasures in sensor networks: a survey.
\newblock In {\em Network security}, pages 251--272. Springer, 2010.

\bibitem{sensorsurvey2}
P.~{Sinha}, V.~K. {Jha}, A.~K. {Rai}, and B.~{Bhushan}.
\newblock Security vulnerabilities, attacks and countermeasures in wireless
  sensor networks at various layers of osi reference model: A survey.
\newblock In {\em 2017 International Conference on Signal Processing and
  Communication (ICSPC)}, pages 288--293, 2017.

\bibitem{sensorsurvey3}
Furrakh Shahzad, Maruf Pasha, and Arslan Ahmad.
\newblock A survey of active attacks on wireless sensor networks and their
  countermeasures.
\newblock {\em arXiv preprint arXiv:1702.07136}, 2017.

\bibitem{sensorsurvey4}
Dr~G Padmavathi, Mrs Shanmugapriya, et~al.
\newblock A survey of attacks, security mechanisms and challenges in wireless
  sensor networks.
\newblock {\em arXiv preprint arXiv:0909.0576}, 2009.

\bibitem{sensorsurvey5}
David Martins and Herv{\'e} Guyennet.
\newblock Wireless sensor network attacks and security mechanisms: A short
  survey.
\newblock In {\em 2010 13th International Conference on Network-Based
  Information Systems}, pages 313--320. IEEE, 2010.

\bibitem{sensorsurvey6}
Shio~Kumar Singh, MP~Singh, and Dharmendra~K Singh.
\newblock A survey on network security and attack defense mechanism for
  wireless sensor networks.
\newblock {\em International Journal of Computer Trends and Technology},
  1(2):9--17, 2011.

\bibitem{sensorsurvey7}
Leela~Krishna Bysani and Ashok~Kumar Turuk.
\newblock A survey on selective forwarding attack in wireless sensor networks.
\newblock In {\em 2011 International Conference on Devices and Communications
  (ICDeCom)}, pages 1--5. IEEE, 2011.

\bibitem{sensorsurvey8}
Majid Meghdadi, Suat Ozdemir, and Inan G{\"u}ler.
\newblock A survey of wormhole-based attacks and their countermeasures in
  wireless sensor networks.
\newblock {\em IETE technical review}, 28(2):89--102, 2011.

\bibitem{sensorsurvey9}
K~Venkatraman, J~Vijay Daniel, and G~Murugaboopathi.
\newblock Various attacks in wireless sensor network: Survey.
\newblock {\em International Journal of Soft Computing and Engineering
  (IJSCE)}, 3(1):208--212, 2013.

\bibitem{sensorsurvey10}
Anthony~D Wood and John~A Stankovic.
\newblock A taxonomy for denial-of-service attacks in wireless sensor networks.
\newblock {\em Handbook of sensor networks: compact wireless and wired sensing
  systems}, pages 739--763, 2004.

\bibitem{sensorsurvey11}
David~R Raymond and Scott~F Midkiff.
\newblock Denial-of-service in wireless sensor networks: Attacks and defenses.
\newblock {\em IEEE Pervasive Computing}, 7(1):74--81, 2008.

\bibitem{sensorphysics}
Wenyuan~Xu Kevin~Fu.
\newblock Risks of trusting the physics of sensors.
\newblock
  hhttps://cacm.acm.org/opinion/articles/224627-risks-of-trusting-the-physics-of-sensors/fulltext,
  2018.
\newblock Online: Accessed 20-June-2019.

\bibitem{zhang2017dolphinattack}
Guoming Zhang, Chen Yan, Xiaoyu Ji, Tianchen Zhang, Taimin Zhang, and Wenyuan
  Xu.
\newblock Dolphinattack: Inaudible voice commands.
\newblock In {\em Proceedings of the 2017 ACM SIGSAC Conference on Computer and
  Communications Security}, pages 103--117, 2017.

\bibitem{uluagac2014sensory}
A~Selcuk Uluagac, Venkatachalam Subramanian, and Raheem Beyah.
\newblock Sensory channel threats to cyber physical systems: A wake-up call.
\newblock In {\em 2014 IEEE Conference on Communications and Network Security},
  pages 301--309. IEEE, 2014.

\bibitem{spreitzer2014pin}
Raphael Spreitzer.
\newblock Pin skimming: Exploiting the ambient-light sensor in mobile devices.
\newblock In {\em Proceedings of the 4th ACM Workshop on Security and Privacy
  in Smartphones \& Mobile Devices}, pages 51--62, 2014.

\bibitem{cai2012practicality}
Liang Cai and Hao Chen.
\newblock On the practicality of motion based keystroke inference attack.
\newblock In {\em International Conference on Trust and Trustworthy Computing},
  pages 273--290. Springer, 2012.

\bibitem{al2013keystrokes}
Ahmed Al-Haiqi, Mahamod Ismail, and Rosdiadee Nordin.
\newblock Keystrokes inference attack on android: A comparative evaluation of
  sensors and their fusion.
\newblock {\em Journal of ICT Research and Applications}, 7(2):117--136, 2013.

\bibitem{huang2019risk}
Yan Huang, Xin Guan, Hongyang Chen, Yi~Liang, Shanshan Yuan, and Tomoaki
  Ohtsuki.
\newblock Risk assessment of private information inference for motion sensor
  embedded iot devices.
\newblock {\em IEEE Transactions on Emerging Topics in Computational
  Intelligence}, 2019.

\bibitem{owusu2012accessory}
Emmanuel Owusu, Jun Han, Sauvik Das, Adrian Perrig, and Joy Zhang.
\newblock Accessory: password inference using accelerometers on smartphones.
\newblock In {\em Proceedings of the Twelfth Workshop on Mobile Computing
  Systems \& Applications}, pages 1--6, 2012.

\bibitem{marquardt2011sp}
Philip Marquardt, Arunabh Verma, Henry Carter, and Patrick Traynor.
\newblock (sp) iphone: Decoding vibrations from nearby keyboards using mobile
  phone accelerometers.
\newblock In {\em Proceedings of the 18th ACM conference on Computer and
  communications security}, pages 551--562, 2011.

\bibitem{narain2014single}
Sashank Narain, Amirali Sanatinia, and Guevara Noubir.
\newblock Single-stroke language-agnostic keylogging using stereo-microphones
  and domain specific machine learning.
\newblock In {\em Proceedings of the 2014 ACM conference on Security and
  privacy in wireless \& mobile networks}, pages 201--212, 2014.

\bibitem{lin2019motion}
Jessy Lin and Jason Seibel.
\newblock Motion-based side-channel attack on mobile keystrokes.
\newblock http://css.csail.mit.edu/6.858/2019/projects/lnj-jseibel.pdf, 2019.

\bibitem{xu2012taplogger}
Zhi Xu, Kun Bai, and Sencun Zhu.
\newblock Taplogger: Inferring user inputs on smartphone touchscreens using
  on-board motion sensors.
\newblock In {\em Proceedings of the fifth ACM conference on Security and
  Privacy in Wireless and Mobile Networks}, pages 113--124, 2012.

\bibitem{miluzzo2012tapprints}
Emiliano Miluzzo, Alexander Varshavsky, Suhrid Balakrishnan, and Romit~Roy
  Choudhury.
\newblock Tapprints: your finger taps have fingerprints.
\newblock In {\em Proceedings of the 10th international conference on Mobile
  systems, applications, and services}, pages 323--336, 2012.

\bibitem{nguyen2015using}
Trang Nguyen.
\newblock Using unrestricted mobile sensors to infer tapped and traced user
  inputs.
\newblock In {\em 2015 12th International Conference on Information
  Technology-New Generations}, pages 151--156. IEEE, 2015.

\bibitem{hodges2018reconstructing}
Duncan Hodges and Oliver Buckley.
\newblock Reconstructing what you said: Text inference using smartphone motion.
\newblock {\em IEEE Transactions on Mobile Computing}, 18(4):947--959, 2018.

\bibitem{liang2018deep}
Yi~Liang, Zhipeng Cai, Jiguo Yu, Qilong Han, and Yingshu Li.
\newblock Deep learning based inference of private information using embedded
  sensors in smart devices.
\newblock {\em IEEE Network}, 32(4):8--14, 2018.

\bibitem{roy2016Motors}
Nirupam Roy and Romit Roy~Choudhury.
\newblock Listening through a vibration motor.
\newblock In {\em Proceedings of the 14th Annual International Conference on
  Mobile Systems, Applications, and Services}, MobiSys '16, page 57–69, New
  York, NY, USA, 2016. Association for Computing Machinery.

\bibitem{vuagnoux2009compromising}
Martin Vuagnoux and Sylvain Pasini.
\newblock Compromising electromagnetic emanations of wired and wireless
  keyboards.
\newblock In {\em USENIX security symposium}, pages 1--16, 2009.

\bibitem{han2017pitchin}
Jun Han, Albert~Jin Chung, and Patrick Tague.
\newblock Pitchln: Eavesdropping via intelligible speech reconstruction using
  non-acoustic sensor fusion.
\newblock In {\em Proceedings of the 16th ACM/IEEE International Conference on
  Information Processing in Sensor Networks}, IPSN '17, page 181–192, New
  York, NY, USA, 2017. Association for Computing Machinery.

\bibitem{supplychaindefenses}
{European Union for Cybersecurity}.
\newblock {Guidelines for Securing the Internet of Things}.
\newblock
  https://www.enisa.europa.eu/publications/guidelines-for-securing-the-internet-of-things,
  {2020}.
\newblock Online: Accessed 20-November-2020.

\bibitem{ehrensvard2007tamper}
Jakob Ehrensv{\"a}rd, Stina Ehrensv{\"a}rd, Leif Eriksson, and Fredrik Einberg.
\newblock Tamper evident packaging, January~30 2007.
\newblock US Patent 7,170,409.

\bibitem{tamperevidentpack}
{Rob Helmke}.
\newblock {Tamper-Evident Packaging and Functionality}.
\newblock
  https://www.plasticingenuity.com/blog/tamper-evident-packaging-functionality,
  {2019}.
\newblock Online: Accessed 25-September-2020.

\bibitem{agrawal2007trojan}
Dakshi Agrawal, Selcuk Baktir, Deniz Karakoyunlu, Pankaj Rohatgi, and Berk
  Sunar.
\newblock Trojan detection using ic fingerprinting.
\newblock In {\em 2007 IEEE Symposium on Security and Privacy (SP'07)}, pages
  296--310. IEEE, 2007.

\bibitem{abramovici2009integrated}
Miron Abramovici and Paul Bradley.
\newblock Integrated circuit security: new threats and solutions.
\newblock In {\em Proceedings of the 5th Annual Workshop on Cyber Security and
  Information Intelligence Research: Cyber Security and Information
  Intelligence Challenges and Strategies}, pages 1--3, 2009.

\bibitem{chakraborty2008hardware}
Rajat~Subhra Chakraborty and Swarup Bhunia.
\newblock Hardware protection and authentication through netlist level
  obfuscation.
\newblock In {\em 2008 IEEE/ACM International Conference on Computer-Aided
  Design}, pages 674--677. IEEE, 2008.

\bibitem{chakraborty2009harpoon}
Rajat~Subhra Chakraborty and Swarup Bhunia.
\newblock Harpoon: an obfuscation-based soc design methodology for hardware
  protection.
\newblock {\em IEEE Transactions on Computer-Aided Design of Integrated
  Circuits and Systems}, 28(10):1493--1502, 2009.

\bibitem{chakraborty2009security}
Rajat~Subhra Chakraborty and Swarup Bhunia.
\newblock Security through obscurity: An approach for protecting register
  transfer level hardware ip.
\newblock In {\em 2009 IEEE International Workshop on Hardware-Oriented
  Security and Trust}, pages 96--99. IEEE, 2009.

\bibitem{control4TPRemove}
{Control4}.
\newblock 7in and 10in t3 series in-wall touch screen installation guide.
\newblock
  {https://www.manualslib.com/manual/1436607/Control-4-C4-Wall7-Wh.htmlf},
  2015.
\newblock Online: Accessed 22-September-2020.

\bibitem{petracca2015audroid}
Giuseppe Petracca, Yuqiong Sun, Trent Jaeger, and Ahmad Atamli.
\newblock Audroid: Preventing attacks on audio channels in mobile devices.
\newblock In {\em Proceedings of the 31st Annual Computer Security Applications
  Conference}, pages 181--190, 2015.

\bibitem{strikos2007full}
Andreas~A Strikos.
\newblock A full approach for intrusion detection in wireless sensor networks.
\newblock {\em School of Information and Communication Technology}, 2007.

\bibitem{ioannis2007towards}
Krontiris Ioannis, Tassos Dimitriou, and Felix~C Freiling.
\newblock Towards intrusion detection in wireless sensor networks.
\newblock In {\em Proc. of the 13th European Wireless Conference}, pages 1--10.
  Citeseer, 2007.

\bibitem{farooqi2013novel}
Ashfaq~Hussain Farooqi, Farrukh~Aslam Khan, Jin Wang, and Sungyoung Lee.
\newblock A novel intrusion detection framework for wireless sensor networks.
\newblock {\em Personal and ubiquitous computing}, 17(5):907--919, 2013.

\bibitem{pongaliur2008securing}
Kanthakumar Pongaliur, Zubin Abraham, Alex~X Liu, Li~Xiao, and Leo Kempel.
\newblock Securing sensor nodes against side channel attacks.
\newblock In {\em 2008 11th IEEE High Assurance Systems Engineering Symposium},
  pages 353--361. IEEE, 2008.

\bibitem{yu2008framework}
Zhenwei Yu and Jeffrey~JP Tsai.
\newblock A framework of machine learning based intrusion detection for
  wireless sensor networks.
\newblock In {\em 2008 IEEE International Conference on Sensor Networks,
  Ubiquitous, and Trustworthy Computing (sutc 2008)}, pages 272--279. IEEE,
  2008.

\bibitem{cresnetsniffer}
{Stephen Genusa}.
\newblock Crestron cresnet monitor.
\newblock {https://pushstack. wordpress.com/somfy-rts-protocol/}, 2015.
\newblock Online: Accessed 18-May-2020.

\bibitem{mays2017defending}
Caleb Mays, Mason Rice, Benjamin Ramsey, John Pecarina, and Barry Mullins.
\newblock Defending building automation systems using decoy networks.
\newblock In {\em International Conference on Critical Infrastructure
  Protection}, pages 297--317. Springer, 2017.

\bibitem{Volkova:2019}
A.~{Volkova}, M.~{Niedermeier}, R.~{Basmadjian}, and H.~{de Meer}.
\newblock Security challenges in control network protocols: A survey.
\newblock {\em IEEE Communications Surveys Tutorials}, 21(1):619--639, 2019.

\bibitem{BACNetDOC}
David~G. Holmberg.
\newblock Bacnet wide area network security threat assessment.
\newblock {\em NIST, Department of Commerce}, July 2003.

\bibitem{Reachable-Bacnet}
O.~Gasser, Q.~Scheitle, C.~Denis, N.~Schricker, and G.~Carle.
\newblock Security implications of publicly reachable building automation
  systems.
\newblock In {\em 2017 IEEE Security and Privacy Workshops (SPW)}, pages
  199--204, May 2017.

\bibitem{ASHRAE-BACNET}
{ASHRAE}.
\newblock {BACnet®, the ASHRAE building automation and control networking
  protocol}.
\newblock {http://www.bacnet.org/}, {2016}.

\bibitem{borisov2001intercepting}
Nikita Borisov, Ian Goldberg, and David Wagner.
\newblock Intercepting mobile communications: the insecurity of 802.11.
\newblock In {\em Proceedings of the 7th annual international conference on
  Mobile computing and networking}, pages 180--189, 2001.

\bibitem{lashkari2009survey}
Arash~Habibi Lashkari, Mir Mohammad~Seyed Danesh, and Behrang Samadi.
\newblock A survey on wireless security protocols (wep, wpa and wpa2/802.11 i).
\newblock In {\em 2009 2nd IEEE International Conference on Computer Science
  and Information Technology}, pages 48--52. IEEE, 2009.

\bibitem{ftcprotect}
{Federal Trade Commission}.
\newblock {Securing your Wireless Network}.
\newblock
  https://www.consumer.ftc.gov/articles/0013-securing-your-wireless-network,
  {2015}.
\newblock Online: Accessed 10-November-2020.

\bibitem{cisaprotect}
{Cybersecurity \& Infrastructure Security Agency}.
\newblock {Securing Wireless Networks}.
\newblock https://us-cert.cisa.gov/ncas/tips/ST05-003, {2020}.
\newblock Online: Accessed 10-November-2020.

\bibitem{khasawneh2014survey}
Mahmoud Khasawneh, Izadeen Kajman, Rashed Alkhudaidy, and Anwar Althubyani.
\newblock A survey on wi-fi protocols: Wpa and wpa2.
\newblock In {\em International Conference on Security in Computer Networks and
  Distributed Systems}, pages 496--511. Springer, 2014.

\bibitem{vanhoef2017key}
Mathy Vanhoef and Frank Piessens.
\newblock Key reinstallation attacks: Forcing nonce reuse in wpa2.
\newblock In {\em Proceedings of the 2017 ACM SIGSAC Conference on Computer and
  Communications Security}, pages 1313--1328, 2017.

\bibitem{wpahack}
{Kody}.
\newblock {Hack WPA \& WPA2 Wi-Fi Passwords with a Pixie-Dust Attack Using
  Airgeddon}.
\newblock
  https://null-byte.wonderhowto.com/how-to/hack-wpa-wpa2-wi-fi-passwords-with-pixie-dust-attack-using-airgeddon-0183556/,
  {2018}.
\newblock Online: Accessed 10-November-2020.

\bibitem{vanhoef2020dragonblood}
Mathy Vanhoef and Eyal Ronen.
\newblock Dragonblood: Analyzing the dragonfly handshake of wpa3 and eap-pwd.
\newblock In {\em Proceedings of the 2020 IEEE Symposium on Security and
  Privacy-S\&P 2020)}. IEEE, 2020.

\bibitem{kohlios2018comprehensive}
Christopher~P Kohlios and Thaier Hayajneh.
\newblock A comprehensive attack flow model and security analysis for wi-fi and
  wpa3.
\newblock {\em Electronics}, 7(11):284, 2018.

\bibitem{lounis2019bad}
Karim Lounis and Mohammad Zulkernine.
\newblock Bad-token: denial of service attacks on wpa3.
\newblock In {\em Proceedings of the 12th International Conference on Security
  of Information and Networks}, pages 1--8, 2019.

\bibitem{lounis2019wpa3}
Karim Lounis and Mohammad Zulkernine.
\newblock Wpa3 connection deprivation attacks.
\newblock In {\em International Conference on Risks and Security of Internet
  and Systems}, pages 164--176. Springer, 2019.

\bibitem{wang2010practical}
Ying Wang, Zhigang Jin, and Ximan Zhao.
\newblock Practical defense against wep and wpa-psk attack for wlan.
\newblock In {\em 2010 6th international conference on wireless communications
  networking and mobile computing (WiCOM)}, pages 1--4. IEEE, 2010.

\bibitem{lounis2020attacks}
Karim Lounis and Mohammad Zulkernine.
\newblock Attacks and defenses in short-range wireless technologies for iot.
\newblock {\em IEEE Access}, 8:88892--88932, 2020.

\bibitem{whitehurst2014exploring}
Lindsey~N Whitehurst, Todd~R Andel, and J~Todd McDonald.
\newblock Exploring security in zigbee networks.
\newblock In {\em Proceedings of the 9th Annual Cyber and Information Security
  Research Conference}, pages 25--28, 2014.

\bibitem{benzaid2016fast}
Chafika Benzaid, Karim Lounis, Ameer Al-Nemrat, Nadjib Badache, and Mamoun
  Alazab.
\newblock Fast authentication in wireless sensor networks.
\newblock {\em Future Generation Computer Systems}, 55:362--375, 2016.

\bibitem{ZwaveHowSecure}
M.~Knight.
\newblock Wireless security - how safe is z-wave?
\newblock {\em Computing Control Engineering Journal}, 17(6):18--23, Dec 2006.

\bibitem{IOTWireless}
R.~Krejčí, O.~Hujňák, and M.~Švepeš.
\newblock Security survey of the iot wireless protocols.
\newblock In {\em 2017 25th Telecommunication Forum (TELFOR)}, pages 1--4, Nov
  2017.

\bibitem{wang2013zigbee}
Jianfeng Wang.
\newblock Zigbee light link and its applicationss.
\newblock {\em IEEE Wireless Communications}, 20(4):6--7, 2013.

\bibitem{zillner2015zigbee}
Tobias Zillner and Sebastian Strobl.
\newblock Zigbee exploited: The good, the bad and the ugly.
\newblock {\em Black Hat--2015 https://www.blackhat.
  com/docs/us-15/materials/us-15-Zillner-ZigBee-Exploited-The-Good-The-Bad-And-The-Ugly.pdf},
  2015.

\bibitem{Zigbee-Ghost}
X.~Cao, D.~M. Shila, Y.~Cheng, Z.~Yang, Y.~Zhou, and J.~Chen.
\newblock Ghost-in-zigbee: Energy depletion attack on zigbee-based wireless
  networks.
\newblock {\em IEEE Internet of Things Journal}, 3(5):816--829, Oct 2016.

\bibitem{ZigBee-Nuclear}
E.~Ronen, A.~Shamir, A.~O. Weingarten, and C.~O’Flynn.
\newblock Iot goes nuclear: Creating a zigbee chain reaction.
\newblock {\em IEEE Security Privacy}, 16(1):54--62, January 2018.

\bibitem{Zigbee-PracticalAttacks}
O.~Olawumi, K.~Haataja, M.~Asikainen, N.~Vidgren, and P.~Toivanen.
\newblock Three practical attacks against zigbee security: Attack scenario
  definitions, practical experiments, countermeasures, and lessons learned.
\newblock In {\em 2014 14th International Conference on Hybrid Intelligent
  Systems}, pages 199--206, Dec 2014.

\bibitem{KillerBee}
{RiverLoopSec}.
\newblock {Framework and Tools for Attacking ZigBee and IEEE 802.15.4
  networks.}
\newblock https://github.com/riverloopsec/killerbee, {2017}.

\bibitem{fouladi2013security}
Behrang Fouladi and Sahand Ghanoun.
\newblock Security evaluation of the z-wave wireless protocol.
\newblock {\em Black Hat USA}, 24:1--2, 2013.

\bibitem{ZwaveRogue}
J.~D. Fuller and B.~W. Ramsey.
\newblock Rogue z-wave controllers: A persistent attack channel.
\newblock In {\em 2015 IEEE 40th Local Computer Networks Conference Workshops
  (LCN Workshops)}, pages 734--741, Oct 2015.

\bibitem{bluetoothdefense}
{Lynn Tan}.
\newblock {Protect against Bluetooth threats}.
\newblock https://www.zdnet .com/article/protect-against-bluetooth-threats/,
  {2007}.
\newblock Online: Accessed 10-November-2020.

\bibitem{lounis2019bluetooth}
Karim Lounis and Mohammad Zulkernine.
\newblock Bluetooth low energy makes “just works” not work.
\newblock In {\em 2019 3rd Cyber Security in Networking Conference (CSNet)},
  pages 99--106. IEEE, 2019.

\bibitem{shaked2005cracking}
Yaniv Shaked and Avishai Wool.
\newblock Cracking the bluetooth pin.
\newblock In {\em Proceedings of the 3rd international conference on Mobile
  systems, applications, and services}, pages 39--50, 2005.

\bibitem{darroudi2017bluetooth}
Seyed~Mahdi Darroudi and Carles Gomez.
\newblock Bluetooth low energy mesh networks: A survey.
\newblock {\em Sensors}, 17(7):1467, 2017.

\bibitem{minar2012bluetooth}
Nateq Be-Nazir~Ibn Minar and Mohammed Tarique.
\newblock Bluetooth security threats and solutions: a survey.
\newblock {\em International Journal of Distributed and Parallel Systems},
  3(1):127, 2012.

\bibitem{dunning2010taming}
John Dunning.
\newblock Taming the blue beast: A survey of bluetooth based threats.
\newblock {\em IEEE Security \& Privacy}, 8(2):20--27, 2010.

\bibitem{hypponen2007nino}
Konstantin Hypponen and Keijo~MJ Haataja.
\newblock “nino” man-in-the-middle attack on bluetooth secure simple
  pairing.
\newblock In {\em 2007 3rd IEEE/IFIP International Conference in Central Asia
  on Internet}, pages 1--5. IEEE, 2007.

\bibitem{sun2018man}
Da-Zhi Sun, Yi~Mu, and Willy Susilo.
\newblock Man-in-the-middle attacks on secure simple pairing in bluetooth
  standard v5. 0 and its countermeasure.
\newblock {\em Personal and Ubiquitous Computing}, 22(1):55--67, 2018.

\bibitem{haataja2010two}
Keijo Haataja and Pekka Toivanen.
\newblock Two practical man-in-the-middle attacks on bluetooth secure simple
  pairing and countermeasures.
\newblock {\em IEEE Transactions on Wireless Communications}, 9(1):384--392,
  2010.

\bibitem{haataja2008practical}
Keijo Haataja and Pekka Toivanen.
\newblock Practical man-in-the-middle attacks against bluetooth secure simple
  pairing.
\newblock In {\em 2008 4th International Conference on Wireless Communications,
  Networking and Mobile Computing}, pages 1--5. IEEE, 2008.

\bibitem{haataja2008man}
Keijo~MJ Haataja and Konstantin Hypponen.
\newblock Man-in-the-middle attacks on bluetooth: a comparative analysis, a
  novel attack, and countermeasures.
\newblock In {\em 2008 3rd International Symposium on Communications, Control
  and Signal Processing}, pages 1096--1102. IEEE, 2008.

\bibitem{barnickel2012implementing}
Johannes Barnickel, Jian Wang, and Ulrike Meyer.
\newblock Implementing an attack on bluetooth 2.1+ secure simple pairing in
  passkey entry mode.
\newblock In {\em 2012 IEEE 11th International Conference on Trust, Security
  and Privacy in Computing and Communications}, pages 17--24. IEEE, 2012.

\bibitem{hering2004bluetooth}
J~Hering.
\newblock Bluetooth cracking gun: Bluesniper.
\newblock https://www.def
  con.org/html/links/dc\_press/archives/12/esato\_bluetooth\\cracking.htm,
  2004.
\newblock Online: Accessed 11-Feb-2021.

\bibitem{spill2007bluesniff}
Dominic Spill and Andrea Bittau.
\newblock Bluesniff: Eve meets alice and bluetooth.
\newblock {\em WOOT}, 7:1--10, 2007.

\bibitem{lounis2018connection}
Karim Lounis and Mohammad Zulkernine.
\newblock Connection dumping vulnerability affecting bluetooth availability.
\newblock In {\em International Conference on Risks and Security of Internet
  and Systems}, pages 188--204. Springer, 2018.

\bibitem{alsaidi2018security}
Asma Alsaidi and Firdous Kausar.
\newblock Security attacks and countermeasures on cloud assisted iot
  applications.
\newblock In {\em 2018 IEEE International Conference on Smart Cloud
  (SmartCloud)}, pages 213--217. IEEE, 2018.

\bibitem{IRHacking}
{Admin}.
\newblock {How to hack into the TV remote control and understand the IR code}.
\newblock
  https://www.electrodragon.com/how-to-hack-into-the-tv-remote-control-and-understand-the-ir-code/,
  {2011}.
\newblock Online: Accessed 25-September-2020.

\bibitem{IRTrolling}
{Indrek Mandre}.
\newblock Interfacing with a sensor, ir remotes.
\newblock {http://www. mare.ee/indrek/irtroll/}, 2008.
\newblock Online: Accessed 18-May-2020.

\bibitem{jamming1}
S.~D. {Babar}, N.~R. {Prasad}, and R.~{Prasad}.
\newblock Jamming attack: Behavioral modelling and analysis.
\newblock In {\em Wireless VITAE 2013}, pages 1--5, 2013.

\bibitem{jamming2}
S.~M. {MirhoseiniNejad}, A.~{Rahmanpour}, and S.~M. {Razavizadeh}.
\newblock Phase jamming attack: A practical attack on physical layer-based key
  derivation.
\newblock In {\em 2018 15th International ISC (Iranian Society of Cryptology)
  Conference on Information Security and Cryptology (ISCISC)}, pages 1--4,
  2018.

\bibitem{jamming3}
A.~{Mpitziopoulos}, D.~{Gavalas}, C.~{Konstantopoulos}, and G.~{Pantziou}.
\newblock A survey on jamming attacks and countermeasures in wsns.
\newblock {\em IEEE Communications Surveys Tutorials}, 11(4):42--56, 2009.

\bibitem{jamming4}
Kanika Grover, Alvin Lim, and Qing Yang.
\newblock Jamming and anti--jamming techniques in wireless networks: a survey.
\newblock {\em International Journal of Ad Hoc and Ubiquitous Computing},
  17(4):197--215, 2014.

\bibitem{reactivejamming}
Matthias Wilhelm, Ivan Martinovic, Jens~B. Schmitt, and Vincent Lenders.
\newblock Short paper: Reactive jamming in wireless networks: How realistic is
  the threat?
\newblock In {\em Proceedings of the Fourth ACM Conference on Wireless Network
  Security}, WiSec '11, page 47–52, New York, NY, USA, 2011. Association for
  Computing Machinery.

\bibitem{jammingzb}
{Bastian Bloessl}.
\newblock Low-cost zigbee selective jamming.
\newblock {https://www. bastibl.net/reactive-zigbee-jamming/}, 2019.
\newblock Online: Accessed 18-July-2020.

\bibitem{jamming5}
Mingyan Li, Iordanis Koutsopoulos, and Radha Poovendran.
\newblock Optimal jamming attacks and network defense policies in wireless
  sensor networks.
\newblock In {\em IEEE INFOCOM 2007-26th IEEE International Conference on
  Computer Communications}, pages 1307--1315. IEEE, 2007.

\bibitem{jamminganalysis}
S.~D. {Babar}, N.~R. {Prasad}, and R.~{Prasad}.
\newblock Jamming attack: Behavioral modelling and analysis.
\newblock In {\em Wireless VITAE 2013}, pages 1--5, 2013.

\bibitem{jammingdetection}
V.~C. {Manju} and K.~M. {Sasi}.
\newblock Detection of jamming style dos attack in wireless sensor network.
\newblock In {\em 2012 2nd IEEE International Conference on Parallel,
  Distributed and Grid Computing}, pages 563--567, 2012.

\bibitem{jammingstatisticaldef}
Opeyemi Osanaiye, Attahiru~S Alfa, and Gerhard~P Hancke.
\newblock A statistical approach to detect jamming attacks in wireless sensor
  networks.
\newblock {\em Sensors}, 18(6):1691, 2018.

\bibitem{hdmi-walk}
Luis~Puche Rondon, Leonardo Babun, Kemal Akkaya, and A.~Selcuk Uluagac.
\newblock Hdmi-walk: Attacking hdmi distribution networks via consumer
  electronic control protocol.
\newblock In {\em 35th Annual Computer Security Applications Conference}, 2019.

\bibitem{nccgroup}
Andy Davis.
\newblock {What the HEC? Security implications of HDMI Ethernet Channel and
  other related protocols}.
\newblock
  https://www.nccgroup.trust/globalassets/our-research/uk/whitepapers/2013/44con\_hdmi\_ethernet\_channel\_andy\\\_davis
  \_ncc\_group\_wp.pdf, Aug, 2013.

\bibitem{smithcec}
Joshua Smith.
\newblock {High-Def Fuzzing : Exploring Vulnerabilities in HDMI-CEC}.
\newblock https://media.defcon.org/, Nov, 2015.

\bibitem{davisblackhat}
Andy Davis.
\newblock {HDMI : Hacking Displays Made Interesting}.
\newblock
  https://media.blackhat.com/bh-eu-12/Davis/bh-eu-12-Davis-HDMI-Slides.pdf,
  Mar, 2012.

\bibitem{hdmi-watch}
L.~C. {PucheRondon}, L.~{Babun}, K.~{Akkaya}, and A.~S. {Uluagac}.
\newblock Hdmi-watch: Smart intrusion detection system against hdmi attacks.
\newblock {\em IEEE Transactions on Network Science and Engineering}, pages
  1--1, 2020.

\bibitem{ipversions}
FS.
\newblock Ipv4 vs ipv6: What's the difference?
\newblock https://community.
  fs.com/blog/ipv4-vs-ipv6-whats-the-difference.html, 2018.
\newblock Online: Accessed 10-January-2020.

\bibitem{subnetting}
J.~Mogul and J.~Postel.
\newblock Internet standard subnetting procedure.
\newblock STD~5, RFC Editor, August 1985.
\newblock {http://www.rfc-editor.org/rfc/rfc950.txt}.

\bibitem{vlan}
NETGEAR Support.
\newblock What is a vlan?
\newblock 2020.
\newblock Online: Accessed 18-September-2020.

\bibitem{pakedgevlans}
{Pakedge}.
\newblock Pakedge zones.
\newblock {https://pakedge.com/technology/pakedge-zones.php}, 2020.
\newblock Online: Accessed 18-September-2020.

\bibitem{poweroverethernet}
Veracity.
\newblock Ipv4 vs ipv6: What's the difference?
\newblock https://www.
  veracityglobal.com/resources/articles-and-white-papers/poe-explained-part-1.aspx,
  2016.
\newblock Online: Accessed 10-January-2020.

\bibitem{receiverIP}
{Marantz}.
\newblock {AV Surround Receiver Web Manual}.
\newblock http://manuals. marantz.com/SR7009/EU/EN/HJWMSYmehwmguq.php, {2014}.
\newblock Online: Accessed 25-September-2020.

\bibitem{iptopologies}
Martin~W Murhammer, Kok-Keong Lee, Payam Motallebi, Paolo Borghi, and Karl
  Wozabal.
\newblock {\em IP Network Design Guide}.
\newblock IBM Corporation, 1999.

\bibitem{wifi}
{Intel}.
\newblock {Different Wi-Fi Protocols and Data Rates }.
\newblock https://www.
  intel.com/content/www/us/en/support/articles/000005725\\/network-and-i-o/wireless-networking.html,
  {2017}.

\bibitem{baek2004survey}
Kwang-Hyun Baek, Sean~W Smith, and David Kotz.
\newblock A survey of wpa and 802.11 i rsn authentication protocols.
\newblock 2004.

\bibitem{al2020analyzing}
Intisar~Shadeed Al-Mejibli and Nawaf~Rasheed Alharbe.
\newblock Analyzing and evaluating the security standards in wireless network:
  A review study.
\newblock {\em Iraqi Journal for Computers and Informatics}, 46(1):32--39,
  2020.

\bibitem{enterprisewifi}
{Li Yinghua}.
\newblock {What are the Differences Between Enterprise Wi-Fi and Home Wi-Fi? }.
\newblock
  https://e.huawei.com/en/eblog/enterprise-networking/wifi6/What-the-difference-between-corporate-Wi-Fi-and-home-Wi-Fi,
  {2018}.
\newblock Online: Accessed 10-November-2020.

\bibitem{Zwave}
{Z-Wave}.
\newblock {Safer, Smarter, Zwave}.
\newblock {http://www.z-wave.com/}, {2018}.

\bibitem{Zigbee}
{Zigbee Alliance}.
\newblock {Zigbee}.
\newblock {www.zigbee.org/}, {2018}.

\bibitem{ramya2011study}
C~Muthu Ramya, M~Shanmugaraj, and R~Prabakaran.
\newblock Study on zigbee technology.
\newblock In {\em 2011 3rd International Conference on Electronics Computer
  Technology}, volume~6, pages 297--301. IEEE, 2011.

\bibitem{zwaveguide}
{David Mead}.
\newblock {A Comprehensive Guide to Z-Wave}.
\newblock https:// linkdhome.com/articles/What-is-z-wave, {2020}.
\newblock Online: Accessed 25-October-2020.

\bibitem{bluetoothspec}
{Bluetooth}.
\newblock {Bluetooth Core Specifications}.
\newblock {https://www.bluetooth.com/
  specifications/bluetooth-core-specification/}, 2020.
\newblock Online: Accessed 1-March-2020.

\bibitem{bluetoothoperations}
{Bluetooth}.
\newblock {The Global Standard For Connection}.
\newblock {https://www.
  bluetooth.com/learn-about-bluetooth/bluetooth-technology/}, 2020.
\newblock Online: Accessed 1-March-2020.

\bibitem{savantbluetooth}
Savant.
\newblock Bulbs faq.
\newblock https://www.savant.com/bulbs-faq.
\newblock Online: Accessed 20-December-2019.

\bibitem{scarfone2008guide}
Karen Scarfone and John Padgette.
\newblock Guide to bluetooth security.
\newblock {\em NIST Special Publication}, 800(2008):121, 2008.

\bibitem{IRIntro}
Technopedia.
\newblock {Shipments of Products with HDMI Interface Nears 900 Million Devices
  in 2017; Total Installed Base Approaches Seven Billion}.
\newblock https://www.techopedia.com/definition/630/infrared-ir, 2020.
\newblock Online: Accessed 20-June-2020.

\bibitem{IRFlashers}
SnapAV.
\newblock {Episode® Electronics IR Flasher with LED Feedback}.
\newblock
  https://www.snapav.com/shop/en/snapav/episode-reg\%3B-electronics-ir-flasher-with-led-feedback,
  2020.
\newblock Online: Accessed 20-June-2020.

\bibitem{somfyRF}
Somfy.
\newblock Revolutionizing home comfort control: Radio technology somfy.
\newblock
  https://www.somfysystems.com/en-us/discover-somfy/technology/radio-technology-somfy.
\newblock Online: Accessed 10-February-2020.

\bibitem{lutronRF}
Rich Black.
\newblock Clear connect technology whitepaper.
\newblock http://www.
  lutron.com/TechnicalDocumentLibrary/Clear\_Connect\_Tech\\nology\_whitepaper.pdf,
  2020.
\newblock Online: Accessed 10-January-2020.

\bibitem{crestronRF}
Crestron.
\newblock infinet ex® network and er wireless gateway.
\newblock https://www.
  crestron.com/Products/Control-Hardware-Software/Wireless-Communications/Wireless-Gateways/CEN-GWEXER.
\newblock Online: Accessed 10-January-2020.

\bibitem{legrandRF}
Legrand.
\newblock Legrand® announces ultra-secure wireless lighting controls platform.
\newblock
  https://www.legrand.us/aboutus/press-room/news/legrand-announces-wireless-lighting-controls-platform.aspx,
  2019.
\newblock Online: Accessed 10-January-2020.

\bibitem{levitronRF}
Levitron.
\newblock Levnet rf™: Self-powered wireless built on reliability.
\newblock https://www.leviton.com/en/products/brands/levnet-rf.
\newblock Online: Accessed 10-January-2020.

\bibitem{RS232}
{fCoder}.
\newblock {History RS-232-C - Legacy Serial Connector}.
\newblock {https://www. lookrs232.com/rs232/history\_rs232.htm}, {2002}.

\bibitem{SAM-Module}
{Carrier Enterprise}.
\newblock {Carrier SAM Module}.
\newblock {https://www.carriere
  nterprise.com/carrier-infinity-series-ethernet-cat-5-wired-broadband-remote-access-module-systxccrct01\#tab-info},
  {2013}.

\bibitem{Cresnet}
{Cresnet}.
\newblock {Crestron Cresnet Monitor - Cresnet protocol analysis}.
\newblock {https://archive.codeplex.com/?p=cresnet}, {2017}.

\bibitem{Crestron}
{Crestron}.
\newblock {Crestron Automation Products}.
\newblock {https://www.crestron.com/}, {2017}.

\bibitem{litetouchmanual}
LiteTouch.
\newblock {LiteTouch Lighting Control Systems Installation and Troubleshooting
  Manual}.
\newblock http://sav-documentation.s3.amazonaws.
  com/Internal\%20Documentation/LiteTouch\%\\20and\%20Savant\%20Lighting/Troubleshooting\%20Manual.pdf,
  2006.
\newblock Online: Accessed 20-March-2020.

\bibitem{control4keypad}
Control4.
\newblock {Configurable Decora®Wired Keypad Installation Guide}.
\newblock
  https://www.control4.com/docs/product/wired-keypad/installation-guide/english/revision/B/wired-keypad-installation-guide-rev-b.pdf,
  2013.
\newblock Online: Accessed 20-March-2020.

\bibitem{savantsmartdeploy}
Savant.
\newblock {Savant SmartLightning Deployment Guide}.
\newblock
  https://sav-documentation.s3.amazonaws.com/Product\%20Deployment\%20Guides\\/SmartLightingControl\_DeploymentGuide.pdf,
  2014.
\newblock Online: Accessed 15-August-2020.

\bibitem{somfyrts}
Somfy.
\newblock {Control RTS Solutions with Most Automation Systems}.
\newblock
  https://www.somfysystems.com/en-us/products/1810872/universal-rts-interface,
  2020.
\newblock Online: Accessed 1-March-2020.

\bibitem{serialtv}
{Samsung}.
\newblock {RS-232 on Samsung TV's}.
\newblock https://www.samsung.com /us/support/troubleshooting/TSG01201603/,
  {2020}.
\newblock Online: Accessed 25-September-2020.

\bibitem{TI-BACNET}
{Texas Instruments}.
\newblock {Data communication protocol for control networks enabling automated
  buildings}.
\newblock {http://www.ti.com /lit/wp/spry266/spry266.pdf}, {2014}.

\bibitem{HDMIBil}
Steve Venuti.
\newblock {HDMI Interface Extends Exceptional Digital Quality with Single-Cable
  Simplicity to Over 4 Billion Consumer Devices}.
\newblock https://www.hdmi.org/press/press\_release.aspx?prid=137, Jan, 2015.
\newblock Online: Accessed 20-June-2020.

\bibitem{HDMIBil2}
Doug Wright.
\newblock {Shipments of Products with HDMI Interface Nears 900 Million Devices
  in 2017; Total Installed Base Approaches Seven Billion}, Jan, 2018.
\newblock Online: Accessed 20-June-2020.

\bibitem{hdmistandard}
Akihiro Tsutsui.
\newblock Latest trends in home networking technologies.
\newblock {\em IEICE transactions on communications}, 91(8):2470--2476, 2008.

\bibitem{HDMIPinout}
{HDMI Licensing LLC}.
\newblock {Inside an HDMI Cable}.
\newblock https://www.hdmi.org/ installers/insidehdmicable.aspx, 2018.
\newblock Online: Accessed 20-June-2020.

\bibitem{control4wha}
{Control4}.
\newblock {Control4 Solution Multi-Room Audio}.
\newblock https://www.control4. com/solutions/multi-room-audio/, {2020}.
\newblock Online: Accessed 25-September-2020.

\bibitem{crestronwha}
{Crestron}.
\newblock {Crestron Multiroom Audio}.
\newblock https://www.crestron.com /products/audio/multiroom-audio, {2020}.
\newblock Online: Accessed 25-September-2020.

\bibitem{savantwha}
{Savant}.
\newblock {Savant Whole Home Audio}.
\newblock https://www.savant.com/whole-home-audio, {2020}.
\newblock Online: Accessed 25-September-2020.

\bibitem{crestronCEC}
{Crestron}.
\newblock {Crestron Database Release Notes}.
\newblock http://www.crestron.com/
  release\_notes/crestron\_database\_31\_05.html, {2020}.
\newblock Online: Accessed 25-September-2020.

\bibitem{HDMISpec}
{HDMI Licensing LLC}.
\newblock {HDMI Specification}.
\newblock https://www.hdmi.org/ spec/index, Jun, 2009.

\bibitem{tradenames}
{Google}.
\newblock {What is CEC?}
\newblock {https://support.google.com/chromecast/answer/\\7199917?hl=en}, 2018.
\newblock Online: Accessed 10-January-2020.

\bibitem{ICS}
A.~Mirian, Z.~Ma, D.~Adrian, M.~Tischer, T.~Chuenchujit, T.~Yardley,
  R.~Berthier, J.~Mason, Z.~Durumeric, J.~A. Halderman, and M.~Bailey.
\newblock An internet-wide view of ics devices.
\newblock In {\em 2016 14th Annual Conference on Privacy, Security and Trust
  (PST)}, pages 96--103, Dec 2016.

\bibitem{Zmap}
Zakir Durumeric, Eric Wustrow, and J.~Alex Halderman.
\newblock Zmap: Fast internet-wide scanning and its security applications.
\newblock In {\em Proceedings of the 22Nd USENIX Conference on Security},
  SEC'13, pages 605--620, Berkeley, CA, USA, 2013. USENIX Association.

\bibitem{remotecentral}
{Remote Central}.
\newblock {Index of Remote Control File Areas}.
\newblock http://files.remotecentral.com/index.html, {2020}.
\newblock Online: Accessed 25-September-2020.

\bibitem{Dudak:2019}
J.~{Dudak}, G.~{Gaspar}, S.~{Sedivy}, P.~{Fabo}, L.~{Pepucha}, and
  P.~{Tanuska}.
\newblock Serial communication protocol with enhanced properties–securing
  communication layer for smart sensors applications.
\newblock {\em IEEE Sensors Journal}, 19(1):378--390, 2019.

\bibitem{Wilson:2018}
Paul~Lawrence Wilson.
\newblock {ModSec: A Secure Modbus Protocol}.
\newblock Master's thesis, Georgia Institute of Technology, 2018.

\bibitem{SSCP:2019}
Ieee standard for secure scada communications protocol (sscp).
\newblock {\em IEEE Std 1711.2-2019}, pages 1--37, 2020.

\bibitem{BacnetSecurity}
{ASHRAE}.
\newblock {BACnet Network Security Architecture}.
\newblock {http://www.bacnet .org/Addenda/Add-2004-135g-PR1.pdf}, {2017}.

\bibitem{sanatinia2013wireless}
Amirali Sanatinia, Sashank Narain, and Guevara Noubir.
\newblock Wireless spreading of wifi aps infections using wps flaws: An
  epidemiological and experimental study.
\newblock In {\em 2013 IEEE Conference on Communications and Network Security
  (CNS)}, pages 430--437. IEEE, 2013.

\bibitem{IRsee}
{Carolina Staff}.
\newblock {Make the Invisible Visible }.
\newblock https://www.carolina.
  com/knowledge/2020/02/20/make-the-invisible-visible, {2020}.
\newblock Online: Accessed 25-September-2020.

\bibitem{wood2007deejam}
Anthony~D Wood, John~A Stankovic, and Gang Zhou.
\newblock Deejam: Defeating energy-efficient jamming in ieee 802.15. 4-based
  wireless networks.
\newblock In {\em 2007 4th Annual IEEE Communications Society Conference on
  Sensor, Mesh and Ad Hoc Communications and Networks}, pages 60--69. IEEE,
  2007.

\bibitem{wood2003jam}
Anthony~D Wood, John~A Stankovic, and Sang~Hyuk Son.
\newblock Jam: A jammed-area mapping service for sensor networks.
\newblock In {\em RTSS 2003. 24th IEEE Real-Time Systems Symposium, 2003},
  pages 286--297. IEEE, 2003.

\bibitem{ceclessadapter}
{Amazon}.
\newblock Lindy hdmi cec-less adapter.
\newblock {https://www.amazon.com/
  Lindy-HDMI-Adapter-Female-41232/dp/B00DL48KVI}, 2020.
\newblock Online: Accessed 18-July-2020.

\bibitem{rondon2020poisonivy}
Luis~Puche Rondon, Leonardo Babun, Ahmet Aris, Kemal Akkaya, and A.~Selcuk
  Uluagac.
\newblock Poisonivy: (in)secure practices of enterprise iot systems in smart
  buildings, 2020.

\bibitem{LawshaeCrestron}
{Ricky Lawshae}.
\newblock {Who Controls the Controllers - Hacking Crestron IoT Automation
  Systems}.
\newblock https://av.tib.eu/media/39726, {2017}.
\newblock Online: Accessed 25-September-2020.

\bibitem{CVECrestron}
{CVE Details}.
\newblock {CVE Details Security Vulnerabilities Search: Crestron}.
\newblock
  https://www.cvedetails.com/vulnerability-list/vendor\_id-15891/Crestron.html,
  {2020}.
\newblock Online: Accessed 25-September-2020.

\bibitem{CVESavant}
{CVE Details}.
\newblock {CVE Details Security Vulnerabilities Search: Savant}.
\newblock
  https://www.cvedetails.com/vulnerability-list/vendor\_id-1231/Savant.html,
  {2020}.
\newblock Online: Accessed 25-September-2020.

\bibitem{presentationvulns}
{Lindsey O'Donnell}.
\newblock {Wireless presentation systems have an array of critical flaws.}
\newblock https://threatpost.com/bugs-wireless-presentation-systems/144318/,
  {2019}.
\newblock Online: Accessed 25-September-2020.

\bibitem{patchedvulns}
{Jacob Baines}.
\newblock {Eight Devices, One Exploit}.
\newblock
  https://medium.com/tenable-techblog/eight-devices-one-exploit-f5fc28c70a7c,
  {2019}.
\newblock Online: Accessed 25-September-2020.

\bibitem{chovulns}
{Paul Lilly}.
\newblock {Connected homes can be easy targets for hackers, says cybersecurity
  firm}.
\newblock https://www.techhive.com
  /article/2883246/connected-homes-can-be-easy-targets-for-hackers-says-cybersecurity-firm.html,
  {2015}.
\newblock Online: Accessed 25-September-2020.

\bibitem{rollingyourcrypto}
{Susan Morrow}.
\newblock {The Dangers of “Rolling Your Own” Encryption}.
\newblock
  https://resources.infosecinstitute.com/topic/the-dangers-of-rolling-your-own-encryption/,
  {2019}.
\newblock Online: Accessed 10-November-2020.

\bibitem{cybergibbonsdualcom}
{Andrew Tierney}.
\newblock {CSL Dualcom CS2300-R signalling unit vulnerabilities}.
\newblock
  https://cybergibbons.com/security-2/csl-dualcom-cs2300-signalling-unit-vulnerabilities/,
  {2015}.
\newblock Online: Accessed 10-November-2020.

\bibitem{miraibotnet}
Manos Antonakakis~Tim April, Michael Bailey, Matthew Bernhard, Elie Bursztein,
  Jaime Cochran, Zakir Durumeric, J~Alex Halderman, Luca Invernizzi, Michalis
  Kallitsis, Deepak Kumar, et~al.
\newblock Understanding the mirai botnet.
\newblock https://mbernhard.com/papers/mirai.pdf, 2017.

\bibitem{Dahua}
{Chris Brook}.
\newblock {Dahua Patching Backdoor in DVRs, IP Cameras}.
\newblock
  {https://threatpost.com/dahua-patching-backdoor-in-dvrs-ip-cameras/124119/},
  {2017}.

\bibitem{crestronSecurity}
{Crestron}.
\newblock {Crestron Residential Systems Security Reference Guide}.
\newblock
  https://www.crestron.com/getmedia/db0e7324-f7f7-4964-af90-af7a9b05d541/mg\_sr\_crestron-residential-systems,
  {2020}.
\newblock Online: Accessed 25-September-2020.

\bibitem{synacktips}
{Synack}.
\newblock {Home Automation Benchmarking Results}.
\newblock https://www.synack.com/blog/home-automation-benchmarking-results/,
  {2015}.
\newblock Online: Accessed 25-September-2020.

\bibitem{williamsportfor}
{Paul Williams}.
\newblock {Securing your Connected Life}.
\newblock https://www.control4. com/blog/113/securing-your-connected-life/,
  {2015}.
\newblock Online: Accessed 25-September-2020.

\bibitem{control4composerrelease}
{Control4}.
\newblock Control4 operating system os release notes.
\newblock {https://www.
  control4.com/files/dealers/TechDoc00046-ComposerProSoftware-Release-2.0.6-ReleaseNotes.pdf},
  2010.
\newblock Online: Accessed 20-June-2020.

\bibitem{crestronsoftware}
{Crestron}.
\newblock Crestron software - downloading latest versions.
\newblock {https://
  support.crestron.com/app/answers/detail/a\_id/32/~/crestron-software---downloading-latest-versions},
  2020.
\newblock Online: Accessed 22-September-2020.

\bibitem{whatisftp}
{Jon Martindale}.
\newblock What is ftp?
\newblock {https://www.digitaltrends.com/
  computing/what-is-ftp-and-how-do-i-use-it/}, 2020.
\newblock Online: Accessed 22-September-2020.

\bibitem{ftpsftp}
{Educba}.
\newblock Ftp vs sftp.
\newblock {https://www.educba.com/ftp-vs-sftp/}, 2020.
\newblock Online: Accessed 22-September-2020.

\bibitem{control4putty}
{Nitdroid}.
\newblock {How to Access Control4 through Putty}.
\newblock https://nitdroid.
  wordpress.com/2013/07/30/how-to-access-control4-through-putty/, {2013}.
\newblock Online: Accessed 25-September-2020.

\bibitem{putty}
{Simon Tatham}.
\newblock {PuTTY - a free SSH and telnet client for Windows}.
\newblock https://www.putty.org/, {2020}.
\newblock Online: Accessed 27-September-2020.

\bibitem{CCTVweb}
{Mark M.}
\newblock {Types of Remote Access for DVRs}.
\newblock http://polarisusa .com/articles/8/types-of-remote-access-for-dvrs,
  {2020}.
\newblock Online: Accessed 27-September-2020.

\bibitem{busybox}
{Eruc Andersen, Rob Landley, Denys Vlasenko}.
\newblock {BusyBox: The Swiss Army Knife of Embedded Linux}.
\newblock https://busybox.net/about.html, {2020}.
\newblock Online: Accessed 10-November-2020.

\bibitem{busyboxconfig}
Nicholas Wells.
\newblock Busybox: A swiss army knife for linux.
\newblock {\em Linux J.}, 2000(78es):10–es, October 2000.

\bibitem{control4busybox}
{John Ehringer}.
\newblock {Gaining Serial Console Access on the Control4 Mini Touch Screen}.
\newblock
  http://www.5khz.com/2014/07/22/gaining-serial-console-access-on-the-control4-mini-touch-screen/,
  {2014}.
\newblock Online: Accessed 10-November-2020.

\bibitem{control4sight}
{Control4}.
\newblock {Control4 4sight Services}.
\newblock https://www.control4.com /o/4sight-services, {2020}.
\newblock Online: Accessed 25-September-2020.

\bibitem{cctvsetup}
{Verkada}.
\newblock {Securing Your Video Surveillance Network}.
\newblock https://info.verkada.com/security/surveillance-network/, {2020}.
\newblock Online: Accessed 25-September-2020.

\bibitem{ransomwareWired}
{Andy Greenberg}.
\newblock {A Tesla Employee Thwarted an Alleged Ransomware Plot}.
\newblock https://www.wired.com/story/tesla-ransomware-insider-hack-attempt/,
  {2020}.
\newblock Online: Accessed 10-November-2020.

\bibitem{control4driversearch}
Control4.
\newblock Control4 driver search.
\newblock https://drivers.control4.com/solr/ drivers/browse.
\newblock Online: Accessed 20-June-2019.

\bibitem{Shodanio}
{Shodan.io}.
\newblock Shodan: Analyze the internet in seconds.
\newblock {https://www .shodan.io/}, 2020.
\newblock Online: Accessed 22-September-2020.

\bibitem{de2019cyber}
Michele De~Donno, Alberto Giaretta, Nicola Dragoni, Antonio Bucchiarone, and
  Manuel Mazzara.
\newblock Cyber-storms come from clouds: Security of cloud computing in the iot
  era.
\newblock {\em Future Internet}, 11(6):127, 2019.

\bibitem{liu2015survey}
Yuhong Liu, Yan~Lindsay Sun, Jungwoo Ryoo, Syed Rizvi, and Athanasios~V
  Vasilakos.
\newblock A survey of security and privacy challenges in cloud computing:
  solutions and future directions.
\newblock {\em Journal of Computing Science and Engineering}, 9(3):119--133,
  2015.

\bibitem{ryan2013cloud}
Mark~D Ryan.
\newblock Cloud computing security: The scientific challenge, and a survey of
  solutions.
\newblock {\em Journal of Systems and Software}, 86(9):2263--2268, 2013.

\bibitem{shahzad2014state}
Farrukh Shahzad.
\newblock State-of-the-art survey on cloud computing security challenges,
  approaches and solutions.
\newblock {\em Procedia Computer Science}, 37:357--362, 2014.

\bibitem{subashini2011survey}
Subashini Subashini and Veeraruna Kavitha.
\newblock A survey on security issues in service delivery models of cloud
  computing.
\newblock {\em Journal of network and computer applications}, 34(1):1--11,
  2011.

\bibitem{grobauer2010understanding}
Bernd Grobauer, Tobias Walloschek, and Elmar Stocker.
\newblock Understanding cloud computing vulnerabilities.
\newblock {\em IEEE Security \& privacy}, 9(2):50--57, 2010.

\bibitem{modi2013survey}
Chirag Modi, Dhiren Patel, Bhavesh Borisaniya, Avi Patel, and Muttukrishnan
  Rajarajan.
\newblock A survey on security issues and solutions at different layers of
  cloud computing.
\newblock {\em The journal of supercomputing}, 63(2):561--592, 2013.

\bibitem{singh2016survey}
Saurabh Singh, Young-Sik Jeong, and Jong~Hyuk Park.
\newblock A survey on cloud computing security: Issues, threats, and solutions.
\newblock {\em Journal of Network and Computer Applications}, 75:200--222,
  2016.

\bibitem{fernandes2014security}
Diogo~AB Fernandes, Liliana~FB Soares, Jo{\~a}o~V Gomes, M{\'a}rio~M Freire,
  and Pedro~RM In{\'a}cio.
\newblock Security issues in cloud environments: a survey.
\newblock {\em International Journal of Information Security}, 13(2):113--170,
  2014.

\bibitem{polash2014survey}
Fahad Polash, Abdullah Abuhussein, and Sajjan Shiva.
\newblock A survey of cloud computing taxonomies: Rationale and overview.
\newblock In {\em The 9th International Conference for Internet Technology and
  Secured Transactions (ICITST-2014)}, pages 459--465. IEEE, 2014.

\bibitem{singh2017cloud}
Ashish Singh and Kakali Chatterjee.
\newblock Cloud security issues and challenges: A survey.
\newblock {\em Journal of Network and Computer Applications}, 79:88--115, 2017.

\bibitem{xiao2012security}
Zhifeng Xiao and Yang Xiao.
\newblock Security and privacy in cloud computing.
\newblock {\em IEEE communications surveys \& tutorials}, 15(2):843--859, 2012.

\bibitem{ardagna2015security}
Claudio~A Ardagna, Rasool Asal, Ernesto Damiani, and Quang~Hieu Vu.
\newblock From security to assurance in the cloud: A survey.
\newblock {\em ACM Computing Surveys (CSUR)}, 48(1):1--50, 2015.

\bibitem{hashizume2013analysis}
Keiko Hashizume, David~G Rosado, Eduardo Fern{\'a}ndez-Medina, and Eduardo~B
  Fernandez.
\newblock An analysis of security issues for cloud computing.
\newblock {\em Journal of internet services and applications}, 4(1):5, 2013.

\bibitem{kumar2019cloud}
Rakesh Kumar and Rinkaj Goyal.
\newblock On cloud security requirements, threats, vulnerabilities and
  countermeasures: A survey.
\newblock {\em Computer Science Review}, 33:1--48, 2019.

\bibitem{dvrnvr}
{Swann Security}.
\newblock Nvr vs. dvr – what’s the difference?
\newblock {https://www.swann.com/blog/dvr-vs-nvr-whats-the-difference/}, 2018.
\newblock Online: Accessed 18-July-2020.

\bibitem{humanact1}
K.~K. {Htike}, O.~O. {Khalifa}, H.~A. {Mohd Ramli}, and M.~A.~M. {Abushariah}.
\newblock Human activity recognition for video surveillance using sequences of
  postures.
\newblock In {\em The Third International Conference on e-Technologies and
  Networks for Development (ICeND2014)}, pages 79--82, 2014.

\bibitem{humanact2}
M.~{Babiker}, O.~O. {Khalifa}, K.~K. {Htike}, A.~{Hassan}, and M.~{Zaharadeen}.
\newblock Automated daily human activity recognition for video surveillance
  using neural network.
\newblock In {\em 2017 IEEE 4th International Conference on Smart
  Instrumentation, Measurement and Application (ICSIMA)}, pages 1--5, 2017.

\bibitem{pakedgebakpak}
{Pakedge}.
\newblock Bakpak remote management \& monitoring.
\newblock {https://pakedge .com/bakpak/}, 2020.
\newblock Online: Accessed 18-July-2020.

\bibitem{twofactorcloud}
{Ed Bott}.
\newblock {Make your cloud safer: How you can use two-factor authentication to
  protect cloud services}.
\newblock
  https://www.zdnet.com/article/make-your-cloud-safer-how-you-can-use-two-factor-authentication-to-protect-cloud-services/,
  {2017}.
\newblock Online: Accessed 20-November-2020.

\bibitem{acar2018survey}
Abbas Acar, Hidayet Aksu, A~Selcuk Uluagac, and Mauro Conti.
\newblock A survey on homomorphic encryption schemes: Theory and
  implementation.
\newblock {\em ACM Computing Surveys (CSUR)}, 51(4):1--35, 2018.

\bibitem{oxfordfallout}
Oxford Analytica.
\newblock Fallout of solarwinds hack could last for years.
\newblock {\em Emerald Expert Briefings}, (oxan-es).

\bibitem{solarwinds}
{Threat Intelligence Team}.
\newblock {SolarWinds advanced cyberattack: What happened and what to do now}.
\newblock
  https://blog.malwarebytes.com/threat-analysis/2020/12/advanced-cyber-attack-hits-private-and-public-sector-via-supply-chain-software-update/,
  {2020}.
\newblock Online: Accessed 10-January-2021.

\bibitem{litchfield2016rethinking}
Samuel Litchfield, David Formby, Jonathan Rogers, Sakis Meliopoulos, and Raheem
  Beyah.
\newblock Rethinking the honeypot for cyber-physical systems.
\newblock {\em IEEE Internet Computing}, 20(5):9--17, 2016.

\bibitem{honeynetProject}
The honeynet project.
\newblock {https://www.honeynet.org}, 2020.
\newblock Online: Accessed 25-September-2020.

\bibitem{compot}
{Lukas {Rist}, Johnny {Vesterngaard}, Daniel {Haslinger}, {Andrea Pasquale},
  and John {Smith}}.
\newblock Conpot ics/scada honeypot.
\newblock {https://www.compot.org}, 2020.
\newblock Online: Accessed 11-September-2020.

\bibitem{ammar2018internet}
Mahmoud Ammar, Giovanni Russello, and Bruno Crispo.
\newblock Internet of things: A survey on the security of iot frameworks.
\newblock {\em Journal of Information Security and Applications}, 38:8--27,
  2018.

\bibitem{lin2017survey}
Jie Lin, Wei Yu, Nan Zhang, Xinyu Yang, Hanlin Zhang, and Wei Zhao.
\newblock A survey on internet of things: Architecture, enabling technologies,
  security and privacy, and applications.
\newblock {\em IEEE Internet of Things Journal}, 4(5):1125--1142, 2017.

\bibitem{alaba2017internet}
Fadele~Ayotunde Alaba, Mazliza Othman, Ibrahim Abaker~Targio Hashem, and Faiz
  Alotaibi.
\newblock Internet of things security: A survey.
\newblock {\em Journal of Network and Computer Applications}, 88:10--28, 2017.

\bibitem{hassan2019current}
Wan~Haslina Hassan et~al.
\newblock Current research on internet of things (iot) security: A survey.
\newblock {\em Computer networks}, 148:283--294, 2019.

\bibitem{hassija2019survey}
Vikas Hassija, Vinay Chamola, Vikas Saxena, Divyansh Jain, Pranav Goyal, and
  Biplab Sikdar.
\newblock A survey on iot security: application areas, security threats, and
  solution architectures.
\newblock {\em IEEE Access}, 7:82721--82743, 2019.

\bibitem{oracevic2017security}
Alma Oracevic, Selma Dilek, and Suat Ozdemir.
\newblock Security in internet of things: A survey.
\newblock In {\em 2017 International Symposium on Networks, Computers and
  Communications (ISNCC)}, pages 1--6. IEEE, 2017.

\bibitem{deogirikar2017security}
Jyoti Deogirikar and Amarsinh Vidhate.
\newblock Security attacks in iot: A survey.
\newblock In {\em 2017 International Conference on I-SMAC (IoT in Social,
  Mobile, Analytics and Cloud)(I-SMAC)}, pages 32--37. IEEE, 2017.

\bibitem{balte2015security}
Ashvini Balte, Asmita Kashid, and Balaji Patil.
\newblock Security issues in internet of things (iot): A survey.
\newblock {\em International Journal of Advanced Research in Computer Science
  and Software Engineering}, 5(4), 2015.

\bibitem{zhao2013survey}
Kai Zhao and Lina Ge.
\newblock A survey on the internet of things security.
\newblock In {\em 2013 Ninth international conference on computational
  intelligence and security}, pages 663--667. IEEE, 2013.

\bibitem{kraijak2015survey}
Surapon Kraijak and Panwit Tuwanut.
\newblock A survey on iot architectures, protocols, applications, security,
  privacy, real-world implementation and future trends.
\newblock 2015.

\bibitem{yang2017survey}
Yuchen Yang, Longfei Wu, Guisheng Yin, Lijie Li, and Hongbin Zhao.
\newblock A survey on security and privacy issues in internet-of-things.
\newblock {\em IEEE Internet of Things Journal}, 4(5):1250--1258, 2017.

\bibitem{pawar2016survey}
Ankush~B Pawar and Shashikant Ghumbre.
\newblock A survey on iot applications, security challenges and counter
  measures.
\newblock In {\em 2016 International Conference on Computing, Analytics and
  Security Trends (CAST)}, pages 294--299. IEEE, 2016.

\end{thebibliography}

\end{document}